\documentclass[12pt]{article}
\pdfoutput=1
\usepackage{graphicx}
\usepackage[top=2cm,bottom=3cm,left=2cm,right=2cm]{geometry}
\usepackage{amsmath}
\usepackage{amssymb} 
\usepackage{amsthm}
\usepackage{stmaryrd}
\usepackage{hyperref}
\usepackage{wrapfig}
\usepackage{array}
\usepackage{mathtools}
\usepackage[mathscr]{euscript}
\usepackage{bm}

\numberwithin{equation}{section}
\numberwithin{figure}{section}

\def\eq#1{(\ref{eq:#1})}
\def\lineup{\!\!\!\!\!\!\!\! &&}

\newcommand{\Tr}{\mathop{\rm Tr}\nolimits}

\def\d{\partial}
\def\eps{\epsilon}

\def\tv{\mathrm{tv}}
\def\Sigmabar{\overline{\Sigma}}
\def\sigmabar{\overline{\sigma}}

\def\BCFT{\mathrm{BCFT}}
\def\H{\mathcal{H}}
\def\pre{\mathrm{pre}}
\def\flag#1{| #1)}
\def\Aflag#1{(#1|}

\def\2simp{\mathrm{2simp}}

\def\flg{\mathrm{flag}}
\def\pole{\mathrm{pole}}
\def\out{\mathrm{0}}
\def\inn{\mathrm{*}}
\def\matter{\mathrm{matter}}

\def\gd{\mathrm{gd}}

\def\arrowhead{\mathrm{arrowhead}}
\def\flagantiflag{\text{flag-anti-flag}}

\newtheorem*{SFCP}{Fock space coefficient principle}

\def\bs{\overline \sigma}
\def\s{\sigma}

\begin{document}
\begin{titlepage}
\rightline\today

\begin{center}
\vskip 3.5cm

{\large \bf{String Field Theory Solution \\ for Any Open String Background. II.}}

\vskip 1.0cm

{\large {Theodore Erler${^{(a,b)}}$\footnote{tchovi@gmail.com}, Carlo Maccaferri$^{(c)}$\footnote{maccafer@gmail.com}}}

\vskip 1.0cm

$^{(a)}${\it Institute of Physics of the ASCR, v.v.i.}\\
{\it Na Slovance 2, 182 21 Prague 8, Czech Republic}\\
\vskip .5cm
$^{(b)}${\it Institute of Mathematics, the Czech Academy of Sciences}\\ {\it Zinta 25, 11567 Prague 1, Czech Republic}\\
\vskip .5cm
$^{(c)}${\it Dipartimento di Fisica, Universit\'a di Torino, \\INFN  Sezione di Torino and Arnold-Regge Center\\
Via Pietro Giuria 1, I-10125 Torino, Italy}\\

\vskip 2.0cm

{\bf Abstract} 

\end{center}

Generalizing previous work, we give a new analytic solution in Witten's open bosonic string field theory which can describe any open string background. The central idea is to use Riemann surface degenerations as a mechanism for taming OPE singularities. This requires leaving the familiar subalgebra of wedge states with insertions, but the payoff is that the solution makes no assumptions about the reference and target D-brane systems, and is therefore truly general. For example, unlike in previous work, the solution can describe time dependent backgrounds and multiple copies of the reference D-brane within the universal sector. The construction also resolves some subtle issues resulting from associativity anomalies, giving a more complete understanding of the relation between the degrees of freedom of different D-brane systems, and a nonperturbative proof of background independence in classical open bosonic string field theory.

\end{titlepage}

\tableofcontents
 
\section{Motivation and Main Idea}

One of the major questions in string field theory is how much of the full landscape of string theory can be seen from the point of view of strings propagating in a given background. A limited version of this question pertains to classical open bosonic string field theory: Can open bosonic strings attached to a given D-brane rearrange themselves to create other D-branes which share the same closed string background? Even this limited question has occupied a large part of research in string field theory for almost 20 years. Following the formulation of Sen's conjectures \cite{Sen}, much work was devoted to numerical solution of Witten's open bosonic string field theory in the level truncation scheme. This produced convincing evidence that string field theory on a reference D-brane could describe the tachyon vacuum \cite{SZtach,Taylortach,Rastelli,Sch_trunc}---the configuration where all D-branes have disappeared---as well as lower energy vacua, such as tachyon lump solutions \cite{MZlump}, and at least part of the moduli space of the reference D-brane \cite{SZmarg}. After Schnabl's exact solution for the tachyon vacuum \cite{Schnabl}, it was also possible to explore the space of open string field theory vacua using analytic techniques. However, analytic techniques appeared to be less flexible than level truncation. Aside from the tachyon vacuum, for a long time the only known analytic solutions represented parametrically small deformations of the reference D-brane \cite{Schnabl_marg,KORZ,FK,KO}. The upshot is that---in spite of other developments---the range of vacua that could be seen from the point of view of a reference D-brane has been essentially unchanged since early studies in level truncation. 

It was then a surprise when \cite{KOSsing}, following the work of \cite{KOS}, gave a plausible construction of an analytic solution for any time-independent configuration of D-branes. This includes solutions that had long been searched for without success: Solutions representing the whole D-brane moduli space; solutions describing magnetic flux \cite{flux}; solutions describing D-branes of higher dimension; solutions describing multiple copies of the reference D-brane. The construction was based on an expression of the form 
\begin{equation}\Psi_* = \Psi_\tv-\Sigma \Psi_\tv\Sigmabar.\label{eq:solution}\end{equation}
This is a classical solution in the string field theory of a reference D-brane, given by a boundary conformal field theory $\mathrm{BCFT}_0$, describing the formation of a target D-brane in the same closed string background, given by a boundary conformal field theory $\mathrm{BCFT}_*$. The solution involves products of string fields living in different state spaces, as follows:
\begin{equation}
\setlength{\unitlength}{.25cm}
\begin{picture}(18,8)
\put(.25,6.75){$\Psi_* = \Psi_\tv-\Sigma \Psi_\tv\Sigmabar.$}
\put(1,5.9){\vector(0,-1){2.5}}
\put(.25,1.5){$\mathcal{H}_0$}
\put(5,5.9){\vector(0,-1){2.5}}
\put(4.25,1.5){$\mathcal{H}_0$}
\put(10.5,5.9){\oval(2,10)[bl]}
\put(10.5,.9){\vector(1,0){4}}
\put(15,.4){$\mathcal{H}_{0*}$}
\put(12.5,5.9){\oval(2,6)[bl]}
\put(12.5,2.9){\vector(1,0){2.5}}
\put(15.5,2.4){$\mathcal{H}_{*}$}
\put(14.3,5.9){\oval(2,2)[bl]}
\put(14.3,4.9){\vector(1,0){1}}
\put(15.7,4.4){$\mathcal{H}_{*0}$}
\end{picture}
\end{equation}
Here $\mathcal{H}_0$ is the state space of the reference D-brane $\mathrm{BCFT}_0$, $\mathcal{H}_*$ is the state space of the target D-brane $\mathrm{BCFT}_*$, and $\mathcal{H}_{0*}$ and $\mathcal{H}_{*0}$ are state spaces of stretched strings connecting the reference and target D-branes. The structure of the solution has a fairly straightforward interpretation. The first term
\begin{equation}\Psi_\tv\in\mathcal{H}_0\label{eq:1st}\end{equation}
is a solution for the tachyon vacuum in $\mathrm{BCFT}_0$, and represents annihilation of the reference D-brane through tachyon condensation. In the second term, the factor 
\begin{equation}-\Psi_\tv\in \mathcal{H}_*\label{eq:2nd}\end{equation}
is a solution which represents the creation of the target D-brane $\mathrm{BCFT}_*$ out of the tachyon vacuum. The string fields $\Sigma$ and $\Sigmabar$ do not alter the physical meaning of the second term in the solution, but serve a necessary mathematical function: They map a state of the target D-brane into a state of the reference D-brane, so that the two terms \eq{1st} and \eq{2nd} can be added to define a solution in the reference string field theory. The equations of motion are valid provided the following identities are satisfied:
\begin{eqnarray}
\lineup Q_{\Psi_\tv}\Sigma =0,\ \ \ \  Q_{\Psi_\tv}\Sigmabar = 0,\label{eq:QSigma} \\
\lineup\ \ \ \ \ \ \ \ \ \phantom{\Big{)}}\Sigmabar\Sigma = 1. \label{eq:SigmabarSigma}
\end{eqnarray}
We refer to $\Sigma$ and $\Sigmabar$ as {\it intertwining fields.} A solution of the general form \eq{solution} will be called an {\it intertwining solution}.

Since the intertwining fields represent stretched strings, they must be built from BCFT vacuum states containing boundary condition changing operators. For reasons we will review shortly, earlier work assumed that these boundary condition changing operators have regular OPE of the form:
\begin{equation}\lim_{x\to 0}\sigmabar(x)\sigma(0) = 1,\ \ \ \ x>0,\label{eq:nonsing}\end{equation}
where $\sigma,\sigmabar$ are boundary condition changing operators defining respectively $\Sigma,\Sigmabar$. This condition seems unnatural, since boundary condition changing operators typically have nonzero conformal weight, and their OPEs will have singularities. However, this difficulty can be avoided with a trick. Suppose we have boundary condition changing operators $\sigma_\mathrm{bare},\sigmabar_\mathrm{bare}$ connecting $\BCFT_0$ and $\BCFT_*$ which are primaries of weight $h$. Their OPE takes the form
\begin{equation}\sigmabar_\mathrm{bare}(x)\sigma_\mathrm{bare}(0)= \frac{1}{x^{2h}}+\mathrm{less\ singular},\ \ \ \ x>0.\label{eq:sigmabarsigmabare}\end{equation}
Let us further suppose that the reference and target D-brane systems are time-independent. Specifically, we assume that $\BCFT_0$ and $\BCFT_*$ share a common factor given by a noncompact, timelike free boson $X^0(z)$ subject to Neumann boundary conditions, and that $\sigma_\mathrm{bare},\sigmabar_\mathrm{bare}$ act as the identity operator in this factor of the BCFT. Then we can connect $\BCFT_0$ and $\BCFT_*$ using boundary condition changing operators satisfying \eq{nonsing} by defining
\begin{eqnarray}
\sigma(x)\lineup  = \sigma_\mathrm{bare}e^{i\sqrt{h}X^0}(x),\nonumber\\
\sigmabar(x)\lineup = \sigmabar_\mathrm{bare}e^{-i\sqrt{h}X^0}(x).\label{eq:sigmasigmabar}
\end{eqnarray}
The plane wave vertex operators $e^{\pm i\sqrt{h}X^0}$ generate a Wilson line deformation on the target D-brane given by a constant timelike gauge potential $A_0=\sqrt{h}$. However, since the time direction is noncompact, the Wilson line is not observable; in principle it can be removed by gauge transformation.

With boundary condition changing operators of this kind, it is possible to construct an intertwining solution relating any pair of time-independent backgrounds. However, the solution has two major drawbacks: 
\begin{itemize}
\item The solution does not readily generalize to time dependent backgrounds. Even for static backgrounds, the construction breaks Lorentz symmetry and gives expectation values to primaries in the $X^0$ BCFT which are physically irrelevant. Meanwhile, there is nothing in the general form of the intertwining solution which suggests that the time coordinate should play a special role. 
\item Some computations related to the solution are ambiguous due to the appearance of associativity anomalies. One can show that generically
\begin{eqnarray}
\Sigmabar\Sigma \lineup =1,\nonumber\\
\Sigma\Sigmabar \lineup = \mathrm{constant}\times 1. \label{eq:assanom}
\end{eqnarray}
where the constant in the second equation is given by the ratio of disk partition functions in the reference and target BCFTs. This implies that the triple product 
\begin{equation}\Sigma\Sigmabar\Sigma\end{equation}
is ambiguous; its value depends on the order in which the string fields are multiplied. This does not immediately undermine the validity of the solution, since ambiguous products do not appear in essential computations. However, it makes it difficult to meaningfully discuss the relation between open string degrees of freedom on different backgrounds, and, for technical reasons, it complicates generalization to superstring field theory~\cite{taming}. 
\end{itemize}
These difficulties are related to the requirement that $\sigma$ and $\sigmabar$ have regular OPE. In this paper we give a new realization of the solution where this assumption is not necessary. 

\begin{figure}
\begin{center}
\resizebox{5.5in}{1.4in}{\includegraphics{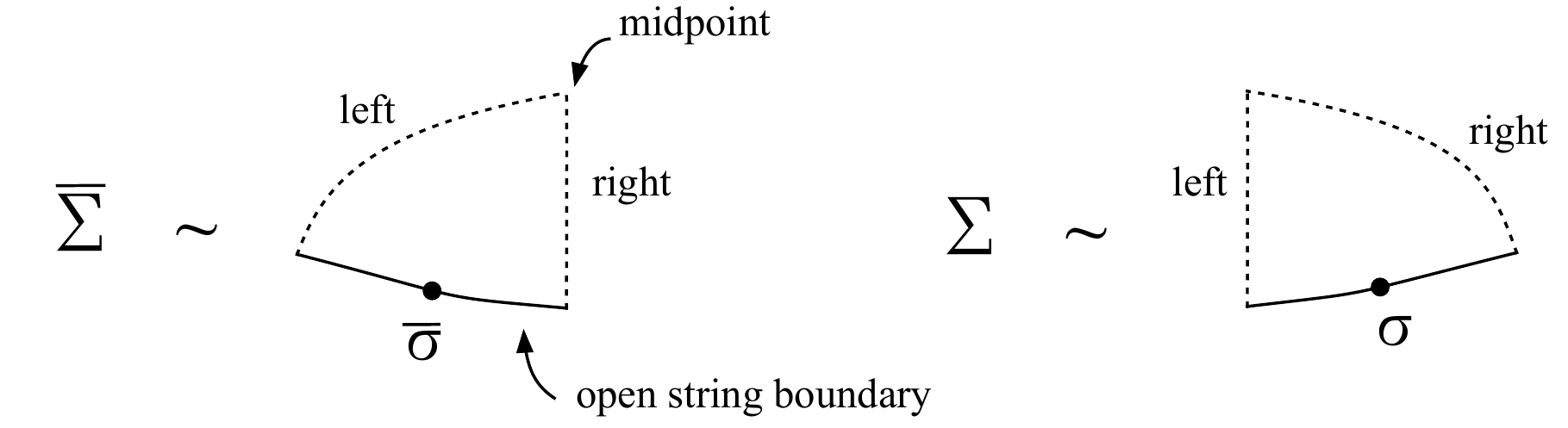}}
\end{center}
\caption{\label{fig:SSb} Schematic representation of $\Sigma$ and $\Sigmabar$ as portions of the open string worldsheet bounded by curves representing the left and right halves of the string.}
\end{figure}

\begin{figure}
\begin{center}
\resizebox{3.7in}{1.4in}{\includegraphics{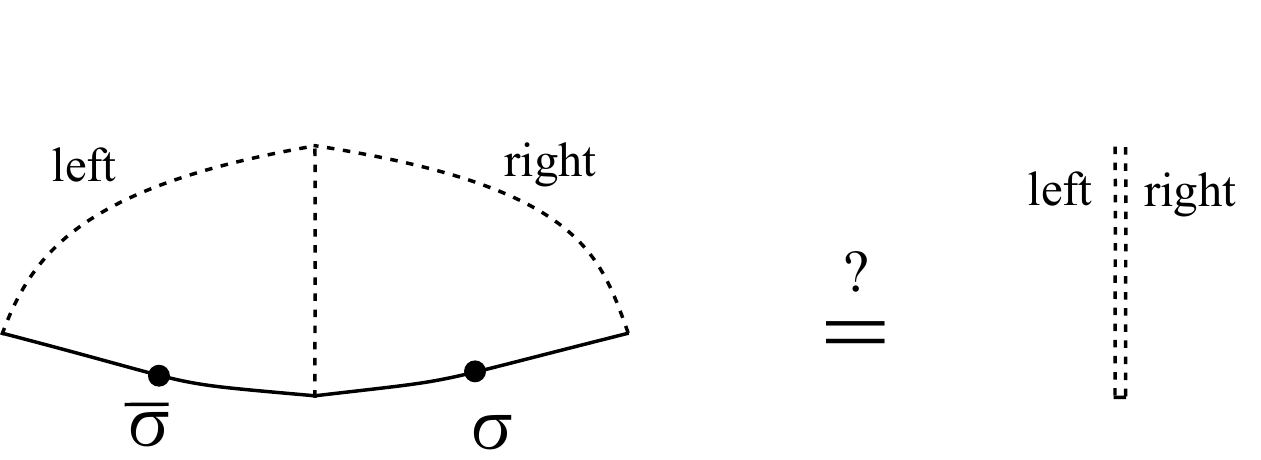}}
\end{center}
\caption{\label{fig:SSb1} Schematic representation of the identity $\Sigmabar\Sigma=1$. This relation appears to be possible only if  $\Sigma$ and $\Sigmabar$ contain no interior surface and if $\sigmabar$ and $\sigma$ have regular OPE as given in \eq{nonsing}.}
\end{figure}

To see the nature of the problem and our proposed resolution, it is useful to schematically visualize the intertwining fields as surface states containing insertions of the respective boundary condition changing operators. The surfaces are portions of the open string worldsheet bounded by two curves representing the left and right halves of the string, as shown in figure \ref{fig:SSb}. In actuality, the intertwining fields will be linear combinations of surface states with additional ghost insertions, but this simplified picture suffices for the present discussion. The product $\Sigmabar\Sigma$ is defined by a surface obtained by gluing the right curve of $\Sigmabar$ to the left curve of $\Sigma$, as shown in figure \ref{fig:SSb1}. On the other hand, we should have 
\begin{equation}\Sigmabar\Sigma = 1.\end{equation}
On the right hand side, the identity string field is characterized by an infinitely thin surface where the left and right curves overlap. Since gluing $\Sigmabar$ to $\Sigma$ should not produce more surface than is present in the identity string field, this seems to imply that $\Sigma$ and $\Sigmabar$ should likewise be infinitely thin surfaces where the left and right curves overlap. Then the product $\Sigmabar\Sigma$ necessarily produces a collision of boundary condition changing operators, and we obtain the identity string field only if their OPE is regular, as postulated in~\eq{nonsing}. 

\begin{figure}
\begin{center}
\resizebox{5.5in}{1.9in}{\includegraphics{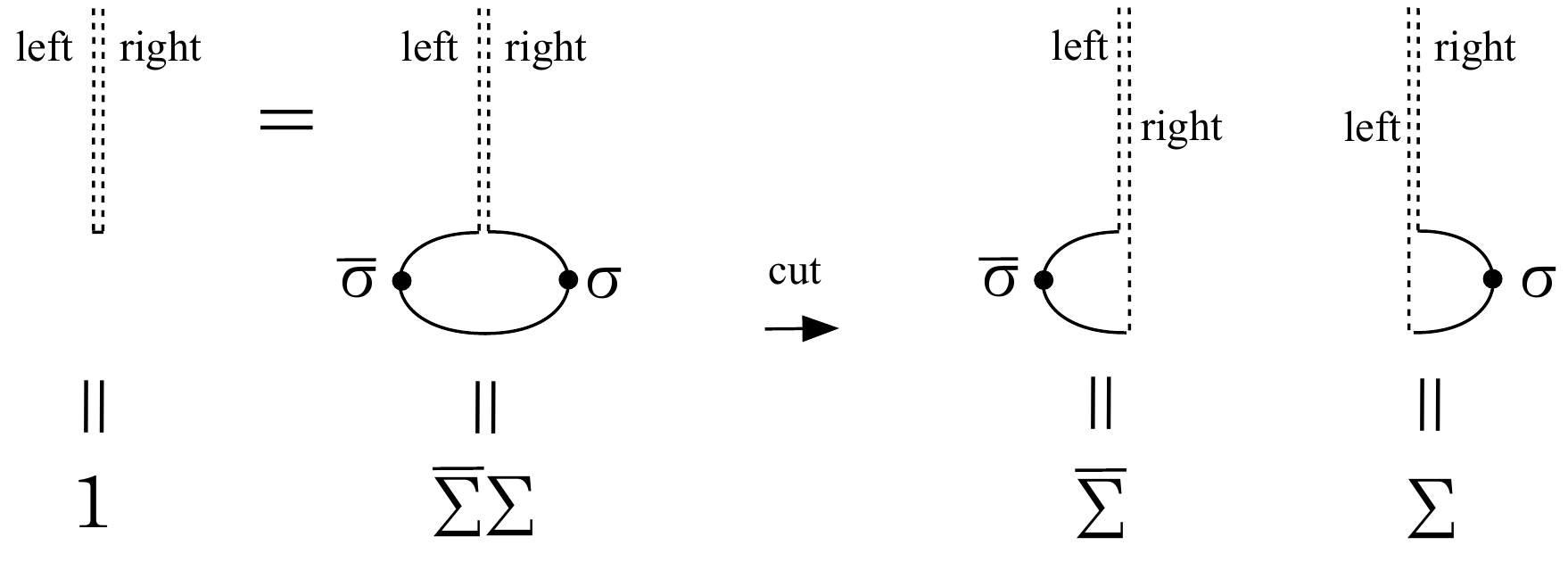}}
\end{center}
\caption{\label{fig:SSb2} The boundary point inside the surface of the identity string field can be replaced with an infinitely narrow neck connecting to a larger region containing $\sigma$ and $\sigmabar$ insertions. Cutting the surface in half gives a new representation of $\Sigmabar$ and $\Sigma$ in terms of ``flag states."}
\end{figure}

The basic tension is that the intertwining fields must contain some surface to avoid collision of boundary condition changing operators, but this surface must not be present after taking the product. The situation is reminiscent of old heuristic arguments against the existence of star algebra projectors \cite{projectors}: gluing surfaces always produces more surface, so it is difficult to imagine a surface state which squares to itself. However in \cite{projectors} it was realized that the star product can effectively destroy surface under a special circumstance: if it produces a surface with degeneration---an infinitely narrow ``neck" that cuts off a region of worldsheet. We can take advantage of this as follows. In the simplest visualization, the identity string field corresponds to a ``needle" of surface containing a single point on the open string boundary. However, we can replace this boundary point with an infinitely narrow neck connecting to a finite region of worldsheet, as shown in figure \ref{fig:SSb2}. In this region we can place two boundary condition changing operators at separated points, and provided that the resulting 2-point function is equal to $1$, we have not changed anything about the identity string field.
This suggests a new definition of $\Sigmabar$ and~$\Sigma$. We can cut between the left and right curves of the identity string field, through the infinitely narrow neck, and into the new surface to a boundary point between $\sigmabar$ and $\sigma$. This produces a pair of ``flag shaped" surfaces which we take to define $\Sigmabar$ and $\Sigma$. Due to the characteristic appearance of the surfaces, we will call them~{\it flag states}.\footnote{The notion of flag states was implicitly discussed in the final section of \cite{Ellwood}, in a technically different but conceptually related context.} 

\begin{figure}
\begin{center}
\resizebox{1.4in}{2in}{\includegraphics{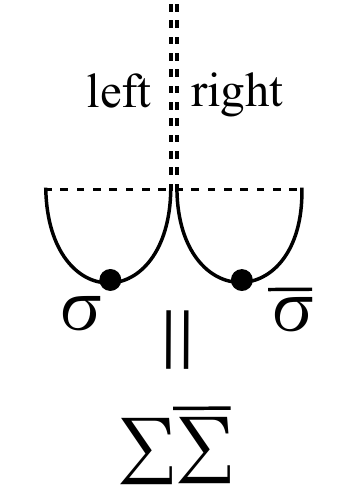}}
\end{center}
\caption{\label{fig:SSb3} Surface state representing the (reverse) product $\Sigma\Sigmabar$. Note that the boundary condition changing insertions are well-separated. This surface can be seen to define a star algebra projector following the arguments given in \cite{projectors}. }
\end{figure}

If the intertwining fields are constructed from flag states, it is clear that the star product $\Sigmabar\Sigma$ will not produce a collision of boundary condition changing operators. Then we do not require any assumption about the boundary condition changing operators aside from normalization; the Wilson line deformation is not necessary, and the construction works equally well for time dependent backgrounds. Perhaps as significant, we no longer have difficulty with associativity anomalies. The star product $\Sigma\Sigmabar$ (in the reverse order) is not proportional to the identity string field. Rather, it produces a nontrivial surface, as shown in figure \ref{fig:SSb3}. In fact $\Sigma\Sigmabar$ is a projector of the star algebra, so the previous inconsistent relations \eq{assanom} are replaced with
\begin{eqnarray}\Sigmabar\Sigma\lineup = 1,\nonumber\\
 (\Sigma\Sigmabar)^2 \lineup = \Sigma\Sigmabar\neq \mathrm{constant}\times 1.\end{eqnarray}
In this way, $\Sigma$ can be thought of as analogous to an isometry of a Hilbert space, while its conjugate $\Sigmabar$ is analogous to a partial isometry.\footnote{The possible relevance of non-unitary isometries to string field theory solutions was pointed out early on in~\cite{Martin}. An {\it isometry} is a linear operator which preserves inner products. A {\it partial isometry} acts as an isometry on the orthogonal complement of its kernel. The adjoint of an isometry is, in general, only a partial isometry. When the adjoint is also an isometry, the operator is {\it unitary}.}

This resolves the two main difficulties with the intertwining solution as realized in \cite{KOSsing}. As a result, we have a completely general construction of D-brane systems in bosonic string field theory, and further, we can establish that the fluctuations of any D-brane in bosonic string theory contain complete information about all other D-brane systems which share the same closed string background.  At a technical level, the flag state solution has a rather different character from previous analytic solutions, mainly due to the nontrivial geometry of the surfaces involved. Explicit calculations with the solution, for example of Fock space coefficients, are a much more serious undertaking. Much of this paper is concerned with describing the solution in a concrete enough form that computations and consistency checks are possible. However, the fundamental idea behind the solution is simple.

\subsection{Organization and Summary}

When proposing a new analytic solution, especially one which is different in essential respects from other solutions, it is necessary to spell out the procedures for explicit calculation in as far as possible. In the study of analytic solutions, the proverbial devil is often hiding in details. For this reason, the paper is long. But we do not wish for this to obscure essential points which may be of wider interest and utility. For those who wish to read more than the introduction and conclusion, but not too much about technical implementation, section \ref{sec:axioms} and \ref{sec:Fock} contain important general ideas and results. We set $\alpha'=1$ and use the left handed star product convention \cite{simple} throughout. In particular, the expectation value of the tachyon is positive at the tachyon vacuum.

The paper is organized as follows. 

In section \ref{sec:axioms} we describe the notion of {\it intertwining solution}, as characterized by a choice of tachyon vacuum in the reference and target backgrounds together with a pair of intertwining fields. We show that variation of the choice of tachyon vacuum and intertwining fields leaves the solution invariant up to gauge transformation provided that the boundary condition of the target background does not change. We also note that the structure of the intertwining solution is preserved by gauge transformation, which leads to the surprising implication that all classical solutions in open bosonic string field theory can be described as intertwining solutions. Next we show that the intertwining solution provides a map between the field variables of the reference and target string field theories which is invertible up to gauge transformation. If an intertwining solution can be found for all D-brane systems in a fixed closed string background, this establishes the background independence of classical open bosonic string field theory. Interestingly, the map between string field theories of different backgrounds is not an isomorphism of field variables, but transforms the open string state space of the target background into a subset of states of the reference background. We also comment on why the tachyon vacuum should be the preferred intermediate background for condensing the reference and target D-brane systems for the purposes of realizing the intertwining solution. 

In section \ref{sec:flags} we describe flag states as a step towards an explicit realization of the algebraic structure of the intertwining solution. There are many possible definitions of flag states which differ by the choice of parameterization of the left and right halves of the string. We fix a parameterization where flag states are characterized as surface states by a region in the sliver coordinate frame consisting of the positive imaginary axis conjoined to an semi-infinite horizontal strip of height $\ell$ above the positive (or negative) real axis. We describe flag states schematically in the split string formalism as operators on the vector space of half-string functionals, which gives a useful understanding of how flag states work and why the map between field variables in different backgrounds is not an isomorphism. To perform concrete calculations with the intertwining solution constructed from flag states, it is necessary to evaluate correlation functions on surfaces formed by gluing flag states and wedge states. The surface which contains only one flag state and its conjugate is called the {\it flag-anti-flag surface}, and in this case correlation functions can be computed by explicitly transforming to the upper half plane with the help of the Schwarz-Christoffel map. With this we show that a flag state with operator insertion multiplies with its conjugate to give the identity string field up to normalization. This is the crucial property needed to realize the intertwining solution.

In section \ref{sec:flag_sol} we give a construction of the intertwining solution using flag states. We describe the solution based on the simple tachyon vacuum \cite{simple} and a more general class of tachyon vacuum solutions of the Okawa form \cite{Okawa}. The primary task in the construction is building intertwining fields out of flag states. This is mostly straightforward. The main subtlety is showing that the string field $B$, corresponding to a vertical line integral insertion of the $b$-ghost in the sliver frame, is preserved by right/left multiplication with a flag state and its conjugate. We write the solution explicitly in terms of flag states and elements of the $KBc$ subalgebra. The case of the simple tachyon vacuum is somewhat special, since it is necessary to lift $c$ ghost insertions off the open string boundary to avoid certain ambiguities. We conclude this section by describing the relation to the solution of \cite{KOSsing}, and a prescription for constructing multiple D-brane solutions, including solutions representing multiple copies of the perturbative vacuum within the universal sector. We also make the curious observation that this construction represents all open string backgrounds by nonperturbative solutions. Even the perturbative vacuum is represented by a nonperturbative state which must be related to the trivial solution $\Psi=0$ by a finite gauge transformation.

In section \ref{sec:consistency} we investigate the consistency of the solution, in particular that the equations of motion hold without anomaly, that the solution is finite, and that gauge invariant observables produce the physically expected results. The most significant subtleties we find are the necessity of lifting $c$ ghost insertions off the open string boundary when using the simple tachyon vacuum, and the necessity of regularization to consistently cancel infrared divergences which appear when $c$ ghosts are cut off by degeneration in the product of intertwining fields. 

Finally, in section \ref{sec:Fock} we consider the coefficients of the solution when expanded into a basis of Fock states. Up to a shift from the tachyon vacuum, the coefficients are given by a canonically normalized 3-point function of two primary boundary condition changing operators and a primary probe vertex operator, multiplied by a universal factor which depends on conformal transformation properties of the test state and boundary condition changing operators which appear inside the solution. The structure generalizes that of the solution of \cite{KOS}. The universal factor is defined by a complicated seven dimensional integral, and we do not attempt its numerical evaluation here. Instead, we consider the coefficients of a ghost number zero toy model of the solution, where the analogous universal factor requires a more manageable three dimensional integral and is expected to give qualitatively similar results. In the context of the toy model, we first consider the Wilson line deformation, and confirm an old proposal \cite{ZwiebachToy} that the expectation value of the gauge field on the reference D-brane is not a global coordinate on the moduli space of the deformation. Second, we consider tachyon lumps, representing the formation of a D$(p-1)$-brane in the field theory of a D$p$-brane compactified on a circle of radius $R$. We analyze lump profiles derived from the toy model using Neumann-Dirichlet twist fields of weight $1/16$ and also excited twist fields of weight $9/16$. Superimposing these profiles gives a first look at the ``double lump" representing a coincident pair of D$(p-1)$-branes. Third, we consider the ghost number zero toy model of the cosh rolling tachyon deformation. The evolution of the tachyon field exhibits the wild oscillation which is typically seen in such solutions, and we find that the oscillation does not disappear when the marginal parameter is taken to the critical value which is supposed to represent placing the tachyon at the local minimum of the potential. However, at the critical marginal parameter a new direction in the moduli space opens up which corresponds to adjusting the imaginary time interval between Wick rotated D-branes. As the imaginary time interval is taken to infinity, the oscillations of the tachyon field disappear, and we are left with the tachyon vacuum. Motivated by the ghost number zero toy model and many other solutions which have been found both analytically and numerically, we propose that the qualitative behavior of coefficients of primary fields in a solution is in large part captured by a 3-point function in the upper half plane consisting of a test vertex operator inserted at the origin and two boundary condition changing operators inserted at opposing points some distance apart on the real axis. This observation is dubbed the {\it Fock space coefficient principle}. We give several supporting examples, and comment on a few exceptional features  which are not captured by the 3-point function. We also note that the 3-point function can sometimes show surprising simplification when the boundary condition changing operators are a certain distance apart; for example, the tachyon profile of a lump representing a D$(p-1)$-brane becomes a delta function at the location of the Dirichlet boundary condition.

We end with concluding remarks followed by appendices. Appendix \ref{app:Sch} shows how the solution for marginal deformations in Schnabl gauge \cite{Schnabl_marg,KORZ} can be expressed as an intertwining solution. Appendix \ref{app:conformal} summarizes a collection of relations pertaining to conformal transformation between the flag-anti-flag surface and the upper half plane. Appendix \ref{app:ghost} gives formulas for correlation functions of ghost operators on the flag-anti-flag surface which appear when multiplying flag states with elements of the $KBc$ subalgebra. Appendix \ref{app:lump} lists some coefficients for the tachyon lump profile derived from the ghost number 0 toy model. In appendix \ref{app:cosine} we derive the three point function of a plane wave vertex operator and two boundary condition changing operators which turn on the cosine deformation on a segment of the open string boundary.

\section{Intertwining Solution}
\label{sec:axioms}

In this section we define the intertwining solution based on a set of ``postulates." Essentially all physical properties of the target background follow from the postulates, without knowledge of how the solution is actually realized. Most things we might hope to learn from string field theory, however, require more. We describe our proposal to construct the solution in later sections. 

The intertwining solution is
\begin{equation}\Psi_* = \Psi_\tv-\Sigma\Psi_\tv\Sigmabar,\end{equation}
where the string fields on the right hand side are characterized by the following postulates:
\begin{description}
\item{{\bf 1) Tachyon Vacuum:}} \hypertarget{anc:1} There are two states, 
\begin{equation}
\Psi_\tv\in \H_0\ \ \ \mathrm{and}\ \ \ \Psi_\tv\in \H_*,
\end{equation}
which represent solutions for the tachyon vacuum respectively in the string field theory of the reference D-brane $\BCFT_0$, and the string field theory of the target D-brane $\BCFT_*$. We do not require that the two tachyon vacuum solutions are ``the same," in the sense that they generate the same coefficients when expanded in a universal basis. We do not even require that the tachyon vacuum solutions are universal.
\item{{\bf 2) Homotopy Operator:}} \hypertarget{anc:2} Each tachyon vacuum solution comes with a respective string field $A$ at ghost number $-1$, called the ``homotopy operator,"
\begin{equation}A\in \H_0\ \ \ \mathrm{and}\ \ \ A\in\H_*,\end{equation}
which satisfies\footnote{We define $Q_{\Psi}\Phi\equiv Q\Phi+\Psi \Phi-(-1)^\Phi\Phi \Psi$ where $\Psi$ is a Grassmann odd string field at ghost number 1. If $\Psi$ is a solution, $Q_\Psi$ is nilpotent.}
\begin{equation}Q_{\Psi_\tv}A = 1\ \ \ \mathrm{and}\ \ \ A^2=0\ \ \ \ \  (\mathrm{in}\ \H_0\ \mathrm{or}\ \H_*).\end{equation}
The existence of a homotopy operator implies that the cohomology of $Q_{\Psi_\tv}$ is empty, so there are no physical open string states around the tachyon vacuum \cite{EllwoodIdentity,EllwoodSchnabl}. We also assume that the homotopy operator squares to zero. There is no particular physical rationale behind this assumption, but nevertheless it is needed. The homotopy operator squares to zero for known analytic solutions for the tachyon vacuum.\footnote{This is true for the trivial reason that the subalgebras of states used to construct most analytic solutions do not contain states with ghost number less than $-1$.}
\item{{\bf 3) Intertwining Fields:}} \hypertarget{anc:3} We have two stretched string states,
\begin{equation}\Sigma\in\H_{0*}\ \ \ \mathrm{and}\ \ \ \Sigmabar\in \H_{*0},\end{equation}
called {\it intertwining fields}. They are assumed to satisfy
\begin{equation}
Q_{\Psi_\tv}\Sigma  = 0, \ \ \ Q_{\Psi_\tv}\Sigmabar = 0,\phantom{\Big{(}},\label{eq:QS}\\
\end{equation}
and 
\begin{equation}\Sigmabar A\Sigma= A. \phantom{\Big{)}}\label{eq:SbAS}\end{equation}
These properties imply the identity
\begin{equation}\Sigmabar\Sigma=1.\end{equation}
discussed in the introduction.
\end{description}
Before going further, let us make a necessary comment about notation. We will often use the same symbol to denote distinct but analogous string fields living in different state spaces. Thus, for example, $\Psi_\tv$ may refer to either the tachyon vacuum of $\BCFT_0$ or the tachyon vacuum of $\BCFT_*$, and $A$ may refer to either of the respective homotopy operators. It is possible to introduce labels to indicate which state space the string field occupies. Thus, for example, the relation $Q_{\Psi_\tv}\Sigma=0$ could be written explicitly as
\begin{equation}Q\Sigma^{(0*)} +\Psi_\tv^{(0)}\Sigma^{(0*)}-\Sigma^{(0*)}\Psi_\tv^{(*)}=0.\end{equation}
However, a moments thought reveals that it is manifest which string field is being referred to simply based on its appearance relative to $\Sigma$ and $\Sigmabar$. Therefore we will drop the labels except when needed for clarity. 

In principle it is not necessary to postulate the existence of a homotopy operator. Instead of \eq{SbAS}, we could take $\Sigmabar\Sigma=1$ as a fundamental identity. However, in practice we do not know how to construct the intertwining solution without the homotopy operator. The reason is that it allows us to solve the $Q_{\Psi_\tv}$-invariance constraints on the intertwining fields. We simply take $\Sigma$ and $\Sigmabar$ of the form 
\begin{eqnarray}
\Sigma \lineup = Q_{\Psi_\tv}\big(A\Sigma_\pre\big),\\
\Sigmabar\lineup = Q_{\Psi_\tv}\big(\Sigmabar_\pre A\big),
\end{eqnarray}
where $\Sigma_\pre$ and $\Sigmabar_\pre$ are stretched string states which we call {\it pre-intertwining fields}. This ansatz does not imply any loss of generality, since we can choose $\Sigma_\pre=\Sigma$ and $\Sigmabar_\pre=\Sigmabar$. Imposing \eq{SbAS} and using $A^2=0$, one can show that 
\begin{equation}\Sigmabar_\pre A\Sigma_\pre = A.\phantom{\Big{(}}\end{equation}
Since analytic expressions for the tachyon vacuum are known, the construction of the intertwining solution reduces to solving this single equation. Note that generally $\Sigmabar_\pre\Sigma_\pre\neq 1$, since we do not assume the pre-intertwining fields are $Q_{\Psi_\tv}$ invariant.

The basic form of the intertwining solution also applies to multiple D-brane systems, but to make this explicit it is helpful to spell out the Chan-Paton structures. Suppose that $\BCFT_0$ represents a system of $n$ D-branes, each individually characterized by boundary conformal field theories $\BCFT_0^i$ for $i=1,...,n$; similarly, $\BCFT_*$ consists of a system of $N$ D-branes, each individually characterized by boundary conformal field theories $\BCFT_*^I,$ for  $I=1,...,N$. In this case, the string field in $\BCFT_0$ is an $n\times n$ matrix whose $(i,j)$-th entry is an element of the state space of a stretched string connecting $\BCFT_0^i$ and $\BCFT_0^j$; similarly, the string field in $\BCFT_*$ is an $N\times N$ matrix whose $(I,J)$-th entry is an element of the state space of a stretched string connecting $\BCFT_*^I$ and $\BCFT_*^J$. The intertwining fields are rectangular matrices
\begin{eqnarray}
\Sigma \lineup = \left(\begin{matrix}\Sigma^{(11)} & \Sigma^{(12)} &  \dots  &\ \ & \ \ &\\
\Sigma^{(21)} & \Sigma^{(22)}& \ \ & \ \ & & \\ 
\vdots & & \ddots &\ \  & &\\
& & & \ \ &\ \ &\Sigma^{(nN)}\end{matrix} \right),\\
\Sigmabar\lineup = \left(\begin{matrix}
\Sigmabar^{(11)} & \Sigmabar^{(12)} & \dots & \\ 
\Sigmabar^{(21)} &\Sigmabar^{(22)}& &\\
\vdots & & \ddots &\\
& & & \\
& & & \\
& & & \Sigmabar^{(Nn)}
\end{matrix}\right),
\end{eqnarray}
where $\Sigma^{(iJ)}$ is an element of the state space of a stretched string connecting $\BCFT_0^i$ to $\BCFT_*^J$, and $\Sigmabar^{(Ij)}$ is an element of the state space of a stretched string connecting $\BCFT_*^I$ and $\BCFT_0^j$. The condition $\Sigmabar\Sigma=1$ can be reexpressed
\begin{equation}\sum_{k=1}^n\Sigmabar^{(Ik)}\Sigma^{(kJ)} = \delta^{(IJ)},\end{equation}
where the ``Kronecker delta" $\delta^{(IJ)}$ is the identity string field in $\BCFT_*^I$ when $I=J$, and is zero otherwise. It is worth emphasizing that the number of D-branes $n$ in $\BCFT_0$ and the number of D-branes $N$ in $\BCFT_*$ need not be equal, and in particular it is possible that $N>n$. So the general form of the intertwining solution does not rule out the possibility of creating new D-branes. We will describe how to do this in subsection \ref{subsec:comments}.

\subsection{Varying the Parameters of the Solution}
\label{subsec:vary}

An explicit realization of the intertwining solution requires various choices, and it is useful to understand how these choices effect the physics. 

First let us see how the solution changes if we vary the choice of tachyon vacuum in $\BCFT_0$ and $\BCFT_*$. Any on-shell deformation of the tachyon vacuum is pure gauge, so we can write
\begin{equation}\delta_\tv \Psi_\tv = Q_{\Psi_\tv}\Lambda,\end{equation}
where $\Lambda$ is an infinitesimal gauge parameter defined correspondingly in $\BCFT_0$ or $\BCFT_*$. As we change the tachyon vacuum the homotopy operator must also change. To maintain $A^2=0$ we assume
\begin{equation}\delta_\tv A = [A,\Lambda].\end{equation}
Finally, the intertwining fields depend on the tachyon vacuum, but also depend on other variables that might appear in the solution of \eq{SbAS}.
So we must define what it means to vary $\Sigma$ and $\Sigmabar$ with respect to the tachyon vacuum while keeping other variables fixed. The natural definition~is 
\begin{eqnarray}
\delta_\tv \Sigma\lineup = [\Sigma,\Lambda],\\
\delta_\tv\Sigmabar\lineup = [\Sigmabar,\Lambda].
\end{eqnarray}
In this way the variation preserves the structure of the intertwining solution. The solution then changes~as
\begin{equation}\delta_\tv\Psi_* = Q_{\Psi_*}\Big(\Lambda - \Sigma\Lambda\Sigmabar\Big),\label{eq:dtvgauge}\end{equation}
which is an infinitesimal gauge transformation. The gauge parameter is essentially the difference in the tachyon vacuum gauge parameters in $\BCFT_0$ and $\BCFT_*$.

Next we can consider varying the intertwining fields $\Sigma$ and $\Sigmabar$ while keeping the tachyon vacuum fixed. The variations 
\begin{equation}\delta\Sigma,\ \ \ \ \delta\Sigmabar,\end{equation}
must be $Q_{\Psi_\tv}$ invariant, and must preserve \eq{SbAS}:
\begin{equation}\delta \Sigmabar A\Sigma + \Sigmabar A \delta\Sigma = 0.\end{equation}
By formal manipulation, one can express the variation of the solution as 
\begin{equation}
\delta \Psi_* = Q_{\Psi_*}\Big(\Sigma\Psi_\tv\delta\Sigmabar A-A\delta\Sigma \Psi_\tv\Sigmabar \Big).\label{eq:dSgauge}
\end{equation}
This appears to show that the solution changes by gauge transformation. This cannot be exactly correct, since a general variation of the intertwining fields will result in a deformation of the target D-brane.\footnote{If we modify the boundary condition, the tachyon vacuum will also change since it lives in a different state space. In the universal subspace, however, there is a natural connection which allows us to say whether the tachyon vacuum is unchanged by a deformation of the boundary condition.} In such a circumstance, however, the above gauge parameter is not well defined. Let us illustrate how this occurs with an example. Suppose that $\BCFT_0$ includes a free boson $X(z,\overline{z})$ subject to Neumann boundary conditions compactified on a circle. We can describe a target $\BCFT_*$ with constant Wilson line on this circle using boundary condition changing operators
\begin{eqnarray}
\sigma(x) \lineup = e^{i \lambda X}(x),\\
\sigmabar(x) \lineup = e^{-i \lambda X}(x),
\end{eqnarray}
where $\lambda$ parameterizes the magnitude of the Wilson line deformation. These are not well-defined operators in $\BCFT_0$ since the zero mode of the free boson $X(z,\overline{z})$ is not single valued on the circle. Of course, we do not expect these operators to be well defined in $\BCFT_0$ since they change the boundary condition. However, the combined nonlocal operator
\begin{equation}\sigma(x)\sigmabar(0)\end{equation}
is well-defined in $\BCFT_0$, since the change of boundary condition implemented by $\sigma$ is undone by~$\sigmabar$. If we vary with respect to the magnitude of the Wilson line  
\begin{equation}
\frac{d}{d\lambda}\Big(\sigma(x)\sigmabar(0)\Big) = \left(\frac{d}{d\lambda}\sigma(x)\right)\sigmabar(0) + \sigma(x)\left(\frac{d}{d\lambda}\sigmabar(0)\right),
\end{equation}
we should still have a sensible operator in $\BCFT_0$.  However, the individual terms on the right hand side are not well-defined; they depend on the zero mode of $X(z,\overline{z})$ in a way that cancels when the terms are added. Returning to the gauge parameter in \eq{dSgauge}, it is clear that such a cancellation will not occur, since the two terms do not add up to the variation of a well-defined state in $\BCFT_0$.  The upshot is that varying the parameters of the intertwining solution produces a gauge transformation {\it provided} we do not change the target D-brane system. This is consistent with the expectation that the physical content of the solution is completely contained in the open string background that the solution describes. 

It is interesting to observe that the algebraic structure of the intertwining solution is gauge invariant. A finite gauge transformation of the intertwining solution takes the form
\begin{equation}
\Psi_*' = \underbrace{U^{-1}(Q+\Psi_\tv)U}_{\displaystyle{\Psi_\tv'}}-\underbrace{(U^{-1}\Sigma)}_{\displaystyle{\Sigma'}} \Psi_\tv \underbrace{(\Sigmabar U)}_{\displaystyle{\Sigmabar'}},
\end{equation}
where $U\in\H_0$ is a finite gauge parameter. As indicated by the underbraces, the gauge transformed solution is still an intertwining solution, but with a new tachyon vacuum $\Psi_\tv'\in \H_0$ and intertwining fields $\Sigma'$ and $\Sigmabar'$. If an intertwining solution can be found in each gauge orbit---which we will argue is possible---this has a surprising consequence: {\it All} classical solutions in open bosonic string field theory are, in some way or another, intertwining solutions. Generally, it can be difficult and awkward to express a solution in intertwining form. To give an example, in appendix \ref{app:Sch} we show how it can be done for marginal deformations in Schnabl gauge. While potentially awkward, it is significant if an intertwining form can be found, since it makes the physical content of the solution completely transparent. 

\subsection{Background Independence}
\label{subsec:BI}

The formulation of string field theory requires a choice of background. A longstanding conjecture is that the choice of background does not matter: string field theories formulated on different backgrounds are related by field redefinition. In \cite{BI1,BI2,BIsuper} it was shown that an infinitesimal deformation of the reference background can be compensated by an infinitesimal field redefinition. But it is not known if large shifts of the background can be similarly compensated. This is the problem of background independence. 

One can postulate on general grounds that the field redefinition relating the string field theories of $\BCFT_0$ and $\BCFT_*$ will take the form
\begin{equation}\Psi^{(0)}=\Psi_* + f(\Psi^{(*)}),\end{equation}
where 
\begin{itemize}
\item $\Psi^{(0)}\in\H_0$ is the dynamical field of the reference D-brane;
\item $\Psi^{(*)}\in\H_*$ is the dynamical field of the target D-brane;
\item $\Psi_*\in\H_0$ is a classical solution of the reference string field theory describing the target D-brane;
\item $f$ is a linear map from $\H_*$ into $\H_0$ which is invertible up to gauge transformation.
\end{itemize}
We can consistently assume that $f$ is linear since both reference and target string field theories are cubic. The problem of background independence then boils down to two questions:
\begin{description}
\item{(1)} Given string field theory formulated on an arbitrary reference D-brane, is it possible to find a solution $\Psi_*$ for every target D-brane system which shares the same closed string background?
\item{(2)} Is it further possible to find an $f$ which transforms the string field of the target D-brane into the fluctuation field around the solution $\Psi_*$ of the reference D-brane?
\end{description}
In later sections we will see that the answer to the first question is affirmative. Here we will show that if $\Psi_*$ is an intertwining solution, the answer to the second question is also affirmative. Together this establishes the background independence of classical open bosonic string field theory. 

We start by noting that $Q_{\Psi_\tv}$ invariance of $\Sigma$ and $\Sigmabar$ implies 
\begin{equation}Q_{\Psi_*0}\Sigma = 0,\ \ \ \ Q_{0\Psi_*}\Sigmabar = 0,
\label{eq:QPsisS}\end{equation}
where we introduce the notation 
\begin{equation}
Q_{\Psi_1\Psi_2}\Phi \equiv Q\Phi + \Psi_1\Phi -(-1)^{\Phi \Psi_2}\Phi \Psi_2.
\end{equation}
This is the BRST operator for a stretched string connecting a D-brane condensed to a classical solution $\Psi_1$ and a D-brane condensed to a classical solution $\Psi_2$. This operator is niltpotent and satisfies a generalized Leibniz rule
\begin{equation}
Q_{\Psi_1\Psi_3}(\Phi\Pi) = (Q_{\Psi_1\Psi_2}\Phi)\Pi +(-1)^{\Phi}\Phi(Q_{\Psi_2\Psi_3}\Pi).
\end{equation}
The relations \eq{QPsisS} have an important interpretation. Since the intertwining fields change the boundary condition, $Q_{\Psi_*0}$ is the BRST operator for a stretched string connecting a $\BCFT_0$ D-brane condensed to the solution $\Psi_*$ and a $\BCFT_*$ D-brane at the perturbative vacuum.   However, both solutions $\Psi_*\in \H_0$ and $0\in\H_*$ represent the same physical background---namely, the target D-brane. Therefore, \eq{QPsisS} is really saying that the intertwining fields are BRST invariant from the point of view of the BRST operator of $\BCFT_*$. Moreover, the intertwining fields cannot be BRST exact. This would be inconsistent with the relation 
\begin{equation}\Sigmabar\Sigma = 1,\end{equation}
since the identity string field is not exact. This means that the intertwining fields must be representatives of the cohomology class of the identity operator in $\BCFT_*$. The identity operator is the only nontrivial element of the cohomology at ghost number zero. 

This suggests that the map $f:\H_*\to\H_0$ defined by
\begin{equation}f(\Phi) = \Sigma\Phi\Sigmabar\end{equation}
should be considered as analogous to left/right multiplication by $1$, at least from the point of view of BRST cohomology in $\BCFT_*$. Therefore $f$ reexpresses a linearized string field of the target D-brane as a linearized fluctuation of the solution $\Psi_*$ without  altering its physical meaning. In fact, this correspondence extends to the nonlinear level, since $f$ defines a homomorphism between the algebraic structures of the two string field theories:
\begin{eqnarray}
f(\Phi\Pi) \lineup = f(\Phi)f(\Pi), \phantom{\Big{)}}\label{eq:alghom}\\
f(Q\Phi)  \lineup = Q_{\Psi_*}f(\Phi), \phantom{\Big{)}}\label{eq:Qhom}\\
\Tr_0[f(\Phi)]\lineup = \Tr_*[\Phi] ,\phantom{\Big{)}}\label{eq:Trhom}
\end{eqnarray}
where $\Tr_0$ and $\Tr_*$ represent the 1-string vertex computed on states in $\H_0$ and $\H_*$, respectively. It~follows that the dynamical fields of the reference and target D-brane systems can be related by 
\begin{equation}\Psi^{(0)} = \Psi_*+\Sigma\Psi^{(*)}\Sigmabar.\phantom{\Bigg(}\label{eq:equivalence}\end{equation}
In particular, the actions of the two string field theories can be written:
\begin{eqnarray}
S_0(\Psi^{(0)}) \lineup = \Tr_0\left[\frac{1}{2}\Psi^{(0)} Q\Psi^{(0)} +\frac{1}{3}(\Psi^{(0)})^3\right],\\
S_*(\Psi^{(*)}) \lineup = \Tr_*\left[\frac{1}{2}\Psi^{(*)} Q\Psi^{(*)} +\frac{1}{3}(\Psi^{(*)})^3\right].
\end{eqnarray}
If the dynamical string fields are related through \eq{equivalence}, the actions are equal up to an additive constant: 
\begin{equation}
S_0(\Psi^{(0)}) -S_0(\Psi_\tv) = S_*(\Psi^{(*)}) - S_*(\Psi_\tv).
\end{equation}
This shows that the string field theories are related by field redefinition. 

Let us mention a few properties of the proposed field redefinition. Suppose that we have three backgrounds $\BCFT_1,$ $\BCFT_2$ and $\BCFT_3$. We have a field redefinition of the form \eq{equivalence} relating $\BCFT_1$ and $\BCFT_2$, and a similar field redefinition relating $\BCFT_2$ and $\BCFT_3$. Composing gives a field redefinition between $\BCFT_1$ and $\BCFT_3$ directly. The composite field redefinition takes the same form as \eq{equivalence} with intertwining fields given by 
\begin{equation}
\Sigma^{(13)} = \Sigma^{(12)}\Sigma^{(23)},\ \ \ \ \Sigmabar^{(31)} = \Sigmabar^{(32)}\Sigmabar^{(21)},
\end{equation}
where $\Sigma^{(12)},\Sigmabar^{(21)}$ and $\Sigma^{(23)},\Sigmabar^{(23)}$ characterize the original field redefinitions. That is to say, field redefinitions of the form \eq{equivalence} define an algebraically closed set under composition. Another useful thing to understand is how the field redefinition changes as we vary the parameters of the intertwining solution. If we vary the tachyon vacuum $\Psi_\tv$ as in the previous section, the field redefinition changes as 
\begin{equation}\delta_\tv\Psi^{(0)} = Q_{\Psi^{(0)}}\Big(\Lambda - \Sigma\Lambda\Sigmabar\Big).\end{equation}
This is an infinitesimal gauge transformation of $\Psi^{(0)}$. If we vary $\Sigma$ and $\Sigmabar$, keeping the target D-brane system unchanged, the field redefinition changes as
\begin{eqnarray}
\delta\Psi^{(0)} \lineup = Q_{\Psi^{(0)}}\Big(A\delta\Sigma\Sigmabar(\Psi^{(0)}-\Psi_\tv) - (\Psi^{(0)}-\Psi_\tv)\Sigma\delta\Sigmabar A\Big)\nonumber\\
\lineup\ \ \ \   +A\delta\Sigma\Sigmabar(Q\Psi^{(0)}+(\Psi^{(0)})^2)+(Q\Psi^{(0)}+(\Psi^{(0)})^2)\Sigma\delta\Sigmabar A.
\end{eqnarray}
This is a combination of an ordinary infinitesimal gauge transformation and a ``trivial" gauge transformation which vanishes on-shell.

\begin{figure}
\begin{center}
\resizebox{5in}{2in}{\includegraphics{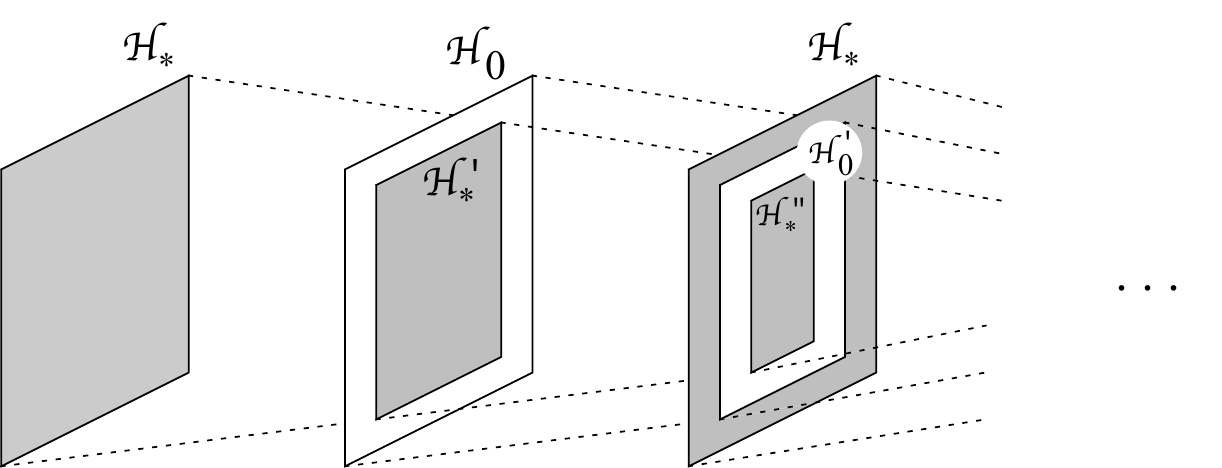}}
\end{center}
\caption{\label{fig:fieldred} The field redefinition between string field theories on different backgrounds is not an isomorphism, but a one-to-one map transforming states of the target D-brane into a subset of states of the reference D-brane. Reversing the role of reference and target backgrounds, it is also possible to map states of the reference D-brane into a subset of states of the target D-brane. Repeated transformation between the reference and target string field theories embeds the state spaces into ever smaller subsets of themselves.}
\end{figure}

Usually it is implied that field redefinitions define an isomorphism between the fields of two theories. Interestingly, this is not what occurs here. The field redefinition \eq{equivalence} defines a one-to-one map from states of the target background into a subset of states of the reference background. To see this, note that if the reference string field $\Psi^{(0)}$  is given by \eq{equivalence} we can recover the target string field $\Psi^{(*)}$ using 
\begin{equation}\Sigmabar(\Psi^{(0)}-\Psi_*)\Sigma = \Psi^{(*)}.\label{eq:ps_inv}\end{equation}
But if $\Psi^{(*)}$ is given by this equation, in general we cannot reconstruct $\Psi^{(0)}$. This is a consequence of the fact that the intertwining fields only multiply to the identity in one direction; $\Sigma$ is analogous to a non-unitary isometry, as mentioned in the introduction. This might suggest that $\H_*$ is ``smaller" than $\H_0$, but this is not right: reversing the role of $\BCFT_0$ and $\BCFT_*$, it is also possible to argue that $\H_0$ can be mapped one-to-one into a proper subset of $\H_*$. This sets the stage for a curious infinite regress, where repeated mapping between the reference and target string field theories sheds ever larger portions of the respective state spaces, as illustrated in figure \ref{fig:fieldred}. In this situation it is not obvious that the field redefinition implies a physical equivalence between the theories. What needs to be seen is that the field redefinition is invertible modulo gauge transformation. This requires two things: 
\begin{enumerate}
\item If $\Phi^{(*)}$ and $\Psi^{(*)}$ are gauge equivalent in the target string field theory, then their images $\Phi^{(0)}$ and $\Psi^{(0)}$ are gauge equivalent in the reference string field theory.
\item If the images $\Phi^{(0)}$ and $\Psi^{(0)}$ are gauge equivalent in the reference string field theory, then $\Phi^{(*)}$ and $\Psi^{(*)}$ are gauge equivalent in the target string field theory.
\end{enumerate}
The first property is an immediate consequence of \eq{alghom} and \eq{Qhom}. The second property is less obvious, since we do not have an inverse map which takes the gauge transformations of $\BCFT_0$ into the gauge transformations of $\BCFT_*$. What we can do, however, is construct a solution in the target string field theory describing the reference string field theory---say using intertwining fields $\Sigma^{(*0)}$ and $\Sigmabar^{(0*)}$---and use this to map $\H_0$ into a subset of $\H_*$. Composing the map from $\H_*$ to $\H_0$, and then from $\H_0$ to $\H_*$, transforms $\Phi^{(*)}$ and 
$\Psi^{(*)}$ into another pair of string fields $\widehat{\Phi}^{(*)}$ and $\widehat{\Psi}^{(*)}$ in the same state space:
\begin{eqnarray}
\widehat{\Phi}^{(*)} \lineup = \Psi_\tv+\Sigma^{(*0)}(\Phi^{(0)}-\Psi_\tv)\Sigmabar^{(0*)}\nonumber\\
\lineup = \Psi_\tv +\Sigma^{(*0)}\Sigma^{(0*)}(\Phi^{(*)}-\Psi_\tv)\Sigmabar^{(*0)}\Sigmabar^{(0*)},\\
\widehat{\Psi}^{(*)} \lineup = \Psi_\tv+\Sigma^{(*0)}(\Psi^{(0)}-\Psi_\tv)\Sigmabar^{(0*)}\nonumber\\
\lineup = \Psi_\tv +\Sigma^{(*0)}\Sigma^{(0*)}(\Psi^{(*)}-\Psi_\tv)\Sigmabar^{(*0)}\Sigmabar^{(0*)}.
\end{eqnarray}
If $\Phi^{(0)}$ and $\Psi^{(0)}$ are gauge equivalent in $\H_0$, it follows that $\widehat{\Phi}^{(*)}$ and $\widehat{\Psi}^{(*)}$ are gauge equivalent in $\H_*$. Now we use the fact that varying the intertwining fields produces a gauge transformation provided that the reference and target boundary conformal field theories stay fixed. Since the intertwining fields $\Sigma^{(*0)}\Sigma^{(0*)}$ and $\Sigmabar^{(*0)}\Sigmabar^{(0*)}$ do not change the boundary condition, they can be continuously deformed to the identity string field, and in the process we demonstrate that $\widehat{\Phi}^{(*)}$ is gauge equivalent to $\Phi^{(*)}$, and $\widehat{\Psi}^{(*)}$ is gauge equivalent to $\Psi^{(*)}$. It follows that $\Psi^{(*)}$ is gauge equivalent to $\Phi^{(*)}$, as we wanted to show. This is enough to establish the physical equivalence of string field theories on different backgrounds. 

The above discussion implies that the dynamical variables of the theories are isomorphic only after integrating out certain pure gauge degrees of freedom from the reference string field theory.\footnote{Integrating the pure gauge variables does not modify the action since it is consistent with the equations of motion to set them equal to zero.} It is natural to ask what these pure gauge degrees of freedom represent. A suggestive interpretation was given~in~\cite{Masuda}.\footnote{The discussion of \cite{Masuda} follows \cite{KOSsing}, where the intertwining fields do not multiply associatively. This is dealt with by requiring that the product of $\Sigma\Sigmabar$ is never explicitly evaluated. The solution of the present paper gives a more well-defined context for their discussion.} Consider the fluctuation field around $\Psi_*$:
\begin{equation}\varphi = \Psi^{(0)}-\Psi_*.\end{equation}
By left and right multiplication by a resolution of the identity
\begin{equation}1=(1-\Sigma\Sigmabar)+\Sigma\Sigmabar, \end{equation}
we can express $\varphi$ as a sum of states in four subsectors of $\H_0$:
\begin{equation}\varphi = \varphi_{**}+\varphi_{*\tv}+\varphi_{\tv*}+\varphi_{\tv\tv},\end{equation}
where 
\begin{eqnarray}
\varphi_{**}\lineup = (\Sigma\Sigmabar)\varphi(\Sigma\Sigmabar),\ \ \ \ \ \ \ \ \ \ \varphi_{*\tv}=(\Sigma\Sigmabar)\varphi(1-\Sigma\Sigmabar),\nonumber\\
\varphi_{\tv*}\lineup = (1-\Sigma\Sigmabar)\varphi(\Sigma\Sigmabar),\ \ \ \ \varphi_{\tv\tv}= (1-\Sigma\Sigmabar)\varphi(1-\Sigma\Sigmabar),
\end{eqnarray}
The subspace of states $\varphi_{**}$ is isomorphic to the state space $\H_*$ of the target D-brane. The remaining three sectors are the degrees of freedom of the reference D-brane which should be integrated out.  One can show that the fluctuation field $\varphi$ multiplies isomorphically to a $2\times 2$ matrix of string fields 
\begin{equation}\varphi\ \leftrightarrow\ \varphi_{2\times 2} = \left(\begin{matrix}\varphi_{**} & \varphi_{*\tv} \\ \varphi_{\tv*} & \varphi_{\tv\tv}\end{matrix}\right).\label{eq:22}\end{equation}
This suggests that the four sectors of $\H_0$ can be naturally interpreted in terms of a double D-brane system consisting of two copies of $\BCFT_0$. Moreover, due to the property
\begin{equation}\Psi_*(1-\Sigma\Sigmabar) = \Psi_\tv(1-\Sigma\Sigmabar),\ \ \ (1-\Sigma\Sigmabar)\Psi_* = (1-\Sigma\Sigmabar)\Psi_\tv,\end{equation}
the action expanded around $\Psi_*$ can be rewritten as the action for two copies of $\BCFT_0$ expanded around the solution
\begin{equation}\Psi_{2\times 2} = \left(\begin{matrix}\Psi_* & 0 \\ 0 &\Psi_\tv\end{matrix}\right).\end{equation}
In particular, one D-brane is condensed to the solution $\Psi_*$ and the other is condensed to the tachyon vacuum. This implies that the ``extra"  degrees of freedom of the reference D-brane represent fluctuations of the tachyon vacuum. Note, however, that they are not arbitrary fluctuations since the entries of the Chan-Paton matrix \eq{22} are projected onto linear subspaces of $\H_0$.  

\subsection{Why the Tachyon Vacuum?}

A notable feature of the intertwining solution is the seeming importance of the tachyon vacuum. In principle one can contemplate a solution where the reference D-brane condenses to a nontrivial intermediate background $\BCFT_{1/2}$ before building the target D-brane. The solution would take the form
\begin{equation}\Psi_* = \Psi_{1/2} -\Sigma\Psi_{1/2}\Sigmabar,\label{eq:int_solution}\end{equation}
where $\Psi_{1/2}$ is a solution respectively in $\H_0$ or $\H_*$ describing $\BCFT_{1/2}$. The intertwining fields will now be $Q_{\Psi_{1/2}}$-invariant and satisfy $\Sigmabar\Sigma=1$. We could, for example choose $\BCFT_{1/2}$ to be $\BCFT_0$ or $\BCFT_*$. In the former case, the first term in the solution could be taken to vanish, and in the later case, the second term (where the ansatz becomes trivial). Given the possibility of a solution of this kind, one might ask why the tachyon vacuum should be the preferred intermediate background.  This question may seem relevant to generalizations to supersymmetric D-brane systems, where the existence of a tachyon vacuum is not established. 

Part of the reason why the tachyon vacuum is special is that otherwise the solution does not seem very constructive. Since a nontrivial intermediate background will support physical open string states, there will be no analogue of the homotopy operator, and we do not have a general approach to solving the BRST invariance constraints of the intertwining fields. Moreover, it is not clear why it should be easier to construct two solutions for $\BCFT_{1/2}$ in two string field theories than it would be to construct a single solution for $\BCFT_*$ in one. These points however are more of an argument of utility than principle.

Under certain assumptions, we can give an example of a solution which passes through a nontrivial background using the original solution which passes through the tachyon vacuum. Suppose we have a solution of $\BCFT_0$ describing $\BCFT_{1/2}$ using intertwining fields $\Sigma^{(0\frac{1}{2})}$ and $\Sigmabar^{(\frac{1}{2}0)}$, and a solution of $\BCFT_{1/2}$ describing $\BCFT_*$ using intertwining fields $\Sigma^{(\frac{1}{2}*)}$ and $\Sigmabar^{(*\frac{1}{2})}$. Using the field redefinition \eq{equivalence}, this implies a solution of $\BCFT_0$ describing $\BCFT_*$:
\begin{equation}\Psi_* = \Psi_\tv-\Sigma^{(0*)}\Psi_\tv\Sigmabar^{(*0)},\end{equation}
where
\begin{equation}
\Sigma^{(0*)} \equiv  \Sigma^{(0\frac{1}{2})}\Sigma^{(\frac{1}{2}*)},\ \ \ \ \Sigmabar^{(*0)}\equiv\Sigmabar^{(*\frac{1}{2})}\Sigmabar^{(\frac{1}{2}0)}.
\end{equation}
Now suppose that the intertwining fields $\Sigma^{(\frac{1}{2}*)}$ and $\Sigmabar^{(\frac{1}{2}*)}$ multiply to $1$ in both directions:
\begin{equation}\Sigmabar^{(*\frac{1}{2})}\Sigma^{(\frac{1}{2}*)} = 1,\ \ \ \  \Sigma^{(\frac{1}{2}*)}\Sigmabar^{(*\frac{1}{2})} =1,\ \ \ \ (\mathrm{assumption}).\end{equation}
This will not be true of the intertwining fields discussed in this paper. However, it is possible to realize this property at least for certain classes of marginal deformations \cite{KOSsing,Maccaferri_marg}. Under this assumption the state spaces of $\BCFT_{1/2}$ and $\BCFT_*$ will be isomorphic. Furthermore we can express the solution in the form
\begin{equation}\Psi_* = \underbrace{\Psi_\tv-\Sigma^{(0\frac{1}{2})}\Psi_\tv\Sigmabar^{(\frac{1}{2}0)}}_{\Psi_{1/2}}- \Sigma^{(0*)}(\underbrace{\Psi_\tv-\Sigmabar^{(*\frac{1}{2})}\Psi_\tv\Sigma^{(\frac{1}{2}*)}}_{\Psi_{1/2}} )\Sigmabar^{(*0)},\label{eq:comp_sol}\end{equation}
where we have added and subtracted a term. As indicated by the underbraces, now the solution can be interpreted as passing through a nontrivial intermediate background, described by $\Psi_{1/2}$, before continuing to the target D-brane.

It is interesting that we need an isomorphism between the state spaces of $\BCFT_{1/2}$ and $\BCFT_*$ for this to work. While it is difficult to make definitive statements based on a single example, we can offer a possible explanation. Note that $\Sigma$ satisfies two BRST invariance conditions:
\begin{equation}Q_{\Psi_{1/2}}\Sigma^{(0*)} = 0,\ \ \ \ Q_{\Psi_*0}\Sigma^{(0*)} = 0.\end{equation}
This implies that the intertwining fields are representatives of the cohomology class of the identity operator simultaneously in two separate backgrounds, namely $\BCFT_{1/2}$ and $\BCFT_*$. In this sense it seems that the intertwining fields are overburdened. The solution may only be possible if $\BCFT_{1/2}$ and $\BCFT_*$ are related in a special way, for example through marginal deformation which implies an isomorphism between the state spaces. From this point of view the tachyon vacuum is unique, since it is the only intermediate background where the intertwining fields are not simultaneously forced to represent the cohomology of disparate D-brane systems. 

\section{Flag States}
\label{sec:flags}

We now proceed in developing a concrete example of an intertwining solution. First we introduce ``flag states." We describe flag states in a specific form which is convenient for gluing with wedge states, as appear in analytic solutions for the tachyon vacuum. 

\subsection{Definition}

\begin{figure}
\begin{center}
\resizebox{4.1in}{2in}{\includegraphics{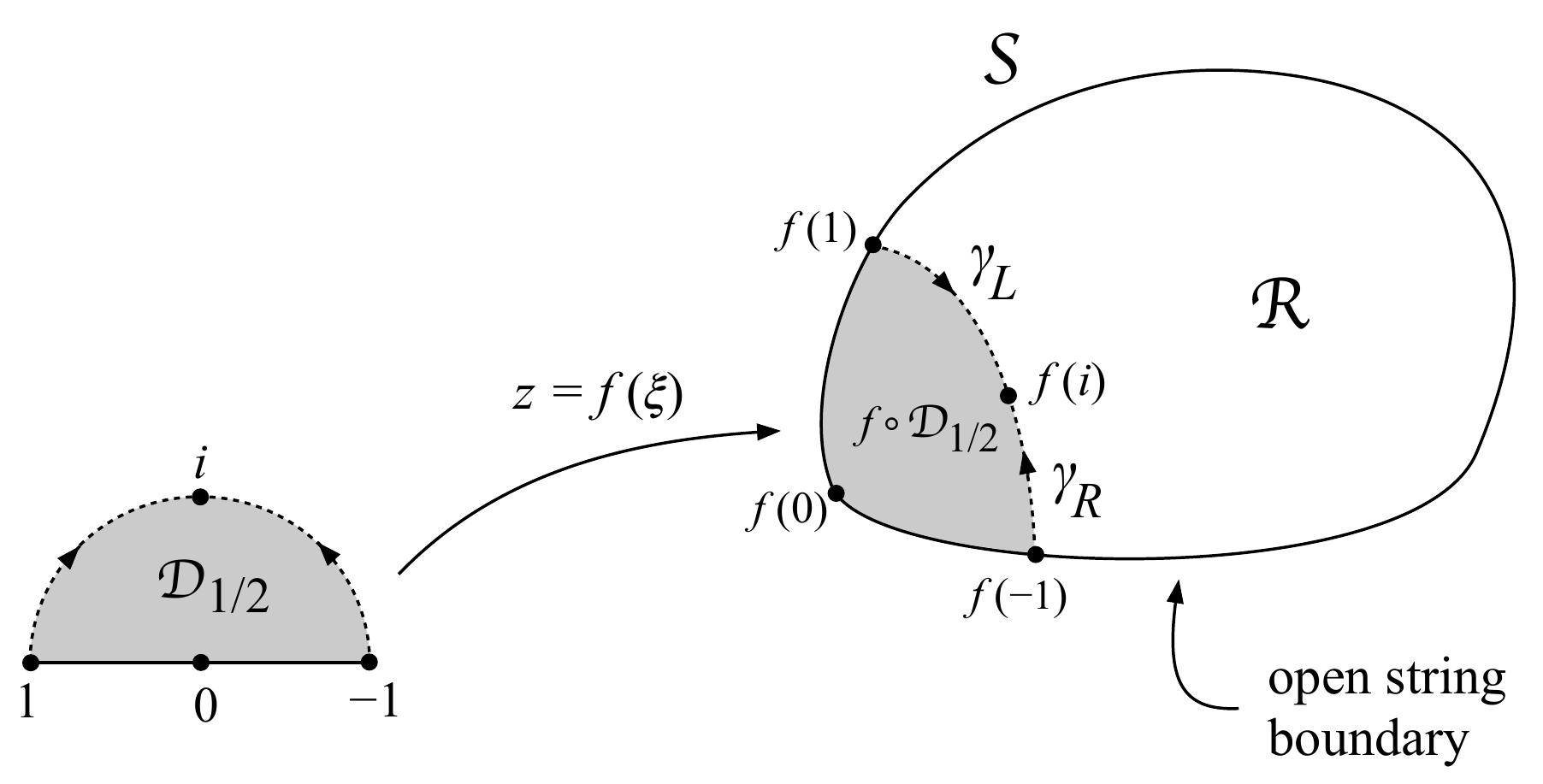}}
\end{center}
\caption{\label{fig:surface} Illustration of the data specifying a generic surface state $|S\rangle$.}
\end{figure}

At first we will describe flag states as surface states; later we will add boundary condition changing operators. Surface states are often characterized by their overlap with a test state $\phi$. We can visualize the test state as the unit half-disk in the complex plane,
\begin{equation}\mathscr{D}_{1/2}:\ \ |\xi|\leq 0,\ \ \mathrm{Im}(\xi)\geq 0,\end{equation}
containing the vertex operator $\phi(0)$ of the state inserted at the origin $\xi=0$. A surface state $|S\rangle$ is defined by a Riemann surface $\mathscr{S}$ with the topology of a disk, a global coordinate $z$ on $\mathscr{S}$, and a conformal map 
\begin{equation}z=f(\xi)\end{equation}
which associates to each point $\xi\in\mathscr{D}_{1/2}$ a corresponding point $z\in \mathscr{S}$. We require that $f$ maps the segment of the real axis $[-1,1]\subset \mathscr{D}_{1/2}$ into a segment of the boundary of $\mathscr{S}$. This implies that the test state vertex operator, after conformal transformation $f\circ\phi(0)$, will be inserted at a boundary point of~$\mathscr{S}$. The BPZ inner product of a surface state $|S\rangle$ with the test state $|\phi\rangle$ is then defined by a correlation function on $\mathscr{S}$:
\begin{equation}\langle \phi ,S\rangle = \langle f\circ\phi(0)\rangle_\mathscr{S},\end{equation}
where operator insertions inside the correlation function are expressed in the coordinate $z$. For our discussion it will be useful to characterize the data of a surface state in a different way. The surface $\mathscr{S}$ is composed of two regions: the image of the unit half disk $f\circ\mathscr{D}_{1/2}$ and the remaining surface~$\mathscr{R}$:
\begin{equation}
\mathscr{S} = f\circ\mathscr{D}_{1/2}\cup \mathscr{R}.
\end{equation}
At the intersection of these regions is a curve $\gamma$:
\begin{equation}\gamma=f\circ \mathscr{D}_{1/2}\cap\mathscr{R}.\end{equation}
This curve is the image of the half circle $\xi = e^{i\sigma}\in \mathscr{D}_{1/2}$, and is naturally parameterized by $\sigma\in [0,\pi]$:
\begin{equation}
z(\sigma) = f(e^{i\sigma})\in\gamma.
\end{equation}
In string field theory we want to think about star products of surface states, and in this context it is useful to think of $\gamma$ as composed of two segments $\gamma_L$ and $\gamma_R$ representing the left and right halves of the string:
\begin{equation}\gamma = \gamma_L\cup\gamma_R.\end{equation}
The curves are respectively parameterized by coordinates $\sigma_L,\sigma_R\in[0,\pi/2]$:
\begin{eqnarray}
z_L(\sigma_L)\lineup = z(\sigma_L)\ \ \ \ \ \, \in \gamma_L,\\
z_R(\sigma_R)\lineup = z(\pi-\sigma_R) \in \gamma_R.
\end{eqnarray}
The coordinates $\sigma_L=0$ and $\sigma_R=0$ represent the two endpoints of the open string, and $\sigma_L=\sigma_R=\pi/2$ represents the midpoint. Specifying the region $\mathscr{R}$ and the parameterized curves $\gamma_L,\gamma_R$ is equivalent to specifying the surface $\mathscr{S}$ and local coordinate map $f$. We illustrate the above discussion in figure \ref{fig:surface}. 

\begin{figure}
\begin{center}
\resizebox{2.9in}{2in}{\includegraphics{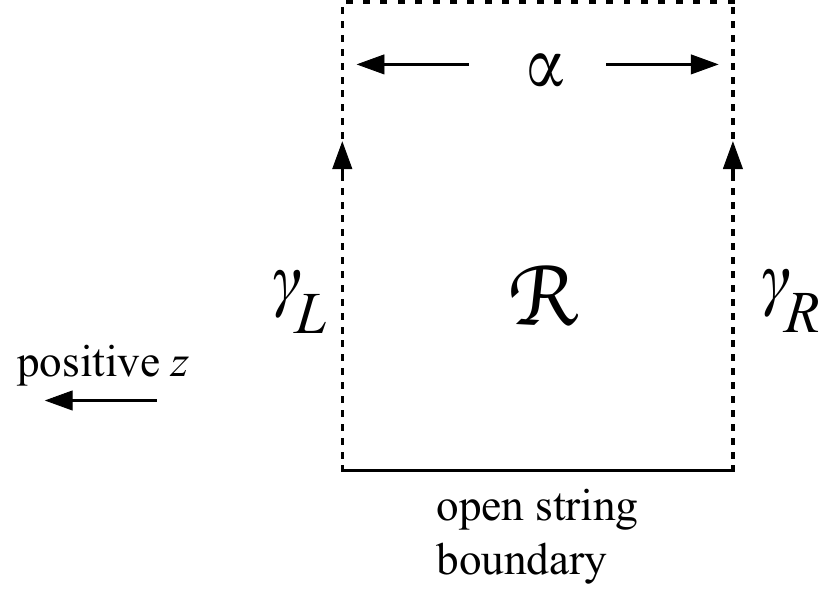}}
\end{center}
\caption{\label{fig:wedge} The region $\mathscr{R}$ and the left and right parameterized curves $\gamma_L$ and $\gamma_R$ characterizing a wedge state $\Omega^\alpha$.}
\end{figure}

Let us first consider wedge states $\Omega^\alpha,\ \alpha\geq 0$, which are star algebra powers of the $SL(2,\mathbb{R})$ vacuum $\Omega = |0\rangle$. In this case, we can choose the region $\mathscr{R}$ to be a semi-infinite vertical strip in the complex plane of width $\alpha$:
\begin{equation}\mathscr{R} = \{z|\ \mathrm{Im}(z)\geq 0;\ \ 0\leq \mathrm{Re}(z)\leq \alpha\}\end{equation}
The segment of the real axis $0\leq z\leq \alpha$ represents the open string boundary. The left curve $\gamma_L$ coincides with the positive-facing vertical boundary of the strip $\mathrm{Re}(z) = \alpha$ and the right curve $\gamma_R$ coincides with the negative-facing vertical boundary $\mathrm{Re}(z) = 0$. The curves are parameterized according to\footnote{Other parameterizations of the curve are possible, but the resulting states are not star algebra powers of the $SL(2,\mathbb{R})$ vacuum. Particularly notable alternatives are given by special projector frames \cite{RZ}.}
\begin{eqnarray}
z_L(\sigma_L) \lineup = \alpha + \frac{i}{\pi}\gd^{-1}\sigma_L\in\gamma_L\\
z_R(\sigma_R)\lineup = \frac{i}{\pi}\gd^{-1}\sigma_R\ \ \ \ \ \in\gamma_R
\end{eqnarray}
where $\gd^{-1}$ is the inverse of the Gudermannian function
\begin{equation}\gd\, x = 2\tan^{-1}\left(\tanh\frac{x}{2}\right)\end{equation}
Through conformal transformation, it is possible to represent wedge states in alternative ways through different regions $\mathscr{R}$ and corresponding parameterized curves $\gamma_L$ and $\gamma_R$. The presentation we are using is standard, and is refered to as the {\it sliver frame}. We denote the sliver coordinate as $z$. Let us mention an issue of visualization. In the conventional picture of the complex plane, the positive real axis increases towards the right and the negative real axis decreases towards the left. This, however, is opposite to the placement of $\gamma_L$ and $\gamma_R$. For this reason, and to simplify the visualization of gluing surfaces in our star product convention, we will henceforth picture the complex plane so that the positive real axis increases towards the left \cite{simple}. Our visualization is related to the usual one through $z\to -z^*$. With this explanation, we give a picture of the wedge state surface in figure~\ref{fig:wedge}.

\begin{figure}
\begin{center}
\resizebox{4.6in}{1.8in}{\includegraphics{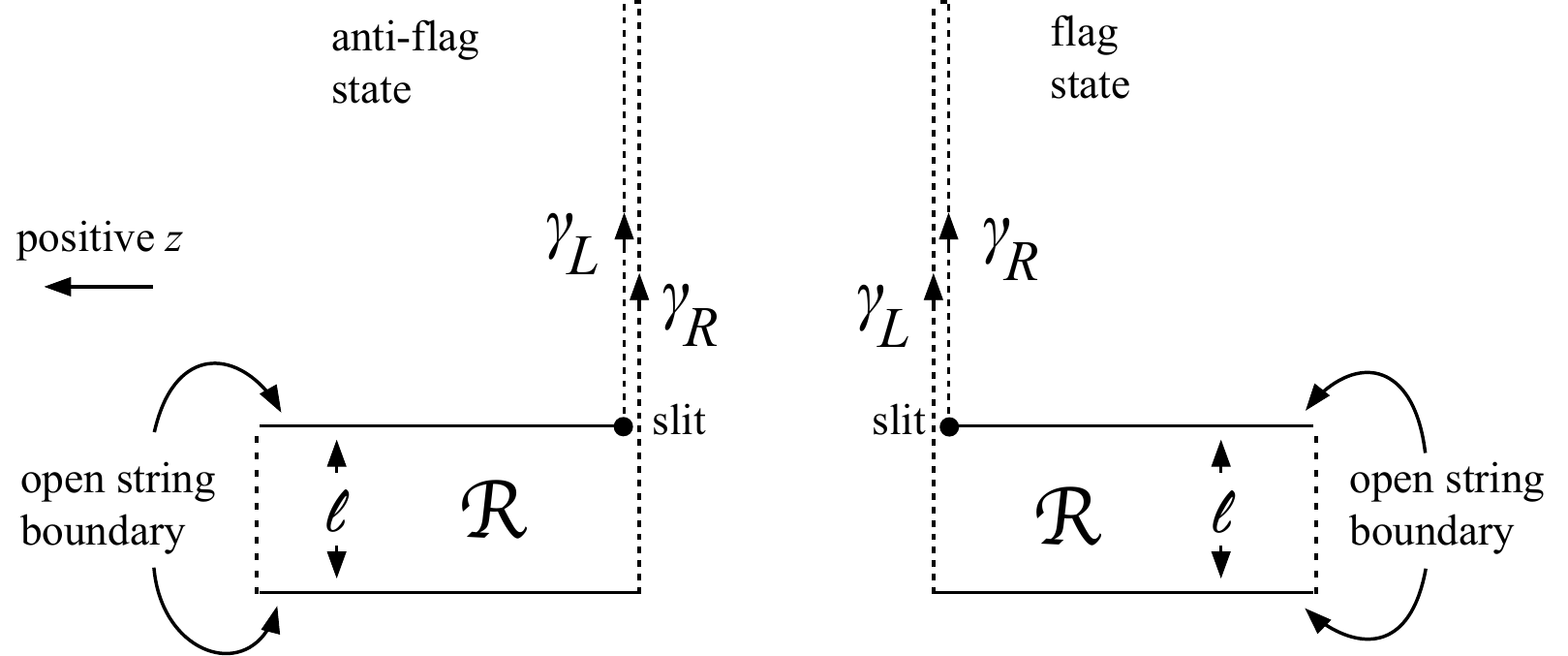}}
\end{center}
\caption{\label{fig:flags} The region $\mathscr{R}$ and the left and right parameterized curves $\gamma_L$ and $\gamma_R$ characterizing the flag state $\flag{1}$ and anti-flag state $\Aflag{1}$.}
\end{figure}

We now define flag states. They come in two kinds: a ``flag state'' and an ``anti-flag state," which will be denoted
\begin{eqnarray}
\text{flag state}:\ \ \flag{1},\ \ \ \text{anti-flag state}:\ \ \Aflag{1}.
\end{eqnarray}
The label $1$ indicates that the flag states carry an ``insertion" of the identity operator. Nontrivial insertions will be discussed in a moment. The flag states are characterized by the following regions:
\begin{eqnarray}
\text{flag state}:\lineup \ \ \ \mathscr{R} = \{z|\mathrm{Re}(z)= 0\}\cup\{z|\mathrm{Re}(z)\leq 0;\ 0\leq\mathrm{Im}(z)\leq\ell\},\\
\text{anti-flag state}:\lineup \ \ \ \mathscr{R} = \{z|\mathrm{Re}(z)= 0\}\cup\{z|\mathrm{Re}(z)\geq 0;\ 0\leq\mathrm{Im}(z)\leq\ell\}.
\end{eqnarray}
This is shown in figure \ref{fig:flags}. The regions are given as a union of two subsets. The first part is the positive imaginary axis, and represents the ``pole" of the flag.  The second part is a horizontal, semi-infinite strip of height $\ell$---this is the ``flag" attached to the pole. For the flag state the strip extends in the negative direction, and for the anti-flag state it extends in the positive direction. The boundary of the open string lies on the top and bottom horizontal edges of the strips. The point $z=i\ell$ is where the top horizontal edge of the strip intersects the pole, and will be called the {\it slit}. The slit is important, since this is where the horizontal strip detaches from the remaining surface at degeneration. The point $z=+\infty$ at the extreme end of the horizontal strip will be called the {\it puncture} on the anti-flag surface; likewise, $z=-\infty$ will be called the {\it puncture} on the flag surface. This is where we place operator insertions on the flag states. If we wish to indicate the dependence of the flag and anti-flag state on the height $\ell$ of the strip, we write
\begin{equation}\flag{1} =\flag{1}_\ell,\ \ \ \ \Aflag{1} = \prescript{}{\ell}{\Aflag{1}}.\end{equation}
A convenient numerical value is $\ell=\pi$, but we would like to be free to adjust this parameter. The left and right parameterized curves of the flag and anti-flag state overlap on the pole, but one curve is shifted upwards from the other by a distance $\ell$. The curves are parameterized according~to
\begin{eqnarray}
\text{flag state}:\lineup\left\lbrace \begin{matrix}\displaystyle{\ \ z_L(\sigma_L) = \frac{i}{\pi}\gd^{-1}\sigma_L\ \ \ \ \ \ \, \in\gamma_L } \\ \displaystyle{\ \ z_R(\sigma_R)=i\ell +\frac{i}{\pi}\gd^{-1}\sigma_R \in \gamma_R}\end{matrix}\right.,\label{eq:flagzLR}\\
\text{anti-flag state}:\lineup\left\lbrace \begin{matrix}\displaystyle{\ \ z_L(\sigma_L) = i\ell +\frac{i}{\pi}\gd^{-1} \sigma_L\in\gamma_L} \\ \displaystyle{\ \ z_R(\sigma_R)=\frac{i}{\pi}\gd^{-1}\sigma_R\ \ \ \ \ \ \,\in\gamma_R   }\end{matrix}\right. .
\end{eqnarray}
Multiplying wedge states and flag states is very easy, at least pictorially. We simply translate the corresponding regions, with a possible vertical shift of $i\ell$, so that the vertical line $\gamma_R$ of the first state is attached to the vertical line $\gamma_L$ of the second state. This gives a new region with left and right parameterized curves defining the product. In passing, we note that flag states are conjugate to each other:
\begin{equation}\flag{1}^\ddag = \Aflag{1},\ \ \ \ \flag{1}^\S = \Aflag{1},\label{eq:reality_twist}\end{equation}
where the first equality concerns reality conjugation \cite{Tensor} and the second concerns twist conjugation, as defined for example in appendix A of \cite{simple}.

We now add operator insertions. Consider an open string state $|\mathcal{O}\rangle$, which we represent as a unit half disk $\mathscr{D}_{1/2}$ with a vertex operator $\mathcal{O}(0)$ inserted at the origin. Writing the local coordinate on the half-disk as $\rho$, we can transform to the horizontal strip of the flag state in the sliver frame using
\begin{equation}
z = r(\rho) = \frac{\ell}{\pi}\ln(\rho).
\end{equation}
The conformal transformation inserts the vertex operator at the puncture of the flag surface. This defines a flag state with operator insertion, which we write as $\flag{\mathcal{O}}$. Similarly, given an open string state $|\widetilde{\mathcal{O}}\rangle$ with unit half-disk coordinate $\widetilde{\rho}$, we can transform to the horizontal strip of the anti-flag state in the sliver frame using
\begin{equation}
z=\widetilde{r}(\widetilde{\rho}) =-\frac{\ell}{\pi}\ln(-\widetilde{\rho}).
\end{equation}
The conformal transformation inserts the vertex operator at the puncture of the anti-flag surface. This defines the anti-flag state with an operator insertion, which we write as $\Aflag{\widetilde{\mathcal{O}}}$. Generally, $\mathcal{O}$ and $\widetilde{\mathcal{O}}$ will be boundary condition changing operators, which implies that the open string boundary condition on the top horizontal edge of the flag will be different from the open string boundary condition on the bottom horizontal edge. Specifically, the bottom edge will carry $\BCFT_0$ boundary conditions, while the top edge will carry $\BCFT_*$ boundary conditions.

We have not given the most general definition of what might be called a ``flag state." The identification between $\sigma_L$ and $\sigma_R$ on the ``pole" could have been different, and there are other ways to shape the flags. The flag states we have chosen to descibe multiply relatively easily with wedge states, but there may be other interesting choices. Other possible flag states can be obtained through midpoint-preserving reparameterizations \cite{general_frames}. In fact, there is a special midpoint-preserving reparameterization which relates our choice of flag states for different values of the strip height $\ell$. It is given by the BPZ odd part of Schnabl's $\mathcal{L}_0$ \cite{RZ}:
\begin{equation}\mathcal{L}^- = \mathcal{L}_0 - \mathcal{L}_0^\star.\end{equation} 
In particular,
\begin{equation}\lambda^{\frac{1}{2}\mathcal{L}^-}\flag{\mathcal{O}}_\ell = \flag{\mathcal{O}}_{\lambda\ell},\ \ \ \ \lambda^{\frac{1}{2}\mathcal{L}^-}\prescript{}{\ell}{\Aflag{\widetilde{\mathcal{O}}}} = \prescript{}{\lambda\ell}{\Aflag{\widetilde{\mathcal{O}}}}.\label{eq:Lmflag}\end{equation}
Note that this does not produce a conformal transformation of the operator insertions in the local coordinates $\rho$ and $\widetilde{\rho}$.

\subsection{Flag State Wavefunctionals}

To develop some intuition for flag states, it is helpful to represent them as functionals of the left and right halves of the string \cite{Gross_Taylor,RSZhalf}. Our discussion is intended to be schematic, since this representation is inconvenient for precise calculation.\footnote{Precise calculations are nevertheless possible. The furthest development in this direction is given in the closely related Moyal formalism of \cite{Bars}.} Let us assume that the matter $\BCFT$ contains a free boson $X(z,\overline{z})$ and that we are only concerned with the dependence of the string field on this factor of the $\BCFT$. In this case, the string field can be expressed as a functional of an open string curve in spacetime $x(\sigma),\sigma\in[0,\pi]$
\begin{equation}\langle x(\sigma)|\Psi\rangle=\Psi[x(\sigma)],\end{equation}
where $|x(\sigma)\rangle$ is an eigenstate of the operator $X(\xi,\overline{\xi})$ with $\xi=e^{i\sigma}$ and $x(\sigma)$ is the eigenvalue. More specifically, we express the string field as a functional of the left and right half curves of the string:
\begin{equation}\Psi[l(\sigma),r(\sigma)],\end{equation}
where
\begin{eqnarray}
l(\sigma) \lineup = x(\sigma),\ \ \ \ \ \ \ \, \ \ \sigma\in[0,\pi/2],\nonumber\\
r(\sigma) \lineup = x(\pi-\sigma),\ \ \ \ \sigma\in[0,\pi/2].
\end{eqnarray}
In this representation, the open string star product formally corresponds to a matrix product with ``half string indices:"
\begin{equation}\Psi*\Phi[l(\sigma),r(\sigma)] = \int \left(\prod_{\sigma=0}^{\pi/2}dw(\sigma)\right)\Psi[l(\sigma),w(\sigma)]\Phi[w(\sigma),r(\sigma)].\end{equation}
Now consider the flag state $\flag{1}$. The parameter $\sigma_L$ on the left curve $\gamma_L$ is in two parts identified either with the parameter $\sigma_\rho$ of the local coordinate on the horizontal strip $\rho = e^{i\sigma_\rho}$, or with the parameter $\sigma_R$ on the right curve $\gamma_R$. Using \eq{flagzLR}, one can show that the precise identification is given by
\begin{eqnarray}
\sigma_L\in[0,\sigma_\mathrm{slit}]:\ \ \sigma_L\lineup = \sigma_\flg(\sigma_\rho) \, = \gd (\ell\sigma_\rho),\ \ \ \ \ \ \ \ \ \ \ \ \ \ \ \ \ \ \  \ \sigma_\rho\in[0,\pi],\ \ \ \ \ \ \ \label{eq:sflag}\\
\sigma_L\in[\sigma_\mathrm{slit},\pi/2]:\ \ \sigma_L\lineup = \sigma_\pole(\sigma_R) = \gd(\pi\ell+\gd^{-1}\sigma_R),\ \ \ \ \ \ \ \sigma_R\in[0,\pi/2]\phantom{\Big)},\ \ \ \ \ \ \ \label{eq:spole}
\end{eqnarray}
where $\sigma_\mathrm{slit}$ is the point where $\sigma_L$ coincides with the slit of the flag surface: 
\begin{equation}\sigma_\mathrm{slit} \equiv \sigma_\pole(0) = \sigma_\flg(\pi) = \gd\, \pi\ell .\end{equation}
Since the ``pole" of the flag surface is infinitely thin, the flag state wavefunctional will vanish unless the left and right curves $l(\sigma)$ and $r(\sigma)$ coincide when the parameters $\sigma$ are appropriately related through \eq{spole}. Meanwhile, the flag region of the surface produces the $SL(2,\mathbb{R})$ vacuum functional
\begin{equation}\Omega[x(\sigma)] = \langle x(\sigma)|0\rangle,\end{equation}
where the curve $x(\sigma)$ must coincide with the left curve $l(\sigma)$ when the parameters $\sigma$ are related through \eq{sflag}. Therefore the flag state wavefunctional takes the form
\begin{equation}\flag{1}[l(\sigma),r(\sigma)] = \Omega\Big[l(\sigma_\flg(\sigma))\Big]\delta\Big[l(\sigma_\pole(\sigma))-r(\sigma)\Big].\end{equation}
The dependence on the left half curve $l(\sigma)$ is distributed between the delta functional and the $SL(2,\mathbb{R})$ vacuum functional. The anti-flag state functional takes the form
\begin{equation}\Aflag{1}[l(\sigma),r(\sigma)] = \delta\Big[l(\sigma)-r(\sigma_\pole(\sigma))\Big]\Omega\Big[r(\sigma_\flg(\sigma))\Big].\end{equation}
We may formally compute the star product of an anti-flag state with a flag state:
\begin{eqnarray}
\lineup \Aflag{1}*\flag{1}[l(\sigma),r(\sigma)] \nonumber\\
\lineup = \!\int\! \left(\prod_{\sigma=0}^{\pi/2}dw(\sigma)\!\!\right)\!\delta\Big[l(\sigma)-w(\sigma_\pole(\sigma))\Big]\Omega\Big[w(\sigma_\flg(\sigma))\Big]\Omega\Big[w(\sigma_\flg(\sigma))\Big]\delta\Big[w(\sigma_\pole(\sigma))-r(\sigma)\Big].\ \ \ \ \ \ \ \ 
\end{eqnarray}
Factorizing the measure,
\begin{equation}\prod_{\sigma=0}^{\pi/2}dw(\sigma) = \prod_{\sigma=0}^{\sigma_\mathrm{slit}}dw(\sigma)\prod_{\sigma=\sigma_{\mathrm{slit}}}^{\pi/2}dw(\sigma),\end{equation}
we can write 
\begin{eqnarray}
\Aflag{1}*\flag{1}[l(\sigma),r(\sigma)] \lineup = \int\left(\prod_{\sigma=\sigma_\mathrm{slit}}^{\pi/2}dw(\sigma)\right)\delta\Big[l(\sigma)-w(\sigma_\pole(\sigma))\Big]\delta\Big[w(\sigma_\pole(\sigma))-r(\sigma)\Big] \nonumber\\
\lineup\ \ \ \ \times\int  \left(\prod_{\sigma=0}^{\sigma_\mathrm{slit}}dw(\sigma)\right)\Omega\Big[w(\sigma_\flg(\sigma))\Big]\Omega\Big[w(\sigma_\flg(\sigma))\Big].\ \ \ \ \ \ \ 
\end{eqnarray}
Integrating the first factor eliminates a delta functional, and integrating the second factor produces a constant which, at the level of the present analysis, can be set to $1$. Therefore
\begin{equation}
 \Aflag{1}*\flag{1}[l(\sigma),r(\sigma)]  = \delta[l(\sigma)-r(\sigma)].
\end{equation}
The right hand side is the half-string functional of the identity string field. We can also multiply the flag states in the opposite order to find
\begin{equation}
\flag{1}*\Aflag{1}[l(\sigma),r(\sigma)] =  \Omega\Big[l(\sigma_\flg(\sigma))\Big]\delta\Big[l(\sigma_\pole(\sigma))-r(\sigma_\pole(\sigma))\Big]\Omega\Big[r(\sigma_\flg(\sigma))\Big].
\end{equation} 
This is a star algebra projector, though it is somewhat different from projectors that have been studied before. It can be seen as an amalgam between the sliver state and the identity string field. For $\sigma\in[0,\sigma_\mathrm{slit}]$ it is a left/right factorized functional of $l(\sigma)$ and $r(\sigma)$, like the sliver state and other rank one projectors \cite{projectors}. For $\sigma\in[\sigma_\mathrm{slit},\pi/2]$, it is a delta functional between $l(\sigma)$ and $r(\sigma)$, like the identity string field.

The above analysis shows that the flag states generates an isometry on the vector space of half string functionals, while the anti-flag state implements a partial isometry. If one wants an analogy, one might think about left/right shift operators on the Hilbert space of square summable sequences. But this is not right, since the flag times the anti-flag is not the identity minus a finite rank projector. A better analogy can be found by considering an infinite tensor product of distinguishable harmonic oscillators, with an orthonormal basis of energy eigenstates
\begin{equation}|n_1\rangle \otimes |n_2\rangle \otimes |n_3\rangle\otimes . . . ,\ \ \ n_i=0,1,2...\ .\end{equation}
The flag state is analogous to an operator which adds an oscillator ground state at the beginning of the chain of tensor products:
\begin{equation}
\text{flag\ state}:\ \ \ \ \ |n_1\rangle\otimes |n_2\rangle\otimes |n_3\rangle\otimes ... \ \longrightarrow\  |0\rangle\otimes |n_1\rangle\otimes |n_2\rangle\otimes |n_3\rangle \otimes ...\ .
\end{equation}
The anti-flag state is analogous to a operator which deletes the oscillator ground state at the beginning, and otherwise gives zero: 
\begin{equation}\text{anti-flag\ state}:\ \ \ \ \  |n_1\rangle \otimes |n_2\rangle \otimes | n_3\rangle \otimes ...\ \longrightarrow\ \delta_{n_1=0}\,  |n_2\rangle \otimes |n_3\rangle \otimes ...\ .\end{equation}
Acting with the anti-flag followed by the flag projects the first oscillator onto the ground state, and leaves the remaining oscillators untouched. 

\begin{figure}
\begin{center}
\resizebox{5.5in}{1.7in}{\includegraphics{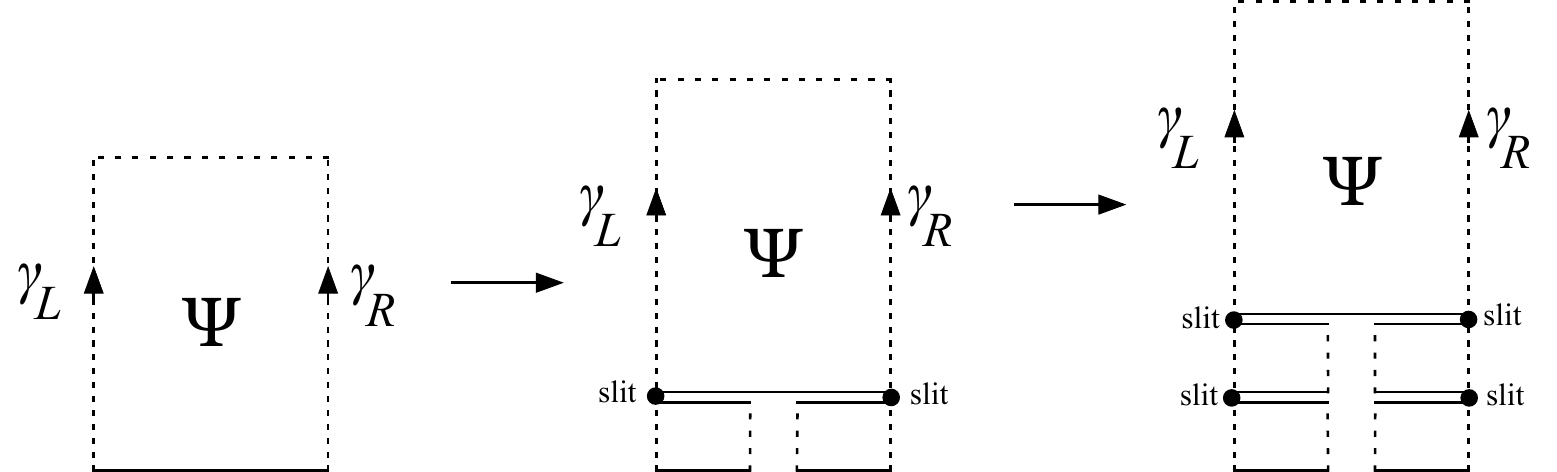}}
\end{center}
\caption{\label{fig:flagfield} The field redefinition \eq{equivalence} effectively attaches a flag and antiflag to the surface of the state of the target D-brane system. Repeated field redefinition attaches more flags to the flags.}
\end{figure}

The intertwining fields are schematically related to flag states through:
\begin{equation}\Sigma\sim\flag{1}\ \ \ \ \Sigmabar\sim\Aflag{1},\end{equation}
where for present purposes we ignore boundary condition changing operators. The field redefinition \eq{equivalence} therefore corresponds to the transformation
\begin{equation}\Psi\to \flag{1}\Psi\Aflag{1}.\end{equation}
The resulting surface in the sliver frame is shown in figure \ref{fig:flagfield}. In terms of half-string functionals, the transformation corresponds to
\begin{equation}
\Psi[l(\sigma),r(\sigma)]\to \Omega\Big[l(\sigma_\flg(\sigma))\Big]\Psi\Big[l(\sigma_\pole(\sigma)),r(\sigma_\pole(\sigma))\Big]\Omega\Big[r(\sigma_\flg(\sigma))\Big].\label{eq:fieldfunc}
\end{equation}
From this we can see why the field redefinition \eq{equivalence} does not map into the whole state space of the reference D-brane. An arbitrary functional in the target background is transformed into a functional which only has arbitrary dependence on a portion of the string which is closer to the midpoint; namely $l(\sigma)$ and $r(\sigma)$ when $\sigma\in[\sigma_\mathrm{slit},\pi/2]$. The dependence on the remainder of the string is determined by a fixed functional. Following subsection \ref{subsec:BI}, we can iterate field redefinitions between string field theories of different backgrounds. In the sliver frame, this creates surfaces with multiple flags, as shown in figure \ref{fig:flagfield}. In terms of functionals, further transformation of \eq{fieldfunc} leads to 
\begin{eqnarray}
\lineup \Omega\Big[ l(\sigma_\flg(\sigma))\Big] \Omega \Big[l(\sigma_\flg(\sigma_\pole(\sigma)))\Big] \Psi \Big[l(\sigma_\pole(\sigma_\pole(\sigma))),r(\sigma_\pole(\sigma_\pole(\sigma))) \Big]\nonumber\\
\lineup\ \ \ \ \ \ \ \ \ \ \ \ \ \ \ \ \ \ \ \ \ \ \ \ \ \ \ \ \ \ \ \ \ \ \ \ \ \ \ \ \ \ \ \ \ \ \ \ \ \ \ \ 
\times \Omega \Big[r(\sigma_\flg(\sigma_\pole(\sigma))) \Big] \Omega\Big[r(\sigma_\flg(\sigma)) \Big]. 
\end{eqnarray}
Now the functional has arbitrary dependence over an even smaller portion of the half-string curves $l(\sigma)$ and $r(\sigma)$ given by $\sigma\in[\sigma_\pole(\sigma_\mathrm{slit}),\pi/2]$. With each iteration of the field redefinition, the freedom of the resulting functional is restricted to an ever smaller neighborhood of the midpoint, and the resulting space of states is an increasingly smaller subset of the full state space. The picture is reminiscent of ``Hilbert's hotel;" there is always room to add more flags without discarding any information contained in the original state.

\subsection{The Flag-anti-flag Surface}

\begin{figure}
\begin{center}
\resizebox{2in}{2.2in}{\includegraphics{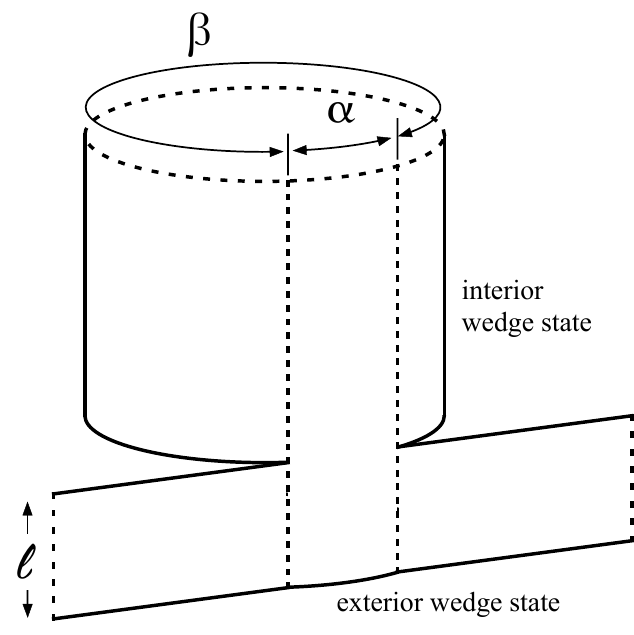}}
\end{center}
\caption{\label{fig:flag_geometry} The flag-anti-flag surface.}
\end{figure}

To perform calculations with the solution, it is necessary to evaluate correlation functions on the surfaces formed by gluing flag states and wedge states. This can be accomplished by finding a conformal transformation which relates correlators on these surfaces to correlators on the upper half plane. This is a nontrivial uniformization problem, and generally it will not be possible to find an explicit solution. Thankfully, it is tractable in the most useful case: the surface formed by gluing two wedge states together with a flag and antiflag state. Modulo operator insertions, this appears when evaluating expressions of the form
\begin{equation}\Tr\Big(\Omega^\alpha\flag{1}\Omega^\beta\Aflag{1}\Big).\label{eq:surface}\end{equation}
We call this the {\it flag-anti-flag surface}. In the sliver frame, it can be visualized as a semi-infinite vertical cylinder whose bottom edge has a segment glued to a horizontal infinite strip. This is shown in figure \ref{fig:flag_geometry}. We will call $\Omega^\alpha$ the {\it reference wedge state}, since it will carry boundary conditions of the reference string field theory. Similarly, we call $\Omega^\beta$ the {\it target wedge state}, since it will carry boundary conditions of the target background described by the solution. Correlators on the flag-anti-flag surface are needed, for example, to compute the coefficients of the solution in the Fock basis. 

We fix the origin of the sliver coordinate $z$ on the flag-anti-flag surface to sit on the open string boundary half way between the vertical edges of the reference wedge state. We look for a conformal transformation $z=\mathcal{F}(u)$ from a coordinate $u$ on the upper half plane such that $u=0$ maps to $z=0$. This can be accomplished in two steps. First we cut the flag-anti-flag surface down a vertical line in the middle of the target wedge state to form a symmetric, seven sided polygon. The polygon can be transformed to the upper half plane using the Schwarz-Christoffel map. The Schwarz-Christoffel map is only determined up to an $SL(2,\mathbb{R})$ transformation of the upper half plane. We fix this ambiguity by requiring that $u_\mathrm{polygon}=\pm 1$ on the upper half plane maps to the points $z = i\ell \pm\frac{\alpha+\beta}{2}$ where the cut reaches the open string boundary, and further that $u_\mathrm{polygon}=\infty$ maps to $z=i\infty$. The coordinate $u_\mathrm{polygon}$ does not represent the upper half plane for the flag-anti-flag surface, since we need to re-glue the cut. This can be accomplished by a further conformal transformation 
\begin{equation}u_\mathrm{polygon} = \frac{u}{\sqrt{1+u^2}},\end{equation}
where $u$ is the upper half plane coordinate for the flag-anti-flag surface. The relation between the sliver coordinate $z$ and the upper half plane coordinate $u$ is finally given by
\begin{equation}
z=\mathcal{F}(u) = \frac{2\ell}{\pi}\left(\frac{p(1+s^2)}{s^2-p^2}\tan^{-1} u +\tanh^{-1}\frac{u}{p}\right),\label{eq:F}
\end{equation}
where 
\begin{equation}s>p>0\end{equation} 

\begin{figure}
\begin{center}
\resizebox{4.7in}{3.5in}{\includegraphics{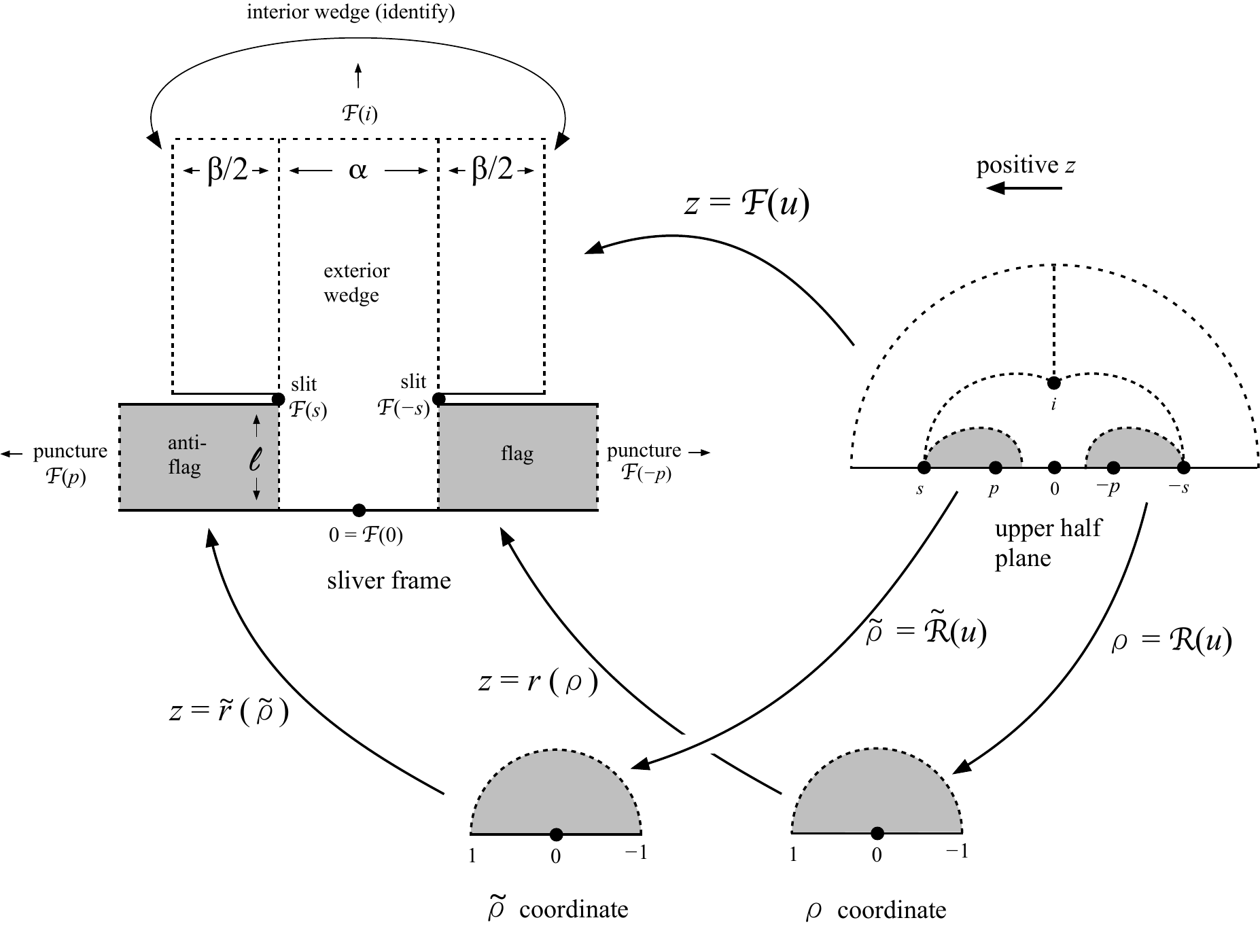}}
\end{center}
\caption{\label{fig:conformal} Conformal transformations relating the upper half plane coordinate $u$, the sliver coordinate $z$, and the local coordinates $\rho$ and $\widetilde{\rho}$ of the punctures of the flag and anti-flag state.}
\end{figure}

\noindent are positive real numbers. The points $u=\pm s$ map to the two slits, while $u=\pm p$ map respectively to the punctures of the anti-flag and flag state. The point $u=i$ maps to the ``top" of the cylinder at $z=+i\infty$. The geometry of the conformal transformation is summarized in figure \ref{fig:conformal}. Considering that the flag-anti-flag surface is rather complicated, the conformal transformation from the upper half plane is fairly simple. Unfortunately, as often happens with the Schwarz-Christoffel map, the inverse transformation cannot be expressed in closed form. The upper half plane moduli $p$ and $s$ can be related to the reference and target wedge parameters through 
\begin{eqnarray}
\alpha \lineup = \frac{4\ell}{\pi}\left[\frac{p(1+s^2)}{s^2-p^2}\tan^{-1}s+\tanh^{-1}\frac{p}{s}\right]\phantom{\Bigg)},\label{eq:alphaps}\\
\beta \lineup = \frac{4\ell}{\pi}\left[\frac{p(1+s^2)}{s^2-p^2}\tan^{-1}\frac{1}{s}-\tanh^{-1}\frac{p}{s}\right]\phantom{\Bigg)}.\label{eq:betaps}
\end{eqnarray}
Again, the transformation cannot be inverted in closed form. 

\begin{figure}
\begin{center}
\resizebox{2.7in}{2.5in}{\includegraphics{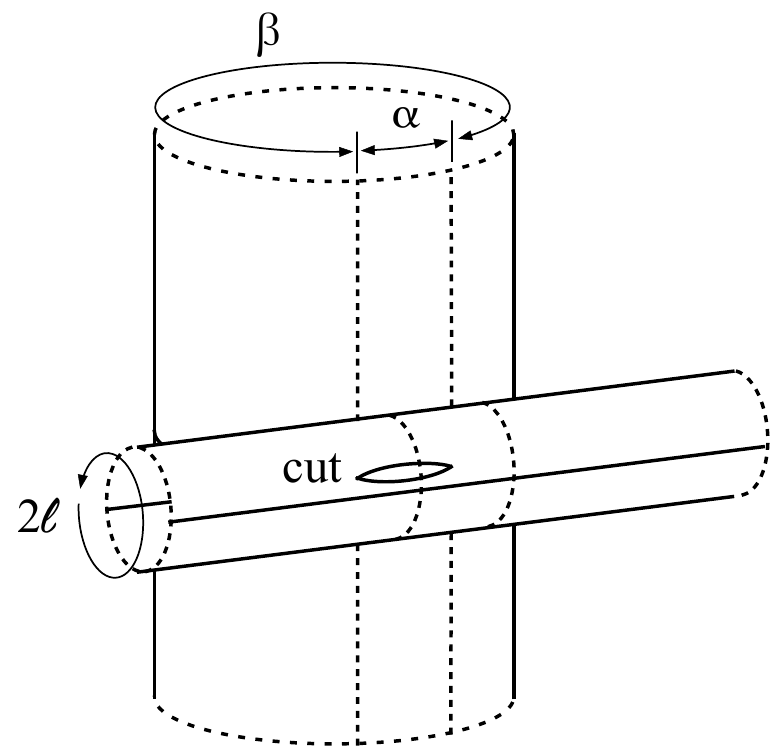}}
\end{center}
\caption{\label{fig:double} Doubled holomorphic representation of the flag-anti-flag surface.}
\end{figure}

Let us describe a few other useful coordinate systems on the flag-anti-flag surface:
\begin{itemize}
\item The local coordinate $\rho$ for the puncture on the flag state, and likewise $\widetilde{\rho}$ for the anti-flag state, cover a pair of regions in the upper half plane surrounding $u=\pm p$. The upper half plane and puncture coordinates are related by
\begin{equation}\rho = \mathcal{R}(u) = r^{-1}\circ\mathcal{F}(u),\ \ \ \ \widetilde{\rho} = \widetilde{\mathcal{R}}(u) = \widetilde{r}^{-1}\circ\mathcal{F}(u).\end{equation}
Often we will want to insert primary operators $\mathcal{O}(0)$ and $\widetilde{\mathcal{O}}(0)$ respectively at the punctures of the flag and anti-flag state. On the upper half plane, this will produce operator insertions
\begin{equation}(\rho')^{-h}\mathcal{O}(-p),\ \ \ \ \ (\rho')^{-\widetilde{h}}\widetilde{\mathcal{O}}(p),\end{equation} 
where $(h,\widetilde{h})$ are the respective weights and the conformal factor is given by
\begin{equation}\rho' = \mathcal{R}'(-p)=\frac{s+p}{2p(s-p)}\exp\left[\frac{2p(1+s^2)}{s^2-p^2}\Big(\tan^{-1}s-\tan^{-1}p\Big)\right].\end{equation}
\item The transformation $\mathcal{F}(u)$ has a branch cut, originating from the inverse tangent, which extends on the imaginary axis from $u=i$ through $i\infty$, and then from $u=-i\infty$ to $u=-i$. The branch cut leads to an identification between the vertical lines $z=\pm \frac{\alpha+\beta}{2} + i(\ell+ y)$ for $y>0$ on the flag-anti-flag surface. In this way, the coordinate $z$ cuts the surface of the target wedge state into two equal parts. Sometimes it is convenient to use a ``dual coordinate" $\widetilde{z}$ which instead cuts the reference wedge state into two equal parts. This coordinate is given~by\footnote{For $\mathrm{Im}(\widetilde{z})<0$ we need to be slightly careful about the branch of the coordinate system, since the flags sit on top of each other; a single coordinate $\widetilde{z}$ may represent two distinct points on the flag-anti-flag surface.}
\begin{equation}\widetilde{z} = z - \frac{\alpha+\beta}{2}\mathrm{sign}(\mathrm{Re}(z))-i\ell.\end{equation}
The origin of $\widetilde{z}$ sits on the open string boundary half way between the vertical edges of the target wedge state. The new coordinate $\widetilde{z}$ is related to a dual coordinate on the upper half plane,
\begin{equation}\widetilde{u} = -\frac{1}{u},\end{equation}
through
\begin{equation}
\widetilde{z}=\widetilde{\mathcal{F}}(\widetilde{u}) = \frac{2\ell}{\pi}\left(\frac{\widetilde{p}(1+\widetilde{s}^2)}{\widetilde{p}^2-\widetilde{s}^2}\tan^{-1} \widetilde{u} -\tanh^{-1}\frac{\widetilde{u}}{\widetilde{p}}\right),\label{eq:dual_coord}
\end{equation}
where
\begin{equation}\widetilde{p}>\widetilde{s}>0\end{equation}
are respectively the pre-images of the puncture and slits on the flag-anti-flag surface.
\item When computing correlators of purely holomorphic and antiholomorphic fields on the disk, it is often useful to employ the doubling trick. The doubled holomorphic worldsheet corresponding to the flag-anti-flag surface is slightly nontrivial because of the complicated shape of the open string boundary. It can be visualized as a vertical and horizontal infinite cylinder sewn together along a small cut. This is illustrated in figure \ref{fig:double}.
\end{itemize}

\noindent This and other related formulas for conformal transformations of the flag-anti-flag surface are summarized in appendix \ref{app:conformal}.

\subsection{Multiplying Flag States}

Consider open string states  $|\mathcal{O}\rangle$ and $|\widetilde{\mathcal{O}}\rangle$ defined by acting matter sector vertex operators $\mathcal{O}(0)$ and $\widetilde{\mathcal{O}}(0)$ on the $SL(2,\mathbb{R})$ vacuum of $\BCFT_0$. These vertex operators may be used to define a flag state $\flag{\mathcal{O}}$ and anti-flag state $\Aflag{\widetilde{\mathcal{O}}}$. The central purpose behind the definition of flag states is to realize the following relation: 
\begin{equation}\Aflag{\widetilde{\mathcal{O}}}*\flag{\mathcal{O}} = \frac{\langle \widetilde{\mathcal{O}},\mathcal{O}\rangle_{\mathrm{matter}}}{\langle 0|0\rangle_\mathrm{matter}}\times\mathrm{identity\ string\ field},\phantom{\Bigg{)}}\label{eq:Aflagflag}\end{equation}
where on the left hand side we have the star product of an anti-flag state and a flag state, and on the right hand side is the matter sector BPZ inner product of the states $|\widetilde{\mathcal{O}}\rangle$ and $|\mathcal{O}\rangle$ divided by the norm of the matter sector $SL(2,\mathbb{R})$ vacuum in $\BCFT_0$. This relation holds assuming that the OPE of $\widetilde{\mathcal{O}}(x)$ and $\mathcal{O}(x)$ does not generate operators of negative conformal dimension, and the only operator which can appear at dimension $0$ is the identity. This assumption is valid generically for matter sector operators. We have already given a somewhat schematic demonstration of this identity using half-string functionals. A more precise justification can be given using worldsheet correlation  functions, where the central mechanism behind \eq{Aflagflag} is that the star product produces a degeneration where the horizontal strip disconnects from the remaining surface. Correlators then factorize into a contribution from the remaining surface, and a contribution from the horizontal strip which simply produces the BPZ inner product of the matter sector states. Further discussion of the factorization of disk correlators at a seperating degeneration can be found in \cite{projectors}. 

Let us confirm \eq{Aflagflag} directly using the formulas we have derived for the flag-anti-flag surface. For simplicity, assume that $\mathcal{O}$ and $\widetilde{\mathcal{O}}$ are primary operators of weight $h$, and let $(...)$ denote a generic test state, which for present purposes can be taken as a wedge state with some operator insertions. We would like to show that 
\begin{equation}\lim_{\alpha\to 0}\Tr\Big[\big(\, ...\,\big)\Aflag{\widetilde{\mathcal{O}}} \Omega^\alpha\flag{\mathcal{O}}\Big]=\frac{\langle \widetilde{\mathcal{O}}|\mathcal{O}\rangle_\matter}{\langle 0|0\rangle_\matter}\Tr[\big(\, ...\,\big)].\label{eq:sep_limit}\end{equation}
The left hand side defines a correlation function on the flag-anti-flag surface, which we transform to the upper half plane 
\begin{equation}\Tr\Big[\big(\, ...\,\big)\Aflag{\widetilde{\mathcal{O}}} \Omega^\alpha\flag{\mathcal{O}}\Big] =(\rho')^{-2h}\Big\langle\mathcal{F}^{-1}\circ\big(\, ...\,\big) \widetilde{\mathcal{O}}(p)\mathcal{O}(-p)\Big\rangle_{\mathrm{UHP}}.\end{equation}
Inside the correlator we use $(...)$ to represent the operator insertions defining the test state in the sliver frame. Now we take the limit $\alpha\to 0$, where the reference wedge state degenerates to the identity string field. To do this, we need to understand how to take the appropriate limit of the upper half plane moduli $p$ and $s$. This is accomplished by 
\begin{equation}p,s\to 0,\ \ \ \ \frac{p}{s^2} =\chi =\mathrm{constant}.\label{eq:sep_deg}\end{equation}
The constant $\chi$ encodes information about the width of the target wedge state. Specifically,
\begin{equation}
\alpha = \frac{8\ell}{\pi}\chi s,\ \ \ \ \ \beta = 2\ell \chi,\ \ \ \ \ (s\to 0).
\end{equation}
Since the test state can be arbitrary, we do not need to commit to a specific value for $\chi$. In the limit $\alpha\to 0$ the operators $\mathcal{O}$ and $\widetilde{\mathcal{O}}$ collide and we can use the OPE
\begin{equation}\widetilde{\mathcal{O}}(\chi s^2 )\mathcal{O}(-\chi s^2) = \frac{\langle\widetilde{\mathcal{O}}|\mathcal{O}\rangle_\mathrm{matter}}{\langle 0|0\rangle_\matter}\times\frac{1}{(2\chi s^2)^{2h}}+...\ .\end{equation}
The OPE divergence is compensated by a vanishing conformal factor
\begin{equation}(\rho')^{-2h}= (2\chi s^2)^{2h}+...\ ,\end{equation}
so we are left with 
\begin{equation}\lim_{\alpha\to 0}\Tr\Big[\big(\, ...\,\big)\Aflag{\widetilde{\mathcal{O}}} \Omega^\alpha\flag{\mathcal{O}}\Big]=\frac{\langle \widetilde{\mathcal{O}}|\mathcal{O}\rangle_\matter}{\langle 0|0\rangle_\matter}\lim_{\alpha\to 0}\Big\langle\mathcal{F}^{-1}\circ\big(\, ...\,\big) \Big\rangle_{\mathrm{UHP}}.\end{equation}
Next we note that
\begin{equation}
\lim_{\alpha\to 0}\mathcal{F}(u) = i\ell + f_\beta(u),
\end{equation}
where
\begin{equation}f_\beta(u) = \frac{\beta}{\pi}\tan^{-1}u\end{equation}
is the familiar conformal transformation from the upper half plane to the cylinder $C_\beta$ of circumference $\beta$. This is saying that once we have taken the limit $\alpha\to 0$, the flag and anti-flag detach from the surface and we are left with a correlation function on the cylinder. The additive constant $i\ell$ shifts the origin of the coordinate system on the cylinder, and is unimportant. Therefore we find
\begin{equation}
\lim_{\alpha\to 0}\Big\langle\mathcal{F}^{-1}\circ\big(\, ...\,\big) \Big\rangle_{\mathrm{UHP}} = \big\langle\big(\,...\,\big)\big\rangle_{C_\beta} = \Tr\big[\big(\,...\,\big)\big].
\end{equation}
which gives the desired result. 

\section{Flag State Solution}
\label{sec:flag_sol}

We now describe a realization of the intertwining solution using flag states. The first step is to pick representatives of the tachyon vacuum. The natural place to look is the $KBc$ subalgebra~\cite{Okawa},\footnote{Our conventions for the $KBc$ subalgebra follow \cite{simple}} the subalgebra of states containing Schnabl's analytic solution \cite{Schnabl}. The nicest solution in this subalgebra is  the ``simple" tachyon vacuum of \cite{simple},
\begin{equation}\Psi_\mathrm{simp} = \frac{1}{\sqrt{1+K}}c(1+K)Bc\frac{1}{\sqrt{1+K}},\label{eq:simp}\end{equation}
with the homotopy operator
\begin{equation}
A = \frac{B}{1+K}.
\end{equation}
The construction based on the simple tachyon vacuum requires special considerations (related to the field $c$) which do not apply to other solutions in the $KBc$ subalgebra. Therefore it is worth considering a more general class of tachyon vacuum solutions of the Okawa form \cite{Okawa}\footnote{A representative class of tachyon vacuum solutions of the Okawa form is given by $F(K) = (1-K/\nu)^\nu$ \cite{taming}, where $\nu<0$ is the leading level in the dual $\mathcal{L}^-$ expansioin \cite{IdSing} of the tachyon vacuum. The most general form of a tachyon vacuum solution in the $KBc$ subalgebra is described in \cite{Jokel}.}
\begin{equation}\Psi_\tv = \sqrt{F} c\frac{B}{H}c\sqrt{F},\end{equation}
where $F=F(K)$ is a suitably well-behaved element of the algebra of wedge states (a real function of $K$) satisfying the conditions
\begin{equation}F(0) = 1,\ \ \ F'(0)<0,\ \ \ F(\infty)=0,\ \ \ F(K)<1.\end{equation}
For convenience we denote
\begin{equation}H \equiv \frac{1-F}{K},\label{eq:gen_tv}\end{equation}
so that homotopy operator is written
\begin{equation}A = BH.\label{eq:gen_hom}\end{equation}
The tachyon vacuum is real 
\begin{equation}\Psi_\tv^\ddag = \Psi_\tv,\end{equation}
and the homotopy operator satisfies 
\begin{equation}Q_{\Psi_\tv}A = 1,\ \ \ \ A^2=0.\end{equation}
In this way we realize postulates \hyperlink{anc:1}{{\bf 1)}} and \hyperlink{anc:2}{{\bf 2)}} of the intertwining solution. What remains is to give  a construction of the intertwining fields, i.e to realize postulate \hyperlink{anc:3}{{\bf 3)}}. For simplicity, we will assume that the tachyon vacuum solutions of $\BCFT_0$ and $\BCFT_*$ are characterized by the same choice of~$F(K)$.

\subsection{Pre-Intertwining Fields}
\label{subsec:pre_intertwining}

\begin{figure}
\begin{center}
\resizebox{4in}{2.3in}{\includegraphics{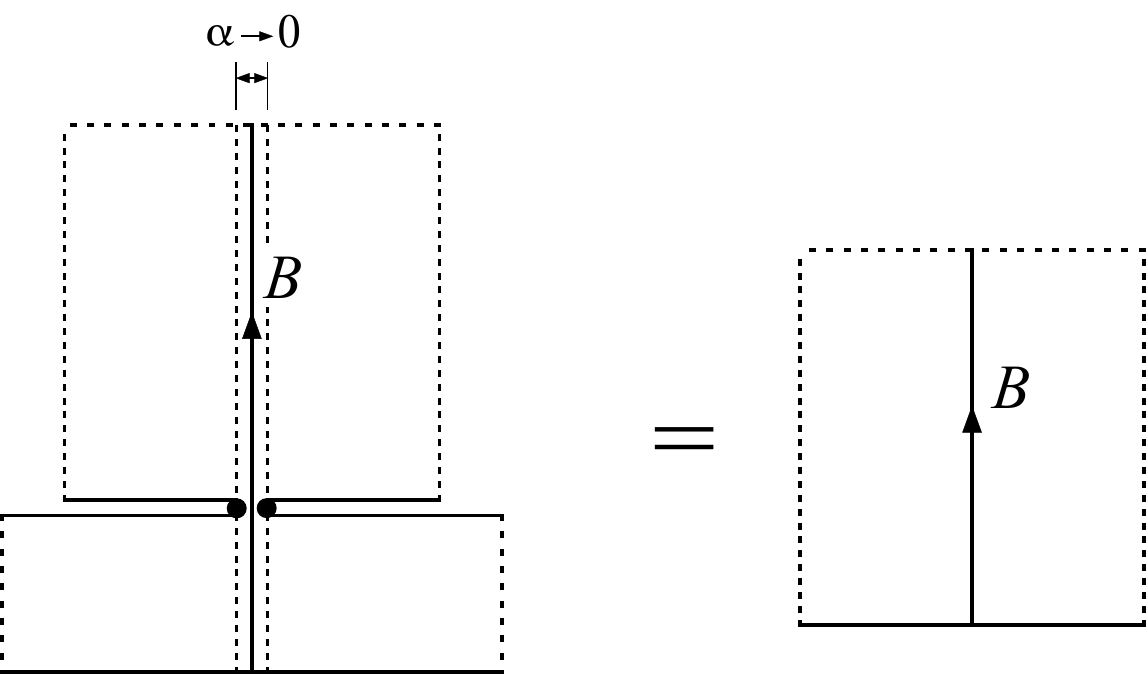}}
\end{center}
\caption{\label{fig:sbBs} If the $b$-ghost contour corresponding to the field $B$ is squeezed between the slits of a flag and anti-flag state, we can ignore the contribution from integration in the horizontal strip.}
\end{figure}

To find intertwining fields, we first look for pre-intertwining fields. They should satisfy 
\begin{equation}\Sigmabar_\pre A\Sigma_\pre = A.\label{eq:sAspre}\end{equation}
We want to solve this equation using flag states. First we introduce stretched string states $|\sigma\rangle\in \H_{0*}$ and $|\sigmabar\rangle\in\H_{*0}$ connecting $\BCFT_0$ and $\BCFT_*$ and vice versa. We assume that the states are built from matter sector vertex operators, and we normalize them according to
\begin{equation}\langle \sigmabar|\sigma\rangle_\mathrm{matter} = \langle 0|0\rangle_\matter.\end{equation}
The flag states corresponding to $|\sigma\rangle$ and $|\sigmabar\rangle$ will then satisfy
\begin{equation}\Aflag{\sigmabar}*\flag{\sigma} = \mathrm{identity\ string\ field}.\end{equation}
We can find pre-intertwining fields in the form 
\begin{eqnarray}
\Sigma_\pre \lineup = \frac{1}{\sqrt{H}}\flag{\sigma}\sqrt{H}, \label{eq:Sigmapre_gen}\\
\Sigmabar_\pre\lineup =\sqrt{H}\Aflag{\sigmabar}\frac{1}{\sqrt{H}}.\label{eq:Sigmabarpre_gen}
\end{eqnarray}
The pre-intertwining fields are reality conjugate to each other,
\begin{equation}\Sigma_\pre^\ddag = \Sigmabar_\pre,\end{equation}
assuming that $|\sigma\rangle$ and $|\sigmabar\rangle$ are likewise reality conjugate to each other. This, in turn, will imply that the intertwining solution is real. The square roots of $H$ in \eq{Sigmapre_gen} and \eq{Sigmabarpre_gen} serve to cancel the factor of $H$ in the homotopy operator of $\BCFT_*$ and create a new factor of $H$ for the homotopy operator of $\BCFT_0$. The pre-intertwining fields will then satisfy \eq{sAspre} if 
\begin{equation}\Aflag{\sigmabar}B\flag{\sigma} = B.\label{eq:sBs}\end{equation}
The field $B$ represents a vertical line integral insertion of the $b$-ghost in correlation functions in the sliver frame. The claim is that when this line integral is pinched between the slits of an anti-flag and a flag, the segment of the $b$-ghost contour integral inside the horizontal strip can be forgotten. See figure \ref{fig:sbBs}. 

\begin{figure}
\begin{center}
\resizebox{5.5in}{1.5in}{\includegraphics{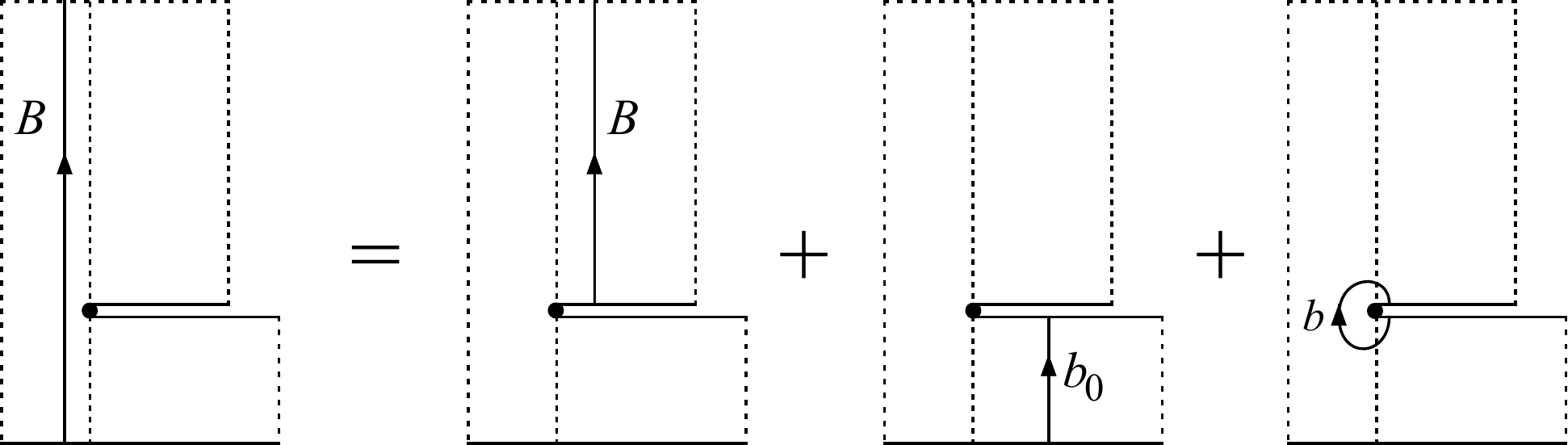}}
\end{center}
\caption{\label{fig:sBcom} Deforming the vertical contour integral corresponding to $B$ through a flag state produces three contributions, including a nontrivial contribution from integrating the $b$-ghost around the slit.}
\end{figure}

This seems like a plausible claim, but is also potentially delicate. To demonstrate \eq{sBs} it is helpful to understand what happens when we commute $B$ through a flag state $\flag{\mathcal{O}}$. Consider
\begin{equation}\Tr\Big[\big(\, ...\,\big) B\flag{\mathcal{O}}\Big],\end{equation}
where $\big(...\big)$ is a generic test state, which can be taken so that the trace represents a correlator on the flag-anti-flag surface.  Commuting $B$ through $\flag{\mathcal{O}}$ entails deforming the $b$-ghost contour from the left to the right of the slit. In doing this we obtain three contributions as shown in figure \ref{fig:sBcom}. The first is a vertical line integral to the right of the slit connecting $\mathrm{Im}(z)=i\ell$ to $\mathrm{Im}(z)=i\infty$. This represents the field $B$ multiplying $\flag{\mathcal{O}}$ from the right. The second contribution is a vertical line integral inside the horizontal strip connecting $\mathrm{Im}(z)=0$ to $\mathrm{Im}(z)=i\ell$. In the local coordinate $\rho$ on the horizontal strip, this is proportional to $b_0$. The third contribution is a contour integral of the $b$-ghost around the slit. If the slit were a generic point on the open string boundary this contour integral would vanish, since the $b$-ghost does not generally diverge on the boundary. However, the slit is special since the surface has a curvature singularity at this point, as can be seen in figure \ref{fig:double}. One way to see that the slit makes a nonzero contribution is to consider the analogous question with the contour integral of the energy momentum tensor. If there was no contribution from integrating the energy-momentum tensor around the slit, this would imply that the slit could be freely moved back and forth parallel to the real axis without changing the surface, which is clearly untrue. Therefore the difference between the $B$ ghost on either side of the slit is given by 
\begin{equation}
\Big[\flag{\mathcal{O}},B\Big] = \flag{\mathcal{O}}b +(-1)^{\mathcal{O}+1}\frac{\pi}{\ell}\flag{b_0\mathcal{O}},\label{eq:Bflag}
\end{equation} 
where commutators are graded and $\flag{\mathcal{O}}b$ is the contribution from the slit. To understand this contribution explicitly, we transform from the flag-anti-flag surface to the upper half plane. Using the doubling trick, the state $\flag{\mathcal{O}}b$ is represented as
\begin{equation}
\Tr\Big[\big(\, ...\,\big)\flag{\mathcal{O}}b\Big]=-\left\langle \mathcal{F}^{-1}\circ\big(\, ...\,\big)\mathcal{R}^{-1}\circ\mathcal{O}(0) \oint_{u=-s} \frac{du}{2\pi i}\frac{1}{\mathcal{F}'(u)}b(u)\right\rangle_\mathrm{UHP}.
\end{equation} 
The contour integral arises from integrating the $b$-ghost around the slit in the sliver frame. The sign appears from the ordering of the commutator in \eq{Bflag}. In the upper half plane the $b$-ghost does not encounter singularity at the boundary, but it turns out that $\mathcal{F}'(u)$ has a zero at $u=-s$. Therefore, the contour integral picks up a residue from a pole, 
\begin{equation}
\Tr[\big(\, ...\,\big)\flag{\mathcal{O}}b]=\mathcal{F}_b\left\langle\big(\, ...\,\big)\mathcal{R}^{-1}\circ\mathcal{O}(0) \, b(-s)\right\rangle_\mathrm{UHP},
\end{equation}
where the residue $\mathcal{F}_b$ is given by
\begin{equation}
\mathcal{F}_b\equiv \frac{\pi}{4\ell ps}\frac{1+s^2}{1+p^2}( s^2-p^2)^2 .
\end{equation}
Therefore, commuting $B$ through the slit produces an insertion of the $b$-ghost at the corresponding position in the upper half plane, multiplied by a residue factor. In a similar way, we can show that
\begin{equation}
\Big[B,\Aflag{\widetilde{\mathcal{O}}}\Big] = b\Aflag{\widetilde{\mathcal{O}}}+\frac{\pi}{\ell}\Aflag{b_0\widetilde{\mathcal{O}}}, \label{eq:BAflag}
\end{equation} 
where by definition $b\Aflag{\widetilde{\mathcal{O}}}$ produces a $b$-ghost insertion at $u=s$ in the upper half plane multiplied by the same factor $\mathcal{F}_b$.

Now we can return to the relation \eq{sBs}. Using \eq{Bflag} to move $B$ outside the flag states, we obtain
\begin{equation}\Aflag{\sigmabar}B\flag{\sigma} = \lim_{\alpha\to 0}\Big(\Aflag{\sigmabar}\Omega^\alpha\flag{\sigma}B-\Aflag{\sigmabar}\Omega^\alpha\flag{\sigma}b\Big).\end{equation}
The term with $b_0$ inside the flag does not contribute since $b_0$ commutes with the matter sector operator $\sigma$ and annihilates the $SL(2,\mathbb{R})$ vacuum. Using \eq{Aflagflag}, the first term reduces to $B$ in the $\alpha\to 0$ limit, which is the desired result. We need to see what happens with the second term. Contracting with a generic test state, 
\begin{equation}\lim_{\alpha\to 0}\Tr\Big[\big(\, ...\,\big)\Aflag{\sigmabar}\Omega^\alpha\flag{\sigma}b\Big]=\lim_{\alpha\to 0}\mathcal{F}_b\Big\langle\mathcal{F}^{-1}\circ\big(\, ...\,\big)\widetilde{\mathcal{R}}^{-1}\circ\sigmabar(0)\mathcal{R}^{-1}\circ\sigma(0)\ b(-s)\Big\rangle_\mathrm{UHP}.\label{eq:sBscor}\end{equation}
The correlator is finite and nonzero, so we have to see what happens with the residue factor. Using \eq{sep_deg} gives
\begin{eqnarray}
\lim_{\alpha\to 0}\mathcal{F}_b \lineup = \lim_{s\to 0} \frac{\pi}{4\ell \chi s^3}\frac{1+s^2}{1+\chi^2 s^4}( s^2-\chi^2 s^4)^2 \nonumber\\
\lineup = \lim_{s\to 0} \frac{\pi s}{4\ell \chi}\nonumber\\
\lineup = 0.
\end{eqnarray}
Therefore \eq{sBs} is satisfied. Note that it was important to assume that $\sigma$ and $\sigmabar$ are matter operators. This was necessary in two respects: First, so we could drop the contribution from $b_0$ acting on $|\sigma\rangle$, and second, so that no potential divergence can appear from the OPE of $b(-s)$ and $\sigma$ and $\sigmabar$ in the correlator of \eq{sBscor}.

\subsection{Solution}
\label{subsec:solution}

Having determined the pre-intertwining fields, the intertwining fields are given by 
\begin{equation}\Sigma = Q_{\Psi_\tv}(A\Sigma_\pre),\ \ \ \ \Sigmabar = Q_{\Psi_\tv}(\Sigmabar_\pre A),\end{equation}
and the construction is in principle complete. Substituting the tachyon vacuum \eq{gen_tv} and pre-intertwining fields \eq{Sigmapre_gen} and \eq{Sigmabarpre_gen} we find explicitly
\begin{eqnarray}
\Sigma \lineup = \sqrt{H}\left(\frac{1}{H}\flag{\sigma}-B\flag{Q\sigma}-\sqrt{\frac{F}{H}}Bc\sqrt{\frac{F}{H}}\flag{\sigma}+ B\flag{\sigma}\sqrt{\frac{F}{H}}H c\frac{B}{H}c\sqrt{\frac{F}{H}}\right)\sqrt{H},\\
\Sigmabar\lineup = \sqrt{H}\left(\Aflag{\sigmabar}\frac{1}{H}+\Aflag{Q\sigmabar}B-\Aflag{\sigmabar}\sqrt{\frac{F}{H}}cB\sqrt{\frac{F}{H}}+\sqrt{\frac{F}{H}}c\frac{B}{H}cH\sqrt{\frac{F}{H}}\Aflag{\sigmabar}B\right)\sqrt{H},
\end{eqnarray}
and the solution can be written as 
\begin{eqnarray}
\lineup\!\!\!\!\! \Psi_* = \sqrt{F}c\frac{B}{H}c\sqrt{F}\nonumber\\
\lineup\ \  -\sqrt{\!H}\!\!\left(\!\frac{1}{H}\flag{\sigma}\!-\! B\flag{Q\sigma}\!-\!\sqrt{\!\frac{F}{H}}Bc\sqrt{\!\frac{F}{H}}\!\flag{\sigma}\!\right)\!\!\sqrt{\!\frac{F}{H}}Hc\frac{B}{H}cH\sqrt{\!\frac{F}{H}}\!\!\left(\!\Aflag{\sigmabar}\frac{1}{H}\!+\!\Aflag{Q\sigmabar}B\!-\!\Aflag{\sigmabar}\!\sqrt{\!\frac{F}{H}}cB\sqrt{\!\frac{F}{H}\!}\right)\!\!\sqrt{\!H}\nonumber\\
\lineup\ \ -\sqrt{\!H}\!\!\left(\!\frac{1}{H}\flag{\sigma}\!-\! B\flag{Q\sigma}\!-\!\sqrt{\!\frac{F}{H}}Bc\sqrt{\!\frac{F}{H}}\!\flag{\sigma}\!\right)\!\sqrt{\!\frac{F}{H}}HcB\frac{1}{H}[c,F]\frac{1}{H}cH\sqrt{\!\frac{F}{H}}\Aflag{\sigmabar}B\sqrt{\! H}\nonumber\\
\lineup\ \ - \sqrt{\! H}B\flag{\sigma}\sqrt{\!\frac{F}{H}}HcB\frac{1}{H}[c,F]\frac{1}{H}cH\sqrt{\!\frac{F}{H}}\!\!\left(\!\Aflag{\sigmabar}\frac{1}{H}\!+\!\Aflag{Q\sigmabar}B\!-\!\Aflag{\sigmabar}\!\sqrt{\!\frac{F}{H}}cB\sqrt{\!\frac{F}{H}\!}\right)\!\!\sqrt{\!H}\nonumber\\
\lineup\ \ -\sqrt{\! H}B\flag{\sigma}\sqrt{\!\frac{F}{H}}HcB\frac{1}{H}[c,F]\frac{1}{H}[c,F]\frac{1}{H}cH\sqrt{\!\frac{F}{H}}\Aflag{\sigmabar}B\sqrt{\! H}.\label{eq:exp_sol}
\end{eqnarray}
Here we multiplied out the cross terms generated by the fourth additive contributions to $\Sigma$ and~$\Sigmabar$, since these are the only cross terms which allow for some reduction in the star product with the tachyon vacuum. If we multiply everything out, the solution contains some 17 terms (or more, depending on whether the commutators are further expanded). The purpose here is to show what the solution looks like in full detail. However, there is not much interesting structure to observe at this level. Unlike the solution of~\cite{KOSsing}, simplifications do not occur since $B$ does not commute through the flag states and $\flag{Q\sigma}$ generally has no useful relation to~$\flag{\sigma}$. The flag states prevent  interesting interplay between the algebraic structures of the reference and target string field theories, and in this sense are somewhat analogous to insertion of a tensor product. 

In section \ref{sec:consistency} we show that the solution appears to work consistently if $F(K)$ falls off faster than $1/K$ for large $K$. For the simple tachyon vacuum it falls off only as $1/K$, and in this case there are difficulties from interactions between the $c$ ghosts and the flags. Since the simple tachyon vacuum is an important solution, it is desirable to find a remedy. We propose that in this case the $c$ ghosts should be lifted off the open string boundary so as to be inserted in the interior of the cylinder of the flag-anti-flag surface.  Specifically, we propose that the field $c$ in the $\BCFT_0$ tachyon vacuum represents an insertion of the $c$ ghost a distance $y_0>\ell $ above the open string boundary in the sliver frame, and in the $\BCFT_*$ tachyon vacuum it represents an insertion of the $c$-ghost a distance $y_*=y_0-\ell>0$ off the boundary. With this definition, $c$ commutes with flag states,
\begin{equation}[c,\flag{\sigma}] = 0,\ \ \ \ [c,\Aflag{\sigmabar}]=0\ \ \ \ \ (\mathrm{assumption\ for\ simple\ tachyon\ vacuum}),\end{equation}
with the understanding that $c$ is defined differently depending on whether it appears to the left or right of the flag or anti-flag state in the commutator. In this way, we can reduce the four terms in $\Sigma$ and $\Sigmabar$ into only three: 
\begin{eqnarray}
\Sigma\lineup = \frac{1}{\sqrt{1+K}}\left((1+K)\flag{\sigma}-B\flag{Q\sigma}+B\flag{\sigma}\frac{B}{1+K}c\d c\right)\frac{1}{\sqrt{1+K}},\label{eq:simpS}\\
\Sigmabar\lineup = \frac{1}{\sqrt{1+K}}\left(\Aflag{\sigmabar}(1+K)+\Aflag{Q\sigmabar}B+c\d c\frac{B}{1+K}\Aflag{\sigmabar}B\right)\frac{1}{\sqrt{1+K}},\label{eq:simpSb}\\
\lineup \ \ \ \ \ \ \ \ \ \ \ \ \ \ \ \ \ \ \ \ \ \ \ \ \ \ \ \ \ \ \  \ \ \ \ \ \ \ \ \ \ \ \ \ \ \ \ \ \ \ \ \ \ \ \ \ \ \ \ \ \ \  \ \ \ \ \ \ \ \ \ \ \ \ (\mathrm{if}\ c \ \mathrm{commutes\ with\ flags}).\nonumber
\end{eqnarray}
This simplification does not occur for generic choice of $F(K)$. The solution becomes 
\begin{eqnarray}
\Psi_*\lineup = \frac{1}{\sqrt{1+K}}c(1+K)Bc\frac{1}{\sqrt{1+K}}\nonumber\\
\lineup\ \ \ -\frac{1}{\sqrt{1+K}}\Big((1+K)\flag{\sigma}-B\flag{Q\sigma}\Big)\frac{1}{1+K}c(1+K)Bc\frac{1}{1+K}\Big(\Aflag{\sigmabar}(1+K)+\Aflag{Q\sigmabar}B\Big)\frac{1}{\sqrt{1+K}}\nonumber\\
\lineup\ \ \ +\frac{1}{\sqrt{1+K}}B \flag{\sigma}B c\frac{1}{1+K}c\d c\frac{1}{1+K}\Big(\Aflag{\sigmabar}(1+K)+\Aflag{Q\sigmabar}B\Big)\frac{1}{\sqrt{1+K}}\nonumber\\
\lineup\ \ \ +\frac{1}{\sqrt{1+K}}\Big((1+K)\flag{\sigma}-B\flag{Q\sigma}\Big)\frac{1}{1+K}c\d c \frac{1}{1+K}c B\Aflag{\sigmabar}B\frac{1}{\sqrt{1+K}}\nonumber\\
\lineup\ \ \ -\frac{1}{\sqrt{1+K}}B\flag{\sigma}B\left(\frac{1}{1+K}c\d c \frac{1}{1+K}c\d c\frac{1}{1+K}\,+\,c\frac{1}{1+K}c\d c\frac{1}{1+K}c\right)\Aflag{\sigmabar}B\frac{1}{\sqrt{1+K}},
\nonumber\\
\lineup \ \ \ \ \ \ \ \ \ \ \ \ \ \ \ \ \ \ \ \ \ \ \ \ \ \ \ \ \ \ \  \ \ \ \ \ \ \ \ \ \ \ \ \ \ \ \ \ \ \ \ \ \ \ \ \ \ \ \ \ \ \  \ \ \ \ \ \ \ \ \ \ \ \ (\mathrm{if}\ c \ \mathrm{commutes\ with\ flags}).\end{eqnarray}
In the last term we can write the factor in parentheses as 
\begin{equation}\frac{1}{1+K}c\d c \frac{1}{1+K}c\d c\frac{1}{1+K}\,+\,c\frac{1}{1+K}c\d c\frac{1}{1+K}c\,=\,\frac{1}{1+K}\left((1+K)\left[c,\frac{1}{1+K}\right]\right)^4, \end{equation}
where as written on the right hand side it is manifest that the factor commutes with the string field $B$. Multiplying everything out, the solution has  $11$ terms. Note that lifting the $c$ ghosts off the boundary does not break the reality condition. It does, however, break twist symmetry, so this form of the simple tachyon vacuum will give imaginary expectation values to twist odd fields which would otherwise vanish. 

\subsection{Comments}
\label{subsec:comments}

Let us discuss a few implications of the construction.

\subsubsection*{Relation to the solution of \cite{KOSsing}}

 The original solution of \cite{KOSsing} can be derived as a limiting case of the flag state solution. Suppose that we have a flag state solution built with a choice of tachyon vacuum 
\begin{equation}F(K) = \left(1-\frac{1}{\nu}K\right)^\nu,\end{equation}
with $\nu\leq-2$ and the $c$ ghosts on the boundary. Further we assume primary boundary condition changing operators $\sigma_\mathrm{bare},\sigmabar_\mathrm{bare}$ of weight $h$ which act as the identity in the timelike free boson factor of the boundary conformal field theory. We can dress these boundary condition changing operators with a timelike Wilson line deformation
\begin{equation}\sigma = \sigma_\mathrm{bare}e^{i\alpha X^0},\ \ \ \sigmabar=\sigmabar_\mathrm{bare}e^{-i\alpha X^0}.\end{equation}
The solution of \cite{KOSsing} can be recovered by sequentially taking the limits:
\begin{equation}\alpha\to\sqrt{h};\ \ \ell\to 0;\ \ \nu\to -1.\end{equation}
First we take the limit where the boundary condition changing operators have vanishing conformal weight, second the limit where the flags disappear, and third the limit to the simple tachyon vacuum. The limits must be applied in this order, otherwise we encounter singularities at intermediate stages. To see why, consider the limit $\ell\to 0$. If we probe the solution with a Fock state and map the resulting correlation functions into the upper half plane, this limit is implemented by 
\begin{equation}\ell\to 0,\ \ p\to s\ \ \ \ s=\mathrm{constant},\ \ \ \frac{s-p}{\ell}= \mathrm{constant},\end{equation}
which keeps the reference and target wedge parameters fixed. In correlators on the upper half plane, the boundary condition changing operators will be multiplied by a conformal factor
\begin{equation}(\rho')^{-2(h-\alpha^2)}\sim \ell^{2(h-\alpha^2)}.\label{eq:flagdeg0}\end{equation}
Therefore, the solution will vanish in the limit $\ell\to 0$ unless we have already taken the limit $\alpha\to\sqrt{h}$. Of course, this confirms that flag states are necessary if we want a well-defined solution with boundary condition changing operators of nonvanishing conformal weight. The limit $\nu\to -1$ cannot be taken before $\ell\to 0$, since the solution based on the simple tachyon vacuum is anomalous when $c$ ghosts are on the open string boundary. 

\subsubsection*{Multiple D-branes}

One of the nicest applications of the flag state solution is the construction of backgrounds containing several D-branes. The idea is similar to that of \cite{KOSsing}, but the implementation is cleaner and more general. We consider a solution representing $N$ D-branes in the string field theory of a single D-brane. If the reference background contains more than one D-brane, we can proceed in the same way after condensing all but one of them to the tachyon vacuum. Suppose that the D-branes of the target background are individually characterized by boundary conformal field theories $\BCFT_*^I$ for $I=1,...,N$. We introduce $N$ pairs of boundary condition changing operators $\sigma^{(0I)}(0)$ and $\sigmabar^{(I0)}(0)$ representing stretched strings connecting the reference D-brane to the D-brane of $\BCFT_*^I$. We require that the boundary condition changing operators are orthogonal for different $I$s, in the sense that 
\begin{equation}\langle\sigmabar^{(I0)}|\sigma^{(0J)}\rangle_\matter = \delta_{IJ}\langle 0|0\rangle_\matter.\label{eq:sigma_orthogonality}\end{equation}
The solution for the multiple D-brane background is given by a row and column vector of flag states:
\begin{equation}
\flag{\sigma} = \bigg(\begin{matrix}\flag{\sigma^{(01)}} \ \ \ ... \ \ \ \flag{\sigma^{(0N)}}\end{matrix}\bigg).\ \ \ \ \ \ \Aflag{\sigmabar} = \left(\begin{matrix}\Aflag{\sigmabar^{(10)}} \\ \vdots \\ \Aflag{\sigmabar^{(N0)}}\phantom{\Big)}\!\!\!\!\end{matrix}\right).
\label{eq:sigma_rowcollumn}\end{equation}
From \eq{sigma_orthogonality} it follows that the star product of the anti-flag column vector and the flag row vector gives the $N\times N$ identity matrix:
\begin{equation}
\Aflag{\sigmabar}*\flag{\sigma} = \left(\begin{matrix}1 & 0 & ... &  \\ 0 & 1 &   &  \\ \vdots &  & \ddots & \\  &  &  & 1\end{matrix}\right),
\end{equation}
where ``$1$" in the $I\text{-}I$th entry is the identity string field of $\BCFT_*^I$. Therefore we have created $N\times N$ Chan-Paton factors out of a string field theory where Chan-Paton factors are absent. Substituting the flag row vector and antiflag column vector into the solution gives explicitly
\begin{equation}\Psi_* = \Psi_\tv - \Sigma^{(01)}\Psi_\tv\Sigmabar^{(10)}-...-\Sigma^{(0N)}\Psi_\tv\Sigmabar^{(N0)},\end{equation}
where $\Sigma^{(0I)}$ and $\Sigmabar^{(I0)}$ are the intertwining fields constructed from the flag states $\flag{\sigma^{(0I)}}$ and $\Aflag{\sigmabar^{(I0)}}$. Remarkably, the solution (relative to the tachyon vacuum) is given by summing the solutions for the constituent D-branes of the target background. Through the Ellwood invariant \cite{tadpole}, this is directly related to the fact that the boundary state of a multiple D-brane system is given by summing the boundary states of the constituent D-branes. It is surprising that adding solutions to a nonlinear field equation creates a new solution. This is possible because the cross terms vanish, due to the assumed orthogonality of the boundary condition changing operators. The boundary  condition changing operators will automatically be orthogonal if each D-brane of the target background is characterized by a distinct open string boundary condition, since then $\sigmabar^{(I0)}$ and $\sigma^{(0J)}$ cannot contain the identity operator in their OPE when $I\neq J$. If the target background contains copies of the same D-brane, so that for example $\BCFT_*^I$ and $\BCFT_*^J$ are identical, the boundary condition changing operators for $I$ and $J$ must be distinct and orthogonal, even though they represent the same change of boundary condition. This can be achieved, for example, by taking them to be distinct matter Virasoro descendants of the boundary condition changing operator of lowest conformal weight.

A special application of this setup is the construction of solutions representing multiple copies of the reference D-brane. This has become a somewhat famous (or infamous) open problem in the subject, and searches in the level expansion continue to this day \cite{Sch_trunc}.\footnote{Another influential idea is to search for multiple D-brane solutions directly in the $KBc$ subalgebra \cite{Murata}. See \cite{Hata} for recent developments in this direction. Despite some early promise the approach has not yet been successful.} We can, for example, construct a solution representing two copies of the reference D-brane using 
\begin{equation}
\ \ \ \ \ \flag{\sigma} = \left(\begin{matrix}\flag{1} \ \ \ \frac{1}{\sqrt{13}}\flag{L_{-2}^m}\end{matrix}\right).\ \ \ \ \ \ \Aflag{\sigmabar} = \left(\begin{matrix}\Aflag{1} \\ \frac{1}{\sqrt{13}}\Aflag{L_{-2}^m}\phantom{\Big)}\!\!\!\!\end{matrix}\right),
\label{eq:2Dbrane}\end{equation}
where $\flag{L_{-2}^m}$ corresponds to the matter Virasoro descendent of the vacuum $L_{-2}^m|0\rangle$. Unlike the proposal of \cite{KOSsing}, this solution lives in the universal subspace generated by acting ghost oscillators and matter Virosoros on the $SL(2,\mathbb{R})$ vacuum, and therefore exists for any reference D-brane system. Note that the ``boundary condition changing operators" in this case do not actually modify the boundary condition. This is because the solution does not change the boundary condition of the reference D-brane;  it only adds Chan-Paton factors.

\subsubsection*{Even the perturbative vacuum is nonperturbative.}

Let us mention an elementary but surprising fact: In the limit that the reference and target D-brane systems become identical, the flag state solution does not vanish. This is clear since if the boundary condition changing operators reduce to the identity, the flag states are nevertheless distinct from the identity string field. Then the two terms in the solution
\begin{equation}\Psi_*=  \Psi_\tv-\Sigma\Psi_\tv\Sigmabar\end{equation}
do not cancel. So it appears that even the perturbative vacuum is a nonpeturbative solution. One consequence is that the proof of local background independence given in \cite{BI1,BI2} is not a limiting case of the argument of subsection~\ref{subsec:BI}. Moreover, solutions for marginal deformations will be somewhat different from those studied before, as they will include a nontrivial shift of the string field at zeroth order in the marginal parameter.  One can try to avoid this situation by adjusting the height of the flags as we vary $\BCFT_*$ in such a way that the flag height approaches zero when $\BCFT_0$ and $\BCFT_*$ coincide. However, the solution cannot be defined perturbatively in this limit. To see this, assume that the boundary condition changing operators have weight $h(\lambda)$ and the horizontal strip has height $\ell(\lambda)$, both of which are approaching zero as a function of a marginal parameter $\lambda\to 0$. Then \eq{flagdeg0} implies that correlators will be proportional to
\begin{equation}\ell(\lambda)^{2h(\lambda)}.\end{equation}
Typically, $h(\lambda)$ will approach zero as $\lambda^2$, so this quantity cannot be analytic unless $\ell(\lambda)$ approaches zero in a nonanalytic fashion. But since correlators will depend on $\ell$ not only in the combination $\ell^h$, there will be no way to expand the solution perturbatively in $\lambda$. If we want perturbative solutions for marginal deformations in the present approach, it seems we have to expand the theory around the nontrivial solution for the perturbative vacuum. 

\section{Consistency of the Solution}
\label{sec:consistency}

Once we introduce flag states, the formal construction of the intertwining solution is straightforward. The nontrivial question is whether the construction leads to something meaningful. For example, is the solution finite? Do the equations of motion hold without anomaly? Of particular concern is the singular geometry of flag states, which in some circumstances lead to divergence which could threaten the consistency of the solution. In this section we investigate this question. We find a few surprises, but nevertheless the solution appears to work. 

\subsection{Equations of Motion}
\label{subsec:SbS1}

First there is the question of whether the solution satisfies the equations of motion. The most delicate point in this respect is the identity
\begin{equation}\Sigmabar\Sigma = 1.\end{equation}
To check whether this holds, we substitute the explicit expressions for the intertwining fields and multiply everything out. In doing this we encounter the following terms:
\begin{equation}\Aflag{\sigmabar}K\flag{\sigma} + \Aflag{Q\sigmabar}B\flag{\sigma} -\Aflag{\sigmabar}B\flag{Q\sigma}.\label{eq:div_terms}\end{equation}
In theory, the terms combine to 
\begin{equation}Q\Big(\Aflag{\sigmabar}B\flag{\sigma}\Big) = QB = K.\end{equation}
In actuality the terms are divergent. If $\sigma$ and $\sigmabar$ are primaries of weight $h$, the leading behavior of $\Aflag{\sigmabar}\Omega^\eps B\flag{Q\sigma}$ in the limit $\eps\to 0$ is given by 
\begin{eqnarray}
\Tr\Big[\big(\, ...\, \big) \Aflag{Q\sigmabar}B\Omega^\eps\flag{\sigma}\Big] \lineup =\frac{1}{\eps^2}\cdot \frac{16\ell h}{\pi}\Tr\Big[\big(\, ...\,\big)(Bc-cB)\Big]  +\mathrm{subleading}.\ \ \ \ 
\end{eqnarray}
The origin of this divergence is that the BRST operator produces a $c$ ghost which is cut off from the remainder of the surface by a degeneration. This is an ``infrared" divergence, similar to those which appear in string amplitudes from corners of the moduli space where a state of negative dimension is forced to propagate over a large Euclidean distance.\footnote{In string amplitudes, such divergences can be dealt with by inverse Wick rotating to a Lorentzian worldsheet near the boundary of moduli space \cite{ie}. It would be interesting if something similar could be done in the present context.}  The field $\Aflag{\sigmabar}K\flag{\sigma}$ may also be divergent if the OPE of $\sigmabar$ and $\sigma$ generates a sufficiently relevant primary operator $\mathcal{O}(x)$ with dimension in the range $h_\mathcal{O}\in[0,1/2]$:
\begin{equation}\Tr\Big[\big(\, ...\, \big) \Aflag{\sigmabar}\Omega^\eps K\flag{\sigma}\Big] = -\eps^{2h_\mathcal{O}-1}\cdot 2h_\mathcal{O} \left(\frac{\pi}{16\ell}\right)^{h_\mathcal{O}}\Tr\Big[\big(\,...\,\big)\mathcal{O}\Big]+ \mathrm{subleading}.\end{equation}
To consistently handle these divergences, we assume that the product of intertwining fields is defined with a limit
\begin{equation}\Sigmabar\Sigma = \lim_{\eps\to 0} \Sigmabar\Omega^\eps\Sigma.\end{equation}
This regularization has the effect of placing a factor of $\Omega^\eps$ between the anti-flag and flag in \eq{div_terms}, so that the singular contributions add up to 
\begin{equation}Q\Big(\Aflag{\sigmabar}B\Omega^\eps\flag{\sigma}\Big).\end{equation}
By the argument of subsection \ref{subsec:pre_intertwining}, this unambiguously approaches $K$ in the $\eps\to 0$ limit, as desired. The remaining question is whether the other contributions work out correctly in the $\eps\to 0$ limit to give $\Sigmabar\Sigma=1$.

We specialize to the case of the simple tachyon vacuum, since here nontrivial complications arise which are absent when $F(K)$ vanishes faster than $1/K$ for large $K$.  For the time-being, we assume that $c$ ghosts are inserted in the usual way on the open string boundary. We compute
\begin{eqnarray}
 \Sigmabar\Omega^\eps\Sigma \lineup = 
 \frac{1}{\sqrt{1+K}}\left[Q\Big(\Aflag{\sigmabar}B\Omega^\eps\flag{\sigma}\Big)+\Aflag{\sigmabar}B[c,\Omega^\eps]\flag{\sigma}\phantom{\bigg)}\right.\nonumber\\
 \lineup\ \ \ \ \ \ \ \ \ \ \ \ \ \ \ \left.+\Aflag{\sigmabar}B\Omega^\eps\flag{\sigma}\frac{1}{1+K}c(1+K)Bc + c(1+K)Bc\frac{1}{1+K}\Aflag{\sigmabar}B\Omega^\eps\flag{\sigma}\right]\frac{1}{\sqrt{1+K}}.\ \ \ \ \ \ \label{eq:simp_SbS}
\end{eqnarray}
Let us explain what is ``supposed" to happen to this expression as we take the $\eps\to 0$ limit. The first term should approach $K$ (which it does). The second term should vanish since $c$ commutes with the identity string field. In the third and fourth terms, we should be able to use $\Aflag{\sigmabar}B\flag{\sigma}=B$ to find
\begin{eqnarray}
\lineup B\frac{1}{1+K}c(1+K)Bc + c(1+K)Bc\frac{1}{1+K}B\,=\,Bc+cB\, = \,1.
\end{eqnarray}
In total we would obtain $K+1$ inside the square brackets, which cancels the square roots outside to give the identity string field.

To see what actually happens, consider the term 
\begin{equation}\Aflag{\sigmabar}B[c,\Omega^\eps]\flag{\sigma} = B\Aflag{\sigmabar}[c,\Omega^\eps]\flag{\sigma} - b\Aflag{\sigmabar}[c,\Omega^\eps] \flag{\sigma}.\label{eq:an1}\end{equation}
Focusing on the last piece, we contract with a test state:
\begin{equation}
\Tr\Big[\big(\, ...\, \big) b\Aflag{1}[c,\Omega^\eps]\flag{1}\Big]=\mathcal{F}_b\Big\langle\mathcal{F}^{-1}\circ\big(\, ...\, \big) b(s) \Big(\mathcal{F}^{-1}\circ c(\eps/2) - \mathcal{F}^{-1}\circ c(-\eps/2)\Big)\Big\rangle_\mathrm{UHP}.
\end{equation}
Note that the $c$ ghosts are inserted on the open string boundary. The boundary condition changing operators do not play a role in the following, so we set them to unity. The $\eps\to 0$ limit is implemented using upper half plane moduli following \eq{sep_deg}, where we find that 
\begin{equation}\mathcal{F}^{-1}(\eps/2) = 2\chi^2 s^3+\mathrm{subleading}.\end{equation}
Then
\begin{eqnarray}
\Tr\Big[\big(\, ...\, \big) b\Aflag{1}[c,\Omega^\eps]\flag{1}\Big] \lineup = \mathcal{F}_b\mathcal{F}'(2\chi^2 s^3)\Big\langle \mathcal{F}^{-1}\circ\big(\,...\,\big)b(s)(c(2\chi^2 s^3)-c(-2\chi^2 s^3))\Big\rangle_\mathrm{UHP}+\mathrm{subleading}\nonumber\\
\lineup = \frac{\pi s}{4\ell \chi}\cdot \frac{2\ell}{\pi\chi s^2} 4\chi^2 s^3\Big\langle\mathcal{F}^{-1}\circ\big(\,...\,\big)b(s)\d c(0)\Big\rangle_\mathrm{UHP}+\mathrm{subleading}\nonumber\\
\lineup = 2\Big\langle\mathcal{F}^{-1}\circ\big(\,...\,\big)\Big\rangle_\mathrm{UHP}+\mathrm{subleading},
\end{eqnarray}
where in the last step we evaluated the OPE. By a similar computation one can show that the first contribution in \eq{an1} is zero. In total we find
\begin{equation}
\lim_{\eps\to 0}\Aflag{\sigmabar}B[c,\Omega^\eps]\flag{\sigma} = -2.
\end{equation}
This does not vanish. We might have expected a subtlety here. The $c$ ghost produces a divergent factor when it is cut off by degeneration. Together with the OPE with the $b$ ghost, this is enough to compensate for the vanishing of the commutator to produce a nonzero result.

The third and fourth terms in \eq{simp_SbS} are also anomalous, but for a different reason. The difficulty here is that the $c$ ghost vanishes when inserted on the slit: 
\begin{equation}c\Aflag{\sigmabar} = 0,\ \ \ \flag{\sigma}c = 0.\end{equation}
This is the flip side of the earlier observation that the $b$ ghost diverges on the slit. However, this property is inconsistent with associativity; products such as
\begin{equation}(\sigmabar|\sigma)c\end{equation}
are ambiguous. A related ambiguity appears in the third and fourth terms of \eq{simp_SbS}, and placing a thin wedge between $\Sigmabar$ and $\Sigma$ resolves the ambiguity in an unfavorable direction. To see how this happens, consider 
\begin{equation}c(1+K)Bc\frac{1}{1+K}\Aflag{\sigmabar}B\Omega^\eps\flag{\sigma}=\Aflag{\sigmabar}\Omega^\eps\flag{\sigma} -(1+K)c\frac{1}{1+K}b\Aflag{\sigmabar}\Omega^\eps\flag{\sigma} + Bc\d c\frac{1}{1+K}b\Aflag{\sigmabar}\Omega^\eps\flag{\sigma}.\label{eq:an2}\end{equation}
We commuted the $B$ insertion between the flags to the left, and in the first term used the fact that $c$ vanishes on the slit. We contract the last term with a test state:
\begin{eqnarray}
\lineup \!\!\!\Tr\left[\big(\,...\,\big)Bc\d c\frac{1}{1+K}b\Aflag{1}\Omega^\eps\flag{1}\right] = \int_0^\infty dt\, e^{-t} \Tr\left[\big(\,...\,\big)Bc\d c\Omega^t b\Aflag{1}\Omega^\eps\flag{1}\right] \nonumber\\
\lineup \ \ \ \ \ \ \ \ \ \ \ \ \ \ \ \ \  = \left(\int_0^\infty+\int_{-\infty}^0\right) dx\, e^{-(\mathcal{F}(x+s)-\mathcal{F}(s))}\mathcal{F}_b\mathcal{F}'(s+x)^2\left\langle \mathcal{F}^{-1}\circ\big(\,...\,B\big) c\d c(s+x) b(s)\right\rangle_\mathrm{UHP}.\ \ \ \ \ \ \ \ \ \ 
\end{eqnarray} 
In the integrand, $s$ and $p$ are determined by the reference wedge angle $\eps$ and the target wedge angle $\beta+t$, where $\beta$ is the width of the test state. Since $t$ is related to the integration variable $x$ through
\begin{equation}t = \mathcal{F}(x+s)-\mathcal{F}(s),\end{equation}
the upper half plane moduli $s$ and $p$ are implicitly functions of $x, \eps,\beta$. The integration over $x$ splits into two pieces with  $x\in [0,\infty]$ corresponding to $t\in[0,\beta]$  and $x\in[-\infty,0]$ corresponding to $t>\beta$. In the $\eps\to 0$ limit the integration vanishes everywhere except at $x=0^+$, where the vanishing of $\mathcal{F}_b$ is compensated by a divergence from the OPE of $c\d c$ and $b$. To display the behavior near $x=0^+$ we change the integration variable
\begin{equation}x=\frac{\pi\eps}{\beta} \zeta,\end{equation}
and evaluate the OPE to find
\begin{eqnarray}
\lineup \lim_{\eps\to 0}\Tr\left[\big(\,...\,\big)Bc\d c\frac{1}{1+K}b\Aflag{1}\Omega^\eps\flag{1}\right] =\nonumber\\
\lineup\ \ \ \ \ \ \ \ \ \ -\frac{\beta}{\pi}\lim_{\eps\to 0}\int_0^\infty d\zeta \frac{e^{-(\mathcal{F}(s+\frac{\pi\eps \zeta}{\beta})-\mathcal{F}(s))}\mathcal{F}_b \mathcal{F}'\left(s+\frac{\pi\eps \zeta}{\beta}\right)^2}{\eps \zeta^2}
 \Big\langle\mathcal{F}^{-1}\circ\big(\,...\,B\big)c(s)\Big\rangle_\mathrm{UHP}.
\end{eqnarray}
To leading order the upper half plane moduli are given by 
\begin{equation}p=\frac{\pi^2}{32\ell \beta}\eps^2+\mathcal{O}(\eps^3),\ \ \ s=\frac{\pi}{4\beta}\eps+\mathcal{O}(\eps^2),\ \ \ \ (\zeta>0),\end{equation}
and are independent of $\zeta$. With this we can evaluate the limit 
\begin{equation}
\lim_{\eps\to 0}\frac{e^{-(\mathcal{F}(s+\frac{\pi\eps \zeta}{\beta})-\mathcal{F}(s))}\mathcal{F}_b \mathcal{F}'\left(s+\frac{\pi\eps \zeta}{\beta}\right)^2}{\eps \zeta^2} = \frac{8(1+2\zeta)^2}{(1+4\zeta)^4}.
\end{equation}
Since this is nonvanishing, we must have a delta function contribution to the integration over $x$  at $x=0^+$. Evaluating the integral over $\zeta$ gives
\begin{equation}
\lim_{\eps\to 0}Bc\d c\frac{1}{1+K}b\Aflag{1}\Omega^\eps\flag{1} = -\frac{7}{6}Bc.
\end{equation}
The second term in \eq{an2} vanishes in the limit.

Bringing all contributions to \eq{simp_SbS} together we find
\begin{eqnarray}
\lim_{\eps\to 0}\Sigmabar\Omega^\eps\Sigma \lineup = \frac{1}{\sqrt{1+K}}\left[K-2+2-\frac{7}{6}\right]\frac{1}{\sqrt{1+K}}\nonumber\\
\lineup = 1-\frac{13}{6}\frac{1}{1+K},
\end{eqnarray}
which is not the desired result. The anomaly is produced through two distinct mechanisms, but the resolution in both cases is the same: we simply lift the $c$-ghost off the open string boundary. For the $\BCFT_0$ tachyon vacuum, we should lift a distance $y_0>\ell$ off the boundary in the sliver frame to ensure that the $c$-ghost is not trapped between the flags at degeneration. In the $\BCFT_*$ tachyon vacuum, we should lift a distance $y_*>0$ to avoid collisions between the $c$-ghost and the slit. If $y_*=y_0-\ell$, the field $c$ will commute with the flag states, and the solution takes the form described in subsection~\ref{subsec:solution}. 

These remarks apply to the simple tachyon vacuum. If $F(K)$ falls off faster than $1/K$, anomalies are absent. The reason is that the anomalous contributions appear as delta functions at the lower limit of integration over wedge states, and if $F(K)$ vanishes faster than $1/K$, the delta function is integrated against a test function which vanishes on the lower limit. Therefore we are not required to lift $c$-ghosts off the boundary in this case. One might consider doing it anyway so that the field $c$ will commute with the flags. But this does not appear to lead to useful simplifications.

We have therefore shown 
\begin{equation}\lim_{\eps\to 0}\Sigmabar\Omega^\eps\Sigma = 1,\end{equation}
assuming that
\begin{eqnarray}
\lineup 1)\  F(K) \text{ vanishes faster than }1/K\text{ for large }K;\nonumber\\
\lineup\ \ \ \ \ \ \ \ \ \ \ \ \ \ \ \ \ \ \ \ \ \ \ \ \text{or}\nonumber\\
\lineup 2)\  F(K) \text{ vanishes as }1/K\text{ and }c\text{-ghosts are lifted sufficiently far off the }\nonumber\\
\lineup \ \ \ \  \text{open string boundary.}\label{eq:EOMcond}
\end{eqnarray}
This suggests that the equations of motion will hold in the Fock space if $\Psi_*^2$ is interpreted as $\lim_{\eps\to 0}\Psi_*\Omega^\eps \Psi_*$. Technically, to fully establish this we would have to carefully analyze the full expression $\Psi_*\Omega^\eps\Psi_*$, which requires correlators on surfaces with at least four flags. The only conceivable subtlety here concerns collisions of three or more slits. However, such collisions can be suppressed as needed by requiring that $F(K)$ falls off fast enough at infinity. Therefore we are confident that the equations of motion will hold with at least some choices of tachyon vacuum. But we have no specific reason to believe that the conditions \eq{EOMcond} are not already enough to avoid problems of this kind. The most significant issue with our present understanding of the equations of motion is the apparent necessity of regularization of products of intertwining fields. While products of intertwining fields are finite, the cancellation of divergent cross terms in intermediate steps leaves open the possibility of residual ambiguities. We have resolved potential ambiguities through a consistent choice of regularization. However, this is a point where the present understanding leaves room for improvement. 

\subsection{Finiteness}

Now we address the question of whether the solution is finite when contracted with a basis of Fock states. The overlap of the solution with a Fock state produces a flag-anti-flag surface without any open string degeneration, so the issues discussed in the previous subsection do not play a role. The concern in the present context are the ghost insertions of the target tachyon vacuum, and how they are effected by the curvature singularities at the slits. The boundary condition changing operators, wedge algebra factors and insertions outside the flag states are not relevant for this, so the question of finiteness boils down to the question of whether the states
\begin{eqnarray}
\lineup \ \ \ \ \ \ \ \ \ \ \flag{1}\sqrt{\!\frac{F}{H}}HcB\frac{1}{H}cH\sqrt{\!\frac{F}{H}}\Aflag{1},\nonumber\\
\lineup \ \ \ \ \ \flag{1}\sqrt{\!\frac{F}{H}}HcB\frac{1}{H}[c,F]\frac{1}{H}cH\sqrt{\!\frac{F}{H}}\Aflag{1},\nonumber\\
\lineup \flag{1}\sqrt{\!\frac{F}{H}}HcB\frac{1}{H}[c,F]\frac{1}{H}[c,F]\frac{1}{H}cH\sqrt{\!\frac{F}{H}}\Aflag{1}\label{eq:fin_fact}
\end{eqnarray}
are finite. Not every contribution to the above states is of equal concern to the question of finiteness. The most singular contribution appears in the first state:
\begin{eqnarray}
\flag{1}\sqrt{\!\frac{F}{H}}HcB\frac{1}{H}cH\sqrt{\!\frac{F}{H}}\Aflag{1}\lineup\to \flag{1}\sqrt{\!\frac{F}{H}}HcKBcH\sqrt{\!\frac{F}{H}}\Aflag{1}\to \flag{1}\sqrt{\!\frac{F}{H}}Hc\d c BH\sqrt{\!\frac{F}{H}}\Aflag{1}\nonumber\\
\lineup\to \flag{1}\sqrt{\!\frac{F}{H}}Hc\d cH\sqrt{\!\frac{F}{H}}b\Aflag{1}.\label{eq:fin_con}
\end{eqnarray} 
We restrict our analysis to this contribution. The additional factors of $F$ and $H$ in the remaining contributions in \eq{fin_fact} imply that singular collisions of ghosts and slits are only further suppressed.

We may expand the wedge algebra factors  in \eq{fin_con} as a superposition of wedge states
\begin{equation}
\flag{1}\sqrt{\!\frac{F}{H}}Hc\d cH\sqrt{\!\frac{F}{H}}b\Aflag{1}=\int_0^\infty dt_1 dt_2\, g(t_1)g(t_2) \flag{1}\Omega^{t_1}c\d c\Omega^{t_2}b\Aflag{1},\label{eq:fin_cond2}
\end{equation}
where $g(t)$ is the inverse Laplace transform of $H\sqrt{F/H}$. The important fact about $g(t)$ for present purposes is that it is bounded and continuous in the vicinity of $t=0$, which follows from the assumption that $F(K)$ falls off as $1/K$ or faster. First we consider the case where $c$ ghosts are inserted on the boundary. This may lead to difficulties when $c\d c$ collides with the slits. If we keep $t_2$ fixed, integration near $t_1=0$ will produce a collision between $c\d c$ and the slit of the flag state. Since $c\d c$ has negative dimension, the correlator will vanish in this limit and the integration over $t_1$ will be finite.  On the other hand, keeping $t_1$ fixed, integration near $t_2=0$ leads to a collision between $c\d c$ and the $b$-ghost, producing a singular OPE. To see what happens, we contract with a test state and transform the integrand to dual upper half plane coordinates \eq{dual_coord}
\begin{eqnarray}\lineup \Tr\left[\big(\,...\,\big)c\d c\frac{1}{H}\sqrt{\frac{F}{H}}b\Aflag{1}\right] = \int_0^\infty dt\, g(t) \Tr\Big[\big(\,...\,\big)c\d c\Omega^tb\Aflag{1}\Big]\nonumber\\
\lineup  \ \  =\int_0^\infty dx\, g\Big(\widetilde{\mathcal{F}}(-\widetilde{s}+x)-\widetilde{\mathcal{F}}(-\widetilde{s})\Big)\widetilde{\mathcal{F}}'(-\widetilde{s}+x)^2\widetilde{\mathcal{F}}_b\Big\langle\widetilde{\mathcal{F}}^{-1}\circ\big(\,...\,\big)c\d c(-\widetilde{s}+x)b(-\widetilde{s})\Big\rangle_{\widetilde{\mathrm{UHP}}}.\ \ \ \ \ \ \ \ \ \ 
\end{eqnarray}
We obtain a double pole towards $x=0$ from the OPE, but $\widetilde{\mathcal{F}}'$ vanishes as $x$. So the singular OPE cancels against a vanishing conformal factor, and the integration over $t_2$ will be finite.

Lastly we consider the limit where $t_1$ and $t_2$ simultaneously approach zero, and the slits collide. Here a different issue arises, related to the fact that the string field $B$ is divergent when pinched between a flag and anti-flag:
\begin{equation}\Tr\Big[\big(\,...\,\big)\flag{1}B\Omega^\eps\Aflag{1}\Big] = \frac{1}{\eps^{1/3}}\cdot \frac{2}{3}\left(\frac{3\pi}{\alpha(4+(\alpha/\ell)^2)}\right)^{2/3}\lim_{\eps\to 0}\Big\langle\widetilde{\mathcal{F}}^{-1}\circ\big(\,...\,\big)b(0)\Big\rangle_{\widetilde{\mathrm{UHP}}}+\mathrm{subleading},\end{equation} 
where $\alpha$ is the width of the test state. To see the effect of this divergence, we write \eq{fin_cond2} changing the integration variables to $\beta= t_1+t_2$ and $\theta=t_1/(t_1+t_2)$:
\begin{equation}
\Tr\left[\!\big(\,...\,\big)\flag{1}\sqrt{\!\frac{F}{H}}Hc\d cH\sqrt{\!\frac{F}{H}}b\Aflag{1}\right]=\int_0^\infty d\beta \!\int_0^1 \!d\theta \, \beta g(\beta\theta)g(\beta(1-\theta)) \Tr\Big[\big(\,...\,\big)\flag{1}\Omega^{\beta\theta}c\d c\Omega^{\beta(1-\theta)}b\Aflag{1}\Big].
\label{eq:fin_cond3}
\end{equation}
The correlator in the integrand can be written 
\begin{equation}
\Tr\Big[\big(\,...\,\big)\flag{1}\Omega^{\beta\theta}c\d c\Omega^{\beta(1-\theta)}b\Aflag{1}\Big]= \widetilde{\mathcal{F}}_b\widetilde{\mathcal{F}}'(x)\Big\langle\widetilde{\mathcal{F}}^{-1}\circ\big(\,...\,\big)c\d c(x)b(-\widetilde{s})\Big\rangle_{\widetilde{\mathrm{UHP}}}.
\end{equation}
We are interested in the behavior near $\beta=0$, which corresponds to small $\widetilde{s}$ with $\widetilde{p}$ fixed on the dual upper half plane. Since $|x|$ is bounded by $\widetilde{s}$ which is tending to zero, in this limit we can substitute $c\d c$ and $b$ with their OPE. This gives the leading behavior for small $\beta$
\begin{eqnarray}
\lineup \Tr\Big[\big(\,...\,\big)\flag{1}\Omega^{\beta\theta}c\d c\Omega^{\beta(1-\theta)}b\Aflag{1}\Big] \nonumber\\
\lineup\ \ =\frac{1}{\beta^{1/3}}\sqrt{\frac{\theta}{1-\theta}}\cdot \frac{1}{2}\left(\frac{\alpha(4+(\alpha/\ell)^2)}{3\pi}\right)^{1/3}\sqrt{\frac{2-X(\theta)}{2+X(\theta)}}\lim_{\beta\to 0}\Big\langle \widetilde{\mathcal{F}}^{-1}\circ\big(\,...\,\big)c(0)\Big\rangle_{\widetilde{\mathrm{UHP}}}+\mathrm{subleading},\label{eq:cor_bdry}
\ \ \ \ \ \ \ \ \ \ \end{eqnarray}
where $X(\theta)$ is the root to $4\theta = (1-X)^2(2+X)$ with absolute value $\leq 1$. We find an inverse cube root divergence of the correlator for small $\beta$ which is canceled against the measure. We also find a square root zero and inverse square root divergence at $\theta=0$ and $1$, respectively, resulting from the collision of $c\d c$ with the slit of the flag and the $b$-ghost on the slit of the anti-flag. The singularity at $\theta=1$ is nevertheless integrable. 

The discussion is slightly different when $c$ ghosts are lifted off the open string boundary.  Instead of \eq{cor_bdry} we will have
\begin{eqnarray}
\lineup \Tr\Big[\big(\,...\,\big)\flag{1}\Omega^{\beta\theta}c\d c\Omega^{\beta(1-\theta)}b\Aflag{1}\Big] \nonumber\\
\lineup\ \ \ \ \ \ \ \ \ \ =\frac{1}{\beta^{1/3}}\cdot \frac{2}{3}\left(\frac{3\pi}{\alpha(4+(\alpha/\ell)^2)}\right)^{2/3}\lim_{\beta\to 0}\Big\langle\widetilde{\mathcal{F}}^{-1}\circ\big(\,...\,c\d c(i y_*)\big)b(0)\Big\rangle_{\widetilde{\mathrm{UHP}}}+\mathrm{subleading},\ \ \ \ \ \ \ 
\end{eqnarray}
where $y_*$ is the distance the $c$-ghost is lifted off the boundary in the dual sliver coordinate $\widetilde{z}$. To leading order the dependence on $\theta$ is absent, but the inverse cube root singularity is still present. Nevertheless, integration over $\beta$ in \eq{fin_cond3} is finite.

Therefore the flag state solution is finite in a basis of Fock states. This isn't an especially close call. To produce a genuine divergence of the flag state solution, we would need to consider a tachyon vacuum where $F(K)$ does not fall to zero at infinity. A tachyon vacuum of this kind is singular for other reasons \cite{IdSing} and is not of much practical interest.

\subsection{Overlaps}
\label{subsec:Ellwood}

\begin{figure}
\begin{center}
\resizebox{1.3in}{2.1in}{\includegraphics{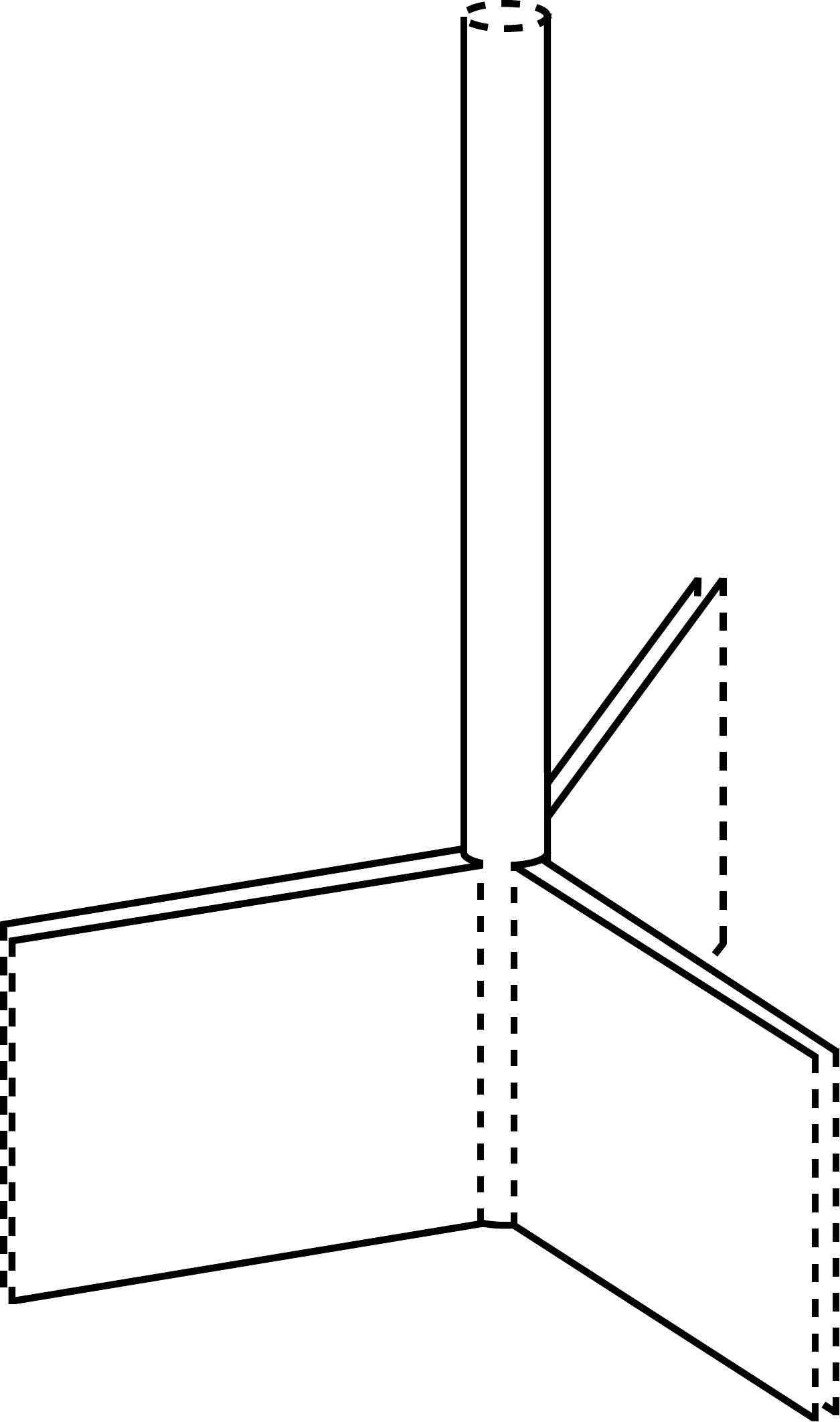}} 
\end{center}
\caption{\label{fig:needle} In computing the cubic vertex evaluated on the solution, at an extreme limit of integration over wedge states one finds singular surfaces where flags are attached to an infinitely thin cylinder.}
\end{figure}

An important test of an analytic solution is whether overlaps of the solution with itself are well-behaved. For example, one should check that the equations of motion are satisfied when contracted with the solution \cite{OkawaVSFT}, and that the action and Ellwood invariant \cite{tadpole} produce the physically expected answers. For the flag state solution, these things follow automatically if we can assume~$\Sigmabar\Sigma=1$. The significant issue which can appear here---beyond the considerations of subsection~\ref{subsec:SbS1}---is that the total width of wedge states in overlaps may not have strictly positive lower bound. Therefore, overlaps of the solution with itself may receive contribution from singular surfaces where flags are attached to a ``needle-like" cylinder, as shown in figure \ref{fig:needle}. It is intuitively clear that in this limit we will not be able to replace anti-flag/flag pairs with the identity operator. Any potential problems here can be controlled by assuming that $F(K)$ falls off sufficiently rapidly towards infinity. But it is interesting to understand the issue in more detail. 

\begin{figure}
\begin{center}
\resizebox{3in}{2in}{\includegraphics{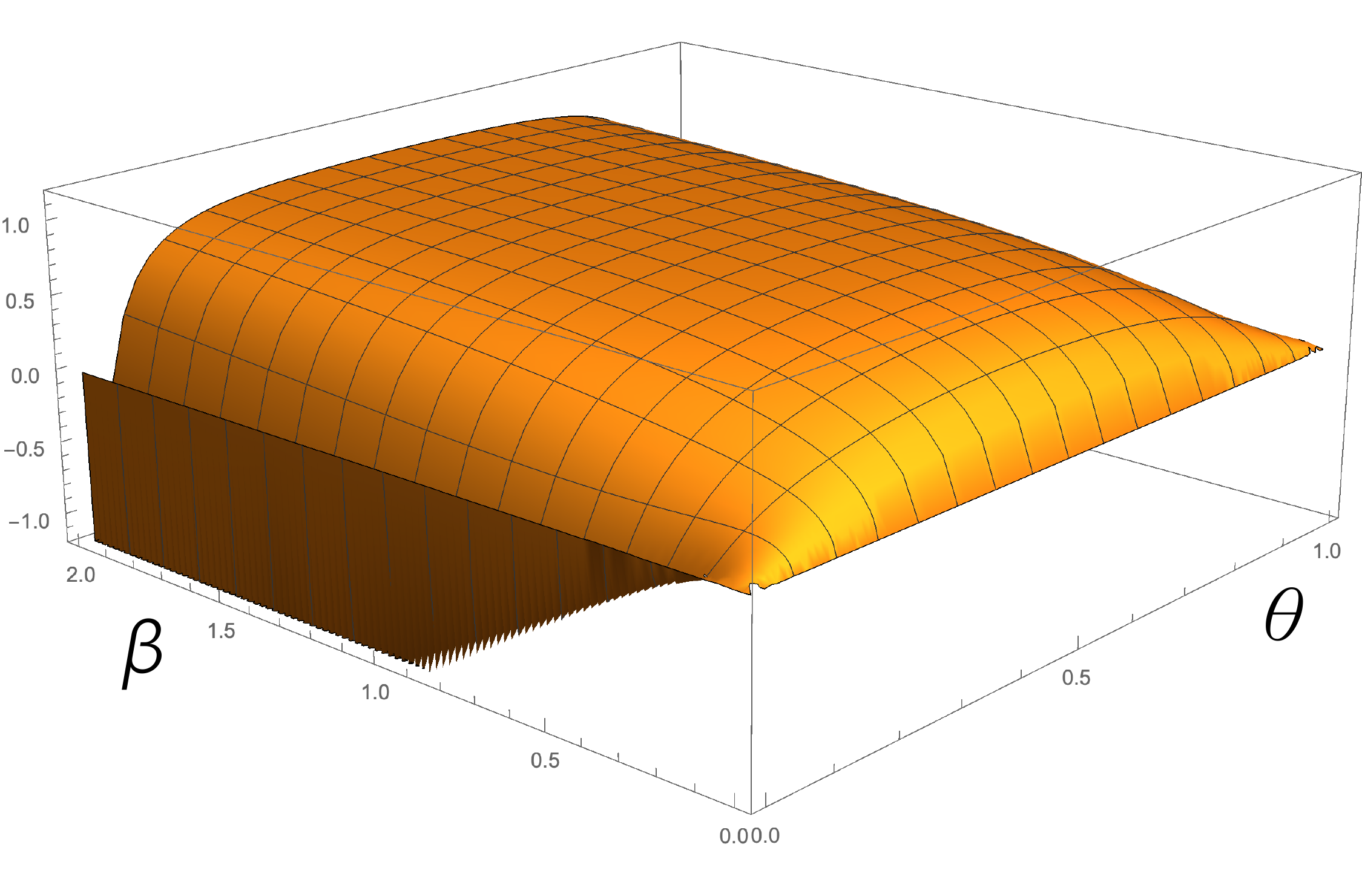}}\ \ \ \ \ \ \ \ \ \ \ \ 
\resizebox{2.5in}{1.8in}{\includegraphics{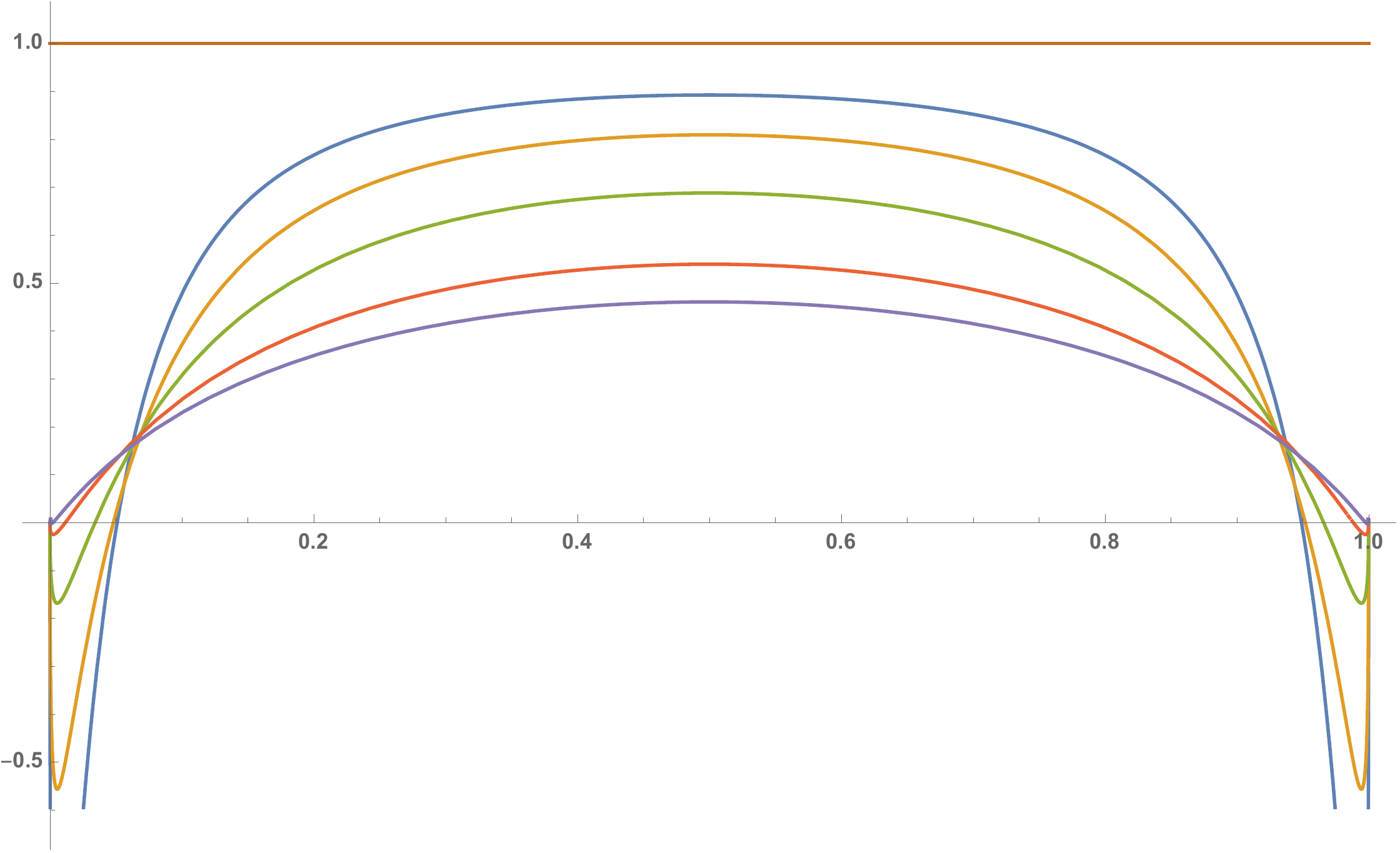}}
\end{center}
\caption{\label{fig:needle_cor} To the left is a plot of the ratio \eq{ex_ratio} for $\eps = 0.1$ as a function of $\beta\in[0,2]$ and $\theta\in[0,1]$. In the middle we see a ``plateau", towards $\theta=0,1$ ``trenches," and towards $\beta=0$ a ``cliff." To the right are cross sections of this plot for $\theta\in [0,1]$ and $\beta = 1,0.5,0.25,0.125$ and $0.0875$.}
\end{figure}

We focus on the Ellwood invariant, since its computation only requires correlators on the flag-anti-flag surface. The Ellwood invariant is given by 
\begin{equation}\lim_{\eps\to 0}\Tr_\mathcal{V}\big[\Sigma\Psi_\tv\Sigmabar \Omega^\eps\big],\end{equation}
where $\Tr_\mathcal{V}$ is the 1-string vertex with a midpoint insertion of an on-shell closed string vertex operator $\mathcal{V}=c\overline{c}V_\matter$ with $V_\matter$ a weight $(1,1)$ matter primary. It is straightforward to expand this out and analyze the contributions term by term, but for illustrative purposes it is enough to focus on a contribution with correlators of the form
\begin{equation}\Tr_\mathcal{V}\big[\Omega^{\beta(1-\theta)}c\d c\Omega^{\beta\theta}\Aflag{1}B\Omega^{\eps}\flag{1}\big],\end{equation}
where for now we assume that the $c$ ghosts are on the boundary. The width of the target wedge state $\beta$ and the parameter $\theta\in[0,1]$ will appear integrated against inverse Laplace transforms of the various wedge algebra factors which appear in the solution. For simplicity we set the boundary condition changing operators equal to unity. Nontrivial boundary condition changing operators would produce a non-universal factor for finite $\eps$ which does not lead to singular behavior in the $\eps\to 0$ limit. For fixed $\beta\neq 0$ and $\theta\neq 0,1$ the correlator simplifies in the $\eps\to 0$ limit to
\begin{equation}\Tr_\mathcal{V}[c\d c B\Omega^\beta] = -\frac{2i}{\pi}\langle V_\mathrm{matter}(i,\overline{i})\rangle^{\mathrm{matter}}_\mathrm{UHP}.\end{equation}
We consider the ratio
\begin{equation}\frac{\Tr_\mathcal{V}\big[\Omega^{\beta(1-\theta)}c\d c\Omega^{\beta\theta}\Aflag{1}B\Omega^{\eps}\flag{1}\big]}{\Tr_\mathcal{V}[c\d c B\Omega^\beta]}\label{eq:ex_ratio}\end{equation}
as a function of $\beta$ and $\theta$, and see in what sense this function approaches unity as $\eps$ gets small. With the help of the correlators given in appendix \ref{app:ghost}, we plot the ratio for small-ish $\eps$ in figure \ref{fig:needle_cor}. In the interior of the region $\beta\geq 0,\theta\in[0,1]$ we have a ``plateau" where the ratio is nearly constant; as $\eps$ approaches zero, the plateau becomes flatter with constant value $1$, and extends over the whole region. The interesting behavior occurs at the boundaries of the region. Near the lines $\theta=0,1$ there is a deep ``trench" in the ratio, and as $\eps$ becomes small the trench becomes thinner and deeper, approaching a delta function. We encountered this delta function in subsection \ref{subsec:SbS1}, where it is responsible for part of the anomaly in the relation $\Sigmabar\Sigma=1$ for the simple tachyon vacuum. Presently we assume that $F(K)$ falls of faster than $1/K$, in which case the delta function is integrated against a function which vanishes at $\theta=0,1$, and therefore does not contribute. Finally we consider the behavior near $\beta=0$, where the cylinder collapses into a needle. Here there is a ``cliff" where the plateau falls from $1$ to $0$, and as $\eps$ gets small the cliff becomes steeper. The fact that the ratio must vanish at $\beta=0$ can be understood since $c\d c$ is inserted on the slits. The relevant point, however, is that there is no divergence in the correlation function in the transition from the plateau towards $\beta=0$. Similar qualitative analysis applies to other contributions to the Ellwood invariant, so we will obtain the physically expected value in the $\eps\to 0$ limit.

For the simple tachyon vacuum, we should consider the ratio \eq{ex_ratio} with the ghosts lifted off the open string boundary. In this case, there are no delta functions at $\theta=0,1$ since $c\d c$ does not produce a singular OPE with the $b$ ghost at the slits. There is also no ``cliff" near $\beta=0$. This can be seen since $c\d c$ will appear inside a very thin cylinder a finite distance above the slits. Applying a scale transformation we can expand the cylinder to unit circumference, and in the process we push the slits and horizontal strip out to infinity, where they become irrelevant. Therefore, even for the simple tachyon vacuum, the computation of the Ellwood invariant will work as expected.

\section{Fock Space Expansion}
\label{sec:Fock}

One of the most unique things we can learn from a string field theory solution is information about the expectation values of D-brane fluctuation fields upon the formation of a new background.  This not only tells us that a new vacuum exists, but gives concrete insight into how the new background is created out of the original configuration, something which is missing from the standard perturbative formulation of string theory.

The fluctuation fields are defined through the Fock space expansion, as the coefficients of an expansion of the dynamical string field into a basis of eigenstates of $L_0$. The Fock space coefficients of the flag state solution schematically have the same structure as those of the Kiermaier, Okawa, Soler solution \cite{KOS}: Up to a shift, they are given by a canonical matter 3-point function of two primary boundary condition changing operators and a primary probe vertex operator, multiplied by a ``universal" factor which only depends on conformal transformation properties of the test state and boundary condition changing operators appearing in the solution. Once the relevant three point functions are known, the main task is the computation of the universal factor. For the Kiermaier, Okawa, Soler solution this has been done for coefficients of primary fields, and is given by a relatively manageable three dimensional integral which can be evaluated numerically. For the flag state solution, the corresponding factor is given by a seven dimensional integral whose integrand can be derived from the formulas of appendices \ref{app:conformal} and \ref{app:ghost}. The numerical evaluation of the integral is a serious undertaking and is beyond the scope of this paper. The good news is that such a calculation only needs to be done once. If good fits for the dependence on the conformal data can be extracted, we would have immediate access to the expectation values of an infinite number of fields representing any background of interest. 

\subsection{Ghost Number Zero Toy Model}
\label{subsec:toy}

In lieu of computing coefficients of the full solution, we look for a simpler computation which can give some insight. Often it is possible to model the behavior of the coefficients of a solution by investigating an analogous and simpler state at ghost number zero which only gives expectation values to total Virasoro descendants of matter primaries \cite{RSZpatterns}. This is often referred to as a ghost number zero toy model. The full solution at ghost number 1 will in addition give expectation values to states created by ghost oscillators, but usually such coefficients are of secondary interest. 

If we assume the simple tachyon vacuum, we propose that a reasonable toy model of the flag state solution is given by 
\begin{eqnarray}\Gamma_*\lineup = 1-\Sigma_\pre\Sigmabar_\pre\nonumber\\
\lineup = 1-\sqrt{1+K}\flag{\sigma}\frac{1}{1+K}\Aflag{\sigmabar}\sqrt{1+K}.\end{eqnarray}
In the subalgebra of wedge states with matter insertions and ghost insertions consisting only of $B$ and $c$, one definition of the ghost number zero toy model $\Gamma$ of a solution $\Psi$ is given by the equation $B\Psi B =B\Gamma$. For example, the ghost number zero toy model for Schnabl's solution~\cite{Schnabl} can be extracted in this way. For the flag state solution, we have instead
\begin{equation}B\Psi_*B = B\Gamma_* + \left({\text{terms with } b \atop \text{insertions on slits}}\right).\end{equation}
Our definition of the ghost number zero toy model ignores contributions from deforming $B$ ghost contours through slits. 

\begin{figure}
\begin{center}
\resizebox{3in}{2.7in}{\includegraphics{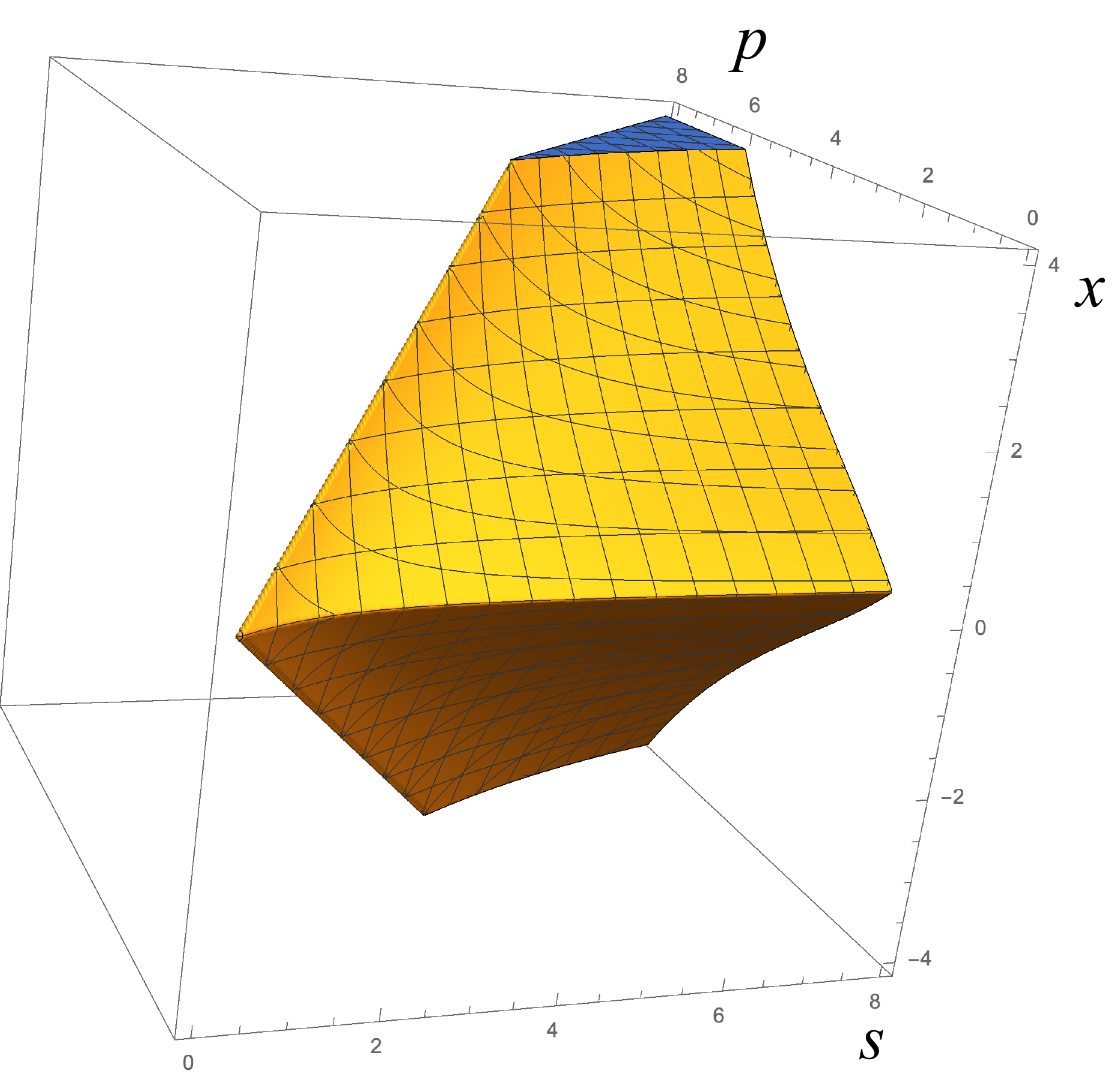}}
\end{center}
\caption{\label{fig:arrowhead} The ``arrowhead" integration region defining the coefficients of the solution. The flag height is taken here at $\ell = .5$.}
\end{figure}

Given a basis for the open string state space, the coefficient of a basis element $|\phi_i\rangle$ is given by computing the BPZ inner product with a dual basis state $|\phi^i\rangle$ satisfying $\langle\phi^i,\phi_j\rangle = \delta^i_j$. For the toy model, the BPZ inner product can be expressed in the form
\begin{equation}
\langle\phi^i,\Gamma_*\rangle = \langle I| \phi^i\rangle - \langle g_3| \Big(|\sigmabar\rangle\otimes|\phi^i\rangle\otimes|\sigma\rangle\Big).\label{eq:coef_gen}
\end{equation}
where $\langle I|$ is the BPZ dual of the identity string field. The nontrivial part is the 3-vertex $\langle g_3|$, which we express as 
\begin{equation}\langle g_3| = \int_\mathcal{\arrowhead}\,d(\mathrm{vol})\, f_{\Gamma_*} \langle \Sigma_\flagantiflag| .\end{equation}
Let us explain the ingredients:
\begin{description}
\item{(1)} The object
\begin{equation}\langle\Sigma_\flagantiflag|: \H_{*0}\otimes \H_0\otimes\H_{0*}\to\mathbb{C}\end{equation}
is the flag-anti-flag surface state. It is a function of three wedge parameters $t_1,t_2,t_3$ and is defined so that
\begin{equation}
\langle\Sigma_\flagantiflag| \Big(|\sigmabar\rangle\otimes|\phi^i\rangle\otimes|\sigma\rangle\Big) = \Tr\big[\sqrt{\Omega}(f_\mathcal{S}\circ \phi^i)\sqrt{\Omega}\,\Omega^{t_1}\flag{\sigma}\Omega^{t_2}\Aflag{\sigmabar}\Omega^{t_3}\big],
\end{equation}
where $f_\mathcal{S}(u)=\frac{2}{\pi}\tan^{-1}u$ is the sliver coordinate map.
\item{(2)} The object $f_{\Gamma_*}$ is a distribution satisfying
\begin{equation} \sqrt{1+K}\flag{\sigma}\frac{1}{1+K}\Aflag{\sigmabar}\sqrt{1+K} =\int_0^\infty dt_1dt_2dt_3\, f_{\Gamma_*}(t_1,t_2,t_3) \,\Omega^{t_1}\flag{\sigma}\Omega^{t_2}\Aflag{\sigmabar}\Omega^{t_3}.\end{equation}
which implies\footnote{To control errors in numerical integration we used an equivalent form of the distribution
\begin{equation}
f_{\Gamma_*}(t_1,t_2,t_3) =e^{-t_1-t_2-t_3} \frac{4\sqrt{t_1t_3}}{\pi} \left(1-\frac{\d}{\d t_1}\right)^2\left(1-\frac{\d}{\d t_3}\right)^2.
\label{eq:4fGamma}\end{equation}
With this Mathematica's numerical integration routine estimated errors of about $1\%$, as opposed to \eq{fGamma} where errors were on the order of $10\%$. However, the additional derivatives in \eq{4fGamma} generate immense formulas which noticeably slowed numerical integration to around 10 minutes per coefficient.}
\begin{equation}
f_{\Gamma_*}(t_1,t_2,t_3) =e^{-t_1-t_2-t_3} \frac{1}{\pi\sqrt{t_1t_3}} \left(1-\frac{\d}{\d t_1}\right)\left(1-\frac{\d}{\d t_3}\right).
\label{eq:fGamma}\end{equation}
\item{(3)} $d(\mathrm{vol})$ is the natural measure for integration over wedge parameters:
\begin{equation}d(\mathrm{vol}) = dt_1dt_2dt_3.\end{equation}
In practice when performing the integration we will need to use upper half plane parameters. We define
\begin{equation}\alpha = t_1+t_3+1,\ \ \ \ \beta = t_2,\ \ \ \ y=(t_1-t_3)/2.\end{equation}
The upper half plane moduli $p,s$ are related to the reference and target wedge parameters through \eq{alphaps}-\eq{betaps}. The vertex operator will be inserted at a point $x$ on the real axis in the upper half plane, related to $y$ through $y=\mathcal{F}(x)$. The measure in these coordinates is given by
\begin{equation}d(\mathrm{vol}) =  dp\, ds\, dx\,\Delta(p,s) \mathcal{F}'(x),\end{equation}
where 
\begin{equation}\Delta(p,s) = \frac{16\ell^2}{\pi}\frac{s^2 p(1+p^2)}{(s^2-p^2)^3}\end{equation}
is the Jacobian for the coordinate transformation from reference/target wedge parameters to upper half plane moduli.
\item{(4)} The integration region, which we call the {\it arrowhead}, is simply over positive wedge parameters:
\begin{equation}0<t_1,t_2,t_3<\infty,\ \ \ \ (t_1,t_2,t_3)\in\arrowhead.\end{equation}
The reason for the name ``arrowhead" comes from the appearance of the region when expressed using upper half plane parameters, as shown in figure \ref{fig:arrowhead}. The region is implicitly defined by the inequalities 
\begin{equation}p>0,\ \ \ s>p,\ \ \ \alpha(p,s)>1,\ \ \ |\mathcal{F}(x)|<\frac{\alpha(p,s)-1}{2},\ \ \ \ \ \ \ \ p,s,x\in\arrowhead,\end{equation}
where $\alpha(p,s)$ is the reference wedge angle expressed as a function of upper half plane moduli.
\end{description}
The 3-vertex is built from surface states, and is therefore independent of the choice of background. The coefficients of the full solution at ghost number $1$ are given by a very similar formula. In this case there will be several terms, with the surface states modified by various ghost insertions together with additional integrations over their positions.

At least initially, the most interesting coefficients are those for which the dual basis state $\phi^i$ and boundary condition changing operators $\sigma,\sigmabar$ are primary operators of weight $h_\phi,h_\sigma$. Evaluating \eq{coef_gen} in this case leads to 
\begin{equation}
\langle\phi^i,\Gamma_*\rangle = \delta_{h_\phi=0} -\Big\langle \big(I\circ\phi^i(0)\big)\sigma(1)\sigmabar(0)\Big\rangle_\mathrm{UHP}g(h_\phi,h_\sigma).\label{eq:primary_coef}
\end{equation}
The nontrivial part is the second term. It is given by a canonical three point function of the vertex operator and two boundary condition changing operators, multiplied by a ``universal" function of the conformal weights:\footnote{Note that $f_{\Gamma_*}$ includes derivatives which act on the remainder of integrand, and in \eq{toycoef} they must be appropriately transformed into derivatives with respect to the upper half plane moduli. The derivatives do not act on the Jacobian factor.}
\begin{equation}
g(h_\phi,h_\sigma) = \int_{\arrowhead}\ dpdsdx\, \Delta(p,s)\mathcal{F}'(x)  f_{\Gamma_*}(p,s,x) \left(\frac{1}{2p\rho'}\right)^{2h_\sigma}\left(\frac{4p}{\pi\mathcal{F}'(x)(p^2-x^2)}\right)^{h_\phi}.\label{eq:toycoef}
\end{equation}
This function is determined by the geometrical data encoded in the 3-vertex $\langle g_3|$ and is the same for all backgrounds.  All of the data about the reference and target D-brane systems has been factored into the three point function. A generalization of \eq{primary_coef} can be found when $\phi^i,\sigma,\sigmabar$ are descendants of a single primary each, which is enough to cover the general situation. In this case, the first term will be modified for descendants of the identity operator; the three point function should be that of the matter primaries from which $\phi^i,\sigma,\sigmabar$ descend; the analogue of $g(h_\phi,h_\sigma)$ will be more complicated, but again will be independent of the choice of background. A similar formula will hold for the coefficients of the full solution at ghost number $1$, but the integrand will be much more complicated and there will be additional integrals over positions of $c$ ghost insertions. Now we compute toy model coefficients in a few examples.

\subsubsection*{Wilson Line}

\begin{figure}
\begin{center}
\resizebox{3in}{1.8in}{\includegraphics{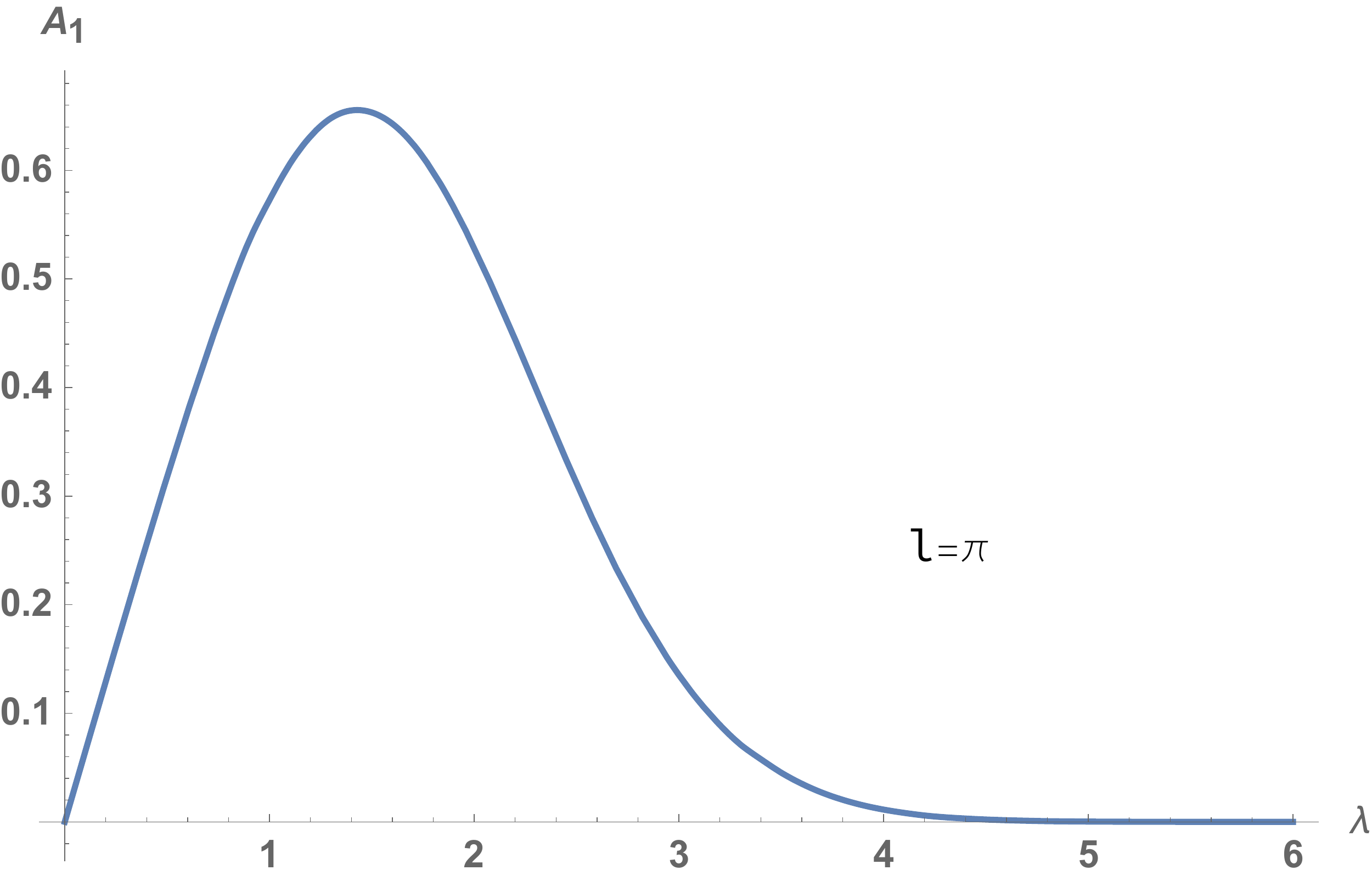}}\ \ \ \ \ 
\resizebox{3in}{1.8in}{\includegraphics{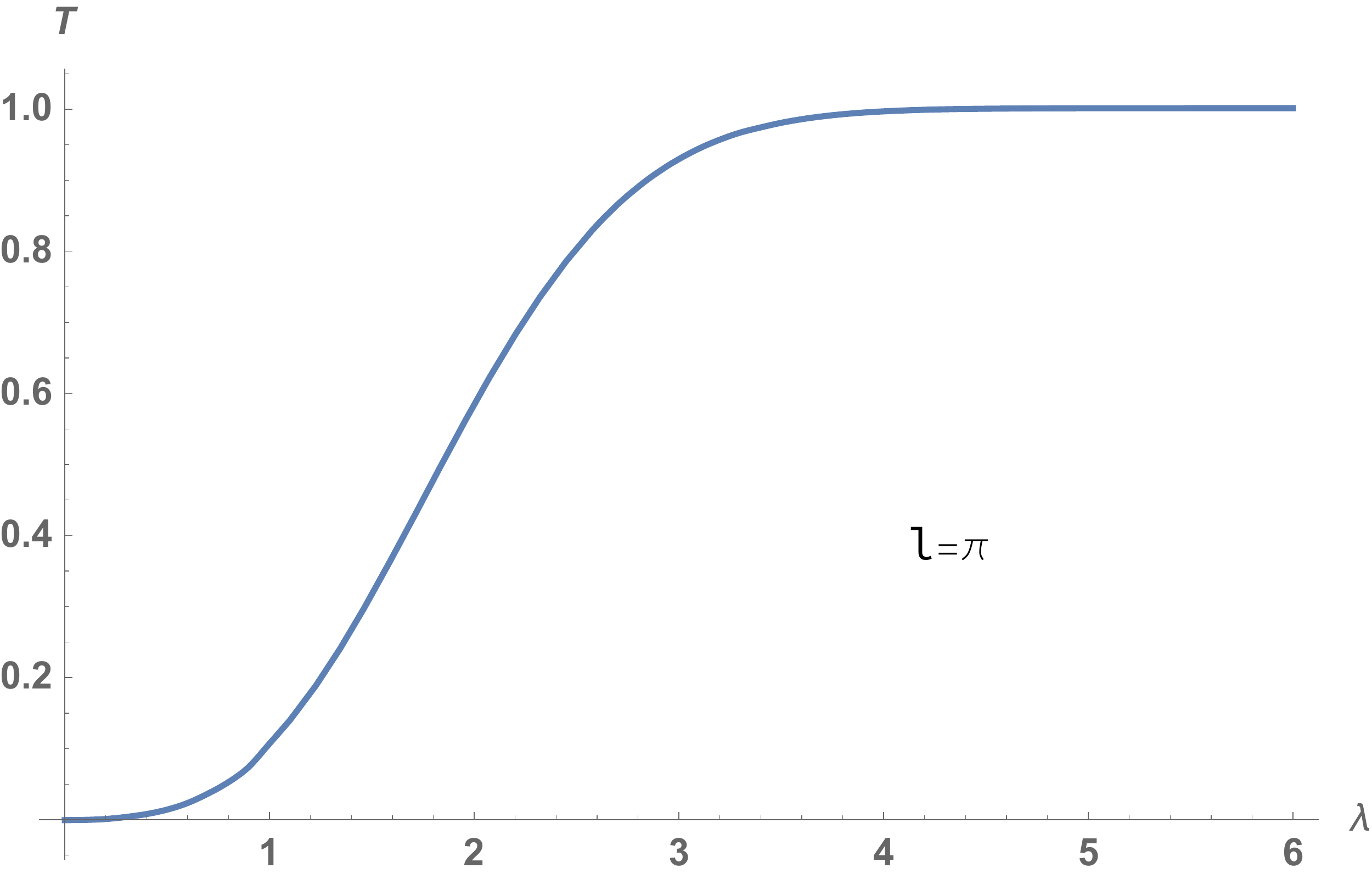}}
\end{center}
\caption{\label{fig:Wilson} Profiles for the gauge field $A_1^\mathrm{toy}$ and the tachyon $T^\mathrm{toy}$ for the toy model as a function of the boundary coupling constant $\lambda$. Here and in subsequent figures we choose $\ell=\pi$. }
\end{figure}

One interesting case is the Wilson line deformation of a free boson $X^1(z,\overline{z})$ subject to Neumann boundary conditions. This will give an expectation value to the gauge field 
\begin{equation}A_1 \frac{i}{\sqrt{2}} c\d_{\parallel} X^1(0)|0\rangle.\end{equation}
A classic question is the relation between the expectation value of $A_1$ and the strength of the coupling $\lambda$ in the conformal boundary deformation
\begin{equation}\exp\left[-\lambda\int_\mathrm{boundary} dt\, \frac{i}{\sqrt{2}} \d_\parallel X^1(t)\right],\end{equation}
which turns on the Wilson line on the worldsheet. In the solution of \cite{KOSsing}, the relation is simple equality. This seems to be a peculiarity of the fact that the boundary condition changing operators of that solution have nonsingular OPE, and so the boundary deformation does not require renormalization. In the singular OPE case, the natural renormalization prescription from the worldsheet point of view \cite{Schomerus} is generally different from that implemented by string field theory~\cite{Maccaferri_ren,Sen_ren}, and we expect a nontrivial relation between $A_1$ and $\lambda$. In fact, in the Siegel gauge solution~\cite{SZmarg}, the gauge field can only reach a finite maximum expectation value even though the boundary coupling is unbounded. This has lead to longstanding questions as to whether the fluctuation fields of the reference D-brane can capture the whole D-brane moduli space. Here we are only working with the toy model, so the analogous question is the relation between the boundary coupling and the expectation value of the field
\begin{equation}A_1^\mathrm{toy}\frac{i}{\sqrt{2}} \d_\parallel X^1(0)|0\rangle.\end{equation}
Also interesting in this context is the expectation value of the ghost number zero analogue of the tachyon:
\begin{equation}T^\mathrm{toy}|0\rangle.\end{equation}
For simplicity we refer to $A_1^\mathrm{toy}$ and $T^\mathrm{toy}$ as the gauge field and tachyon, hopefully without confusion; they are not the physical gauge field and tachyon at ghost number 1. The relevant dual states and boundary condition changing operators defining the solution are given~by
\begin{eqnarray}|\phi^{A_1^\mathrm{toy}}\rangle =-\frac{i}{2\sqrt{2}\langle 0|0\rangle_\mathrm{matter}}c\d c\d^2 c\d_\parallel X(0)|0\rangle,\lineup\ \ \ \ |\phi^{T^\mathrm{toy}}\rangle = -\frac{1}{2\langle 0|0\rangle_\mathrm{matter}}c\d c\d^2 c(0)|0\rangle,\\
\phantom{\bigg]}\sigma(t) = e^{-i\frac{\lambda}{\sqrt{2}}X^1(t)},\lineup\ \ \ \ \sigmabar(t) = e^{i\frac{\lambda}{\sqrt{2}}X^1(t)}.\label{eq:Wilsonbcc}\end{eqnarray}
The three point functions are
\begin{equation}
\Big\langle I\circ\big(\phi^{A_1^\mathrm{toy}}(0)\big)\sigma(1)\sigmabar(0)\Big\rangle_\mathrm{UHP} = -\lambda, \ \ \ \ \ \ \ \Big\langle I\circ\big(\phi^{T^\mathrm{toy}}(0)\big)\sigma(1)\sigmabar(0)\Big\rangle_\mathrm{UHP} = 1.
\end{equation}
We plot these expectation values for the toy model as a function of $\lambda>0$ in figure \ref{fig:Wilson}. We indeed find that the gauge field has a finite maximum expectation value; it represents a turning point where, after initially increasing with the strength of the deformation as one might expect, it begins decreasing. Therefore, the gauge field fails to be a good global coordinate on the moduli space of the Wilson line deformation. This confirms the picture derived from the excitations of a lump in $\phi^3$ theory~\cite{ZwiebachToy}, and also as seen in the TT/KOS solution \cite{Maccaferri_large}.\footnote{The solution of \cite{Maccaferri_marg,Maccaferri_large} can be viewed as an amalgamation between the Takahashi, Tanimoto identity-based solution \cite{TT} and the Kiermaier, Okawa, Soler solution \cite{KOS}.} The expectation value of the tachyon, however, increases monotonically with the strength of the deformation, and is a good coordinate on the moduli space of the Wilson line. This is most likely an artifact of the ghost number zero toy model, but indeed it was noted in \cite{Maccaferri_large,MaccaferriMatjej} that the tachyon at ghost number 1 provides a good coordinate on a significantly larger portion of the Wilson line moduli space than the gauge field. Another important thing to mention is that the universal function $g(h_\phi,h_\sigma)$ vanishes in the limit that the boundary condition changing operators have infinite conformal weight. This should also hold for coefficients of descendant fields. This means that in the limit $\lambda\to\infty$ the toy model will reduce to the ghost number $0$ analogue of the tachyon vacuum. We expect that this will also be true for the full solution at ghost number 1.

One curious fact is that the gauge field is not the same as the boundary coupling constant when the deformation is infinitesimal. Rather, we have
\begin{equation}A_1^\mathrm{toy} = \lambda g(1,0)+\mathcal{O}(\lambda^2).\end{equation}
For $\ell=\pi$ the proportionality factor is $g(1,0)\approx 0.65$. What is surprising is that in the linearized theory one expects an exact equality between the expectation value of fields and the appropriately normalized boundary coupling constants. 
Presently this fails since the flag state solution does not vanish when $\lambda=0$. So we are dealing with a theory of fluctuations around a nontrivial solution for the perturbative vacuum, and the linearized equations of motion are modified from the standard ones. The fluctuation fields around the trivial and nontrivial solutions for the perturbative vacuum are related by a linear field redefinition, which must account for the nontrivial proportionality factor. It is true that the tachyon of the toy model vanishes when $\lambda=0$, but the coefficients of descendants are nonzero. At ghost number 1, even the tachyon will have nonzero expectation value when the boundary coupling vanishes.

\subsubsection*{Tachyon Lumps}

Another example are lump solutions, describing the formation of a D$(p-1)$ brane through inhomogeneous condensation of the tachyon on a D$p$-brane. We consider a free boson $X^1(z,\overline{z})$ compactified on a circle of radius $R$; for the reference D-brane, the free boson is subject to Neumann boundary conditions, and for the target D-brane, it is subject to Dirichlet boundary conditions at the origin of the target space coordinate $x^1$ on the circle. The matter $SL(2,\mathbb{R})$ vacua will be normalized as $2\pi R$ and $2\pi$ for the Neumann and Dirichlet cases, respectively. First we consider the toy model where the boundary condition changing operators are Neumann-Dirichlet twist fields $\sigma_\mathrm{ND},\sigmabar_\mathrm{ND}$ of the lowest conformal weight $=\frac{1}{16}$. The profile of the tachyon lump is defined by the Fourier series
\begin{equation}T^\mathrm{toy}(x^1) = \sum_{n\in\mathbb{Z}}T_n^\mathrm{toy}e^{\frac{in x^1}{R}},\end{equation}
where $T_n^\mathrm{toy}$ are coefficients of the states $e^{in X^1/R}(0)|0\rangle$ in the toy model. These can be extracted by contracting with the dual state
\begin{equation}|\phi^{T_n^\mathrm{toy}}\rangle = -\frac{1}{4\pi R}c\d c\d^2 ce^{-i nX^1(0)/R}|0\rangle\label{eq:phiTntoy}\end{equation}
and using the 3-point function \cite{Mukhopadhyay,Pasando}
\begin{equation}\Big\langle I\circ\big(\phi^{T_n^\mathrm{toy}}(0)\big)\sigma_\mathrm{ND}(1)\sigmabar_\mathrm{ND}(0)\Big\rangle_\mathrm{UHP} = \frac{4^{-n^2/R^2}}{R}.\end{equation}

\begin{figure}
\begin{center}
\resizebox{3.8in}{2in}{\includegraphics{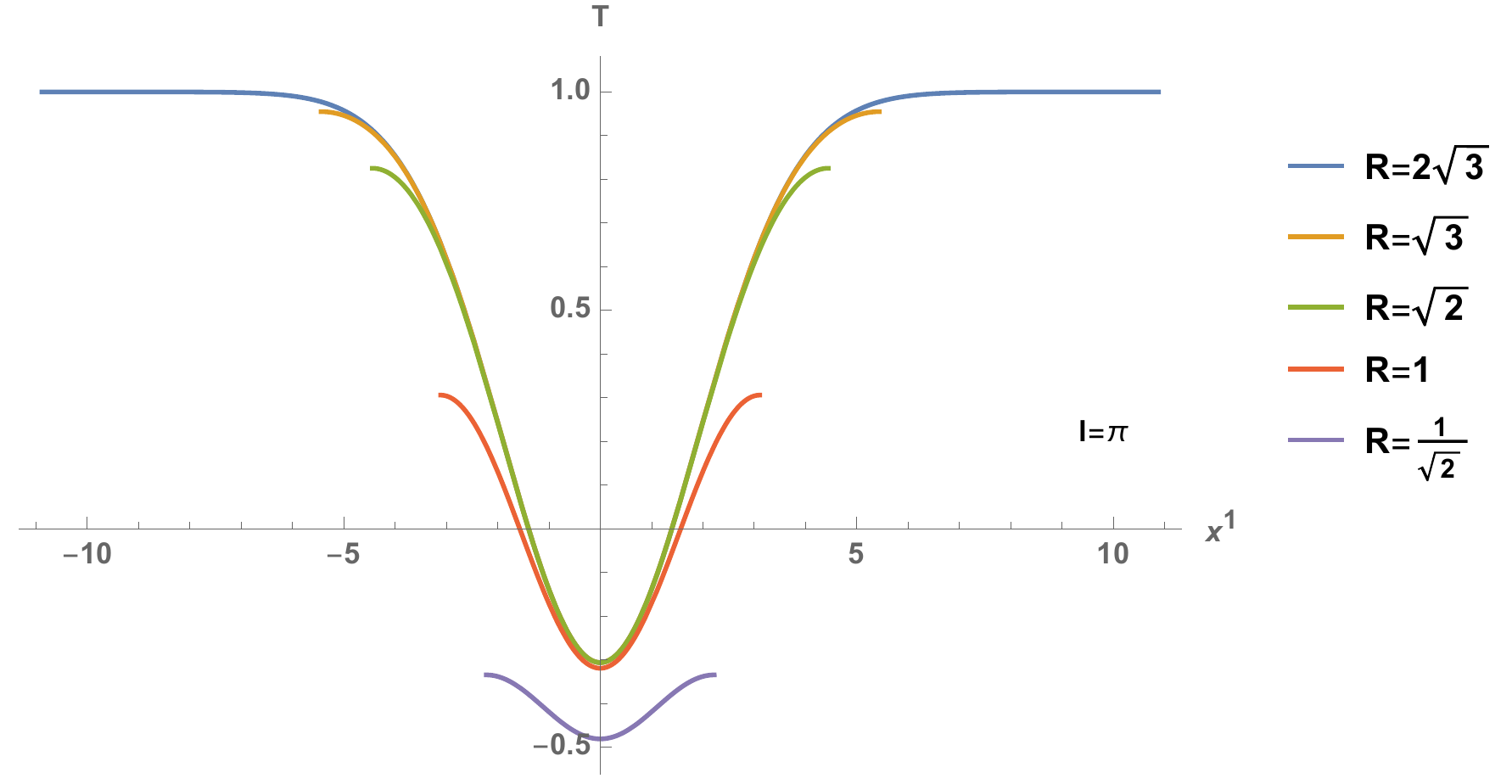}}
\end{center}
\caption{\label{fig:lump} Tachyon lump profiles in the toy model derived from unexcited Neumann-Dirichlet twist fields of weight $1/16$, for various radii.}
\end{figure}
\noindent The lump profile for various radii is shown in figure \ref{fig:lump}; numerical values of the first 14 coefficients at $R=2\sqrt{3}$ can be found in in appendix \ref{app:lump}. For $R$ larger than the self-dual radius $R=1$, the profile can be seen as a ``hole" in the tachyon condensate centered at the origin. The picture is broadly similar to that derived from level truncation studies in Siegel gauge~\cite{MZlump}.  For smaller radii the solution can be understood in the $T$-dual picture as representing the reverse process of forming a D$p$-brane on a circle of radius $1/R$ out of the fluctuations of a D$(p-1)$-brane. In this case the target background has higher energy than the perturbative vacuum. As $R$ decreases we find that the lump rapidly flattens out, and its average value turns negative. This is similar to what was found in~\cite{KOSsing}. 
It is interesting to mention that the expectation value of the zero momentum tachyon for the solution of \cite{KOSsing} depends only on the disk partition functions $g_0 = \langle 0|0\rangle_\mathrm{matter}^{\BCFT_0}$ and $g_*=\langle 0|0\rangle_\mathrm{matter}^{\BCFT_*}$ of the reference and target backgrounds:
\begin{equation}T_0 = T_{\tv}\left(1-\frac{g_*}{g_0}\right),\end{equation}
where $T_{\tv}$ is the tachyon expectation value at the tachyon vacuum. For higher energy solutions, the disk partition function of the target background is necessarily larger than that of the reference background, and the zero momentum tachyon always takes a negative expectation value. For the ghost number zero toy model of the flag state solution, the zero momentum tachyon takes the form
\begin{equation}T_0^\mathrm{toy} = 1-\frac{g_*}{g_0}g(0,h_\sigma).\end{equation}
For boundary condition changing operators with positive nonvanishing conformal weight, the constant $g(0,h_\sigma)$ is positive and strictly less than one. So higher energy solutions may in some cases have positive tachyon expectation value. For fixed conformal weights, however, the expectation value is decreasing linearly with the energy, and will eventually go negative. This is perhaps surprising since a negative tachyon expectation value can give a large negative contribution to the energy, but the total energy derived from the complete set of fields is nevertheless positive. We do not know whether this phenomenon is generic in higher energy solutions or is particular to the intertwining-type solutions which have been derived analytically. For known numerical higher energy solutions in Siegel gauge, the tachyon expectation value appears to be positive \cite{Ising,Matjej}.

\begin{figure}
\begin{center}
\resizebox{3.8in}{2in}{\includegraphics{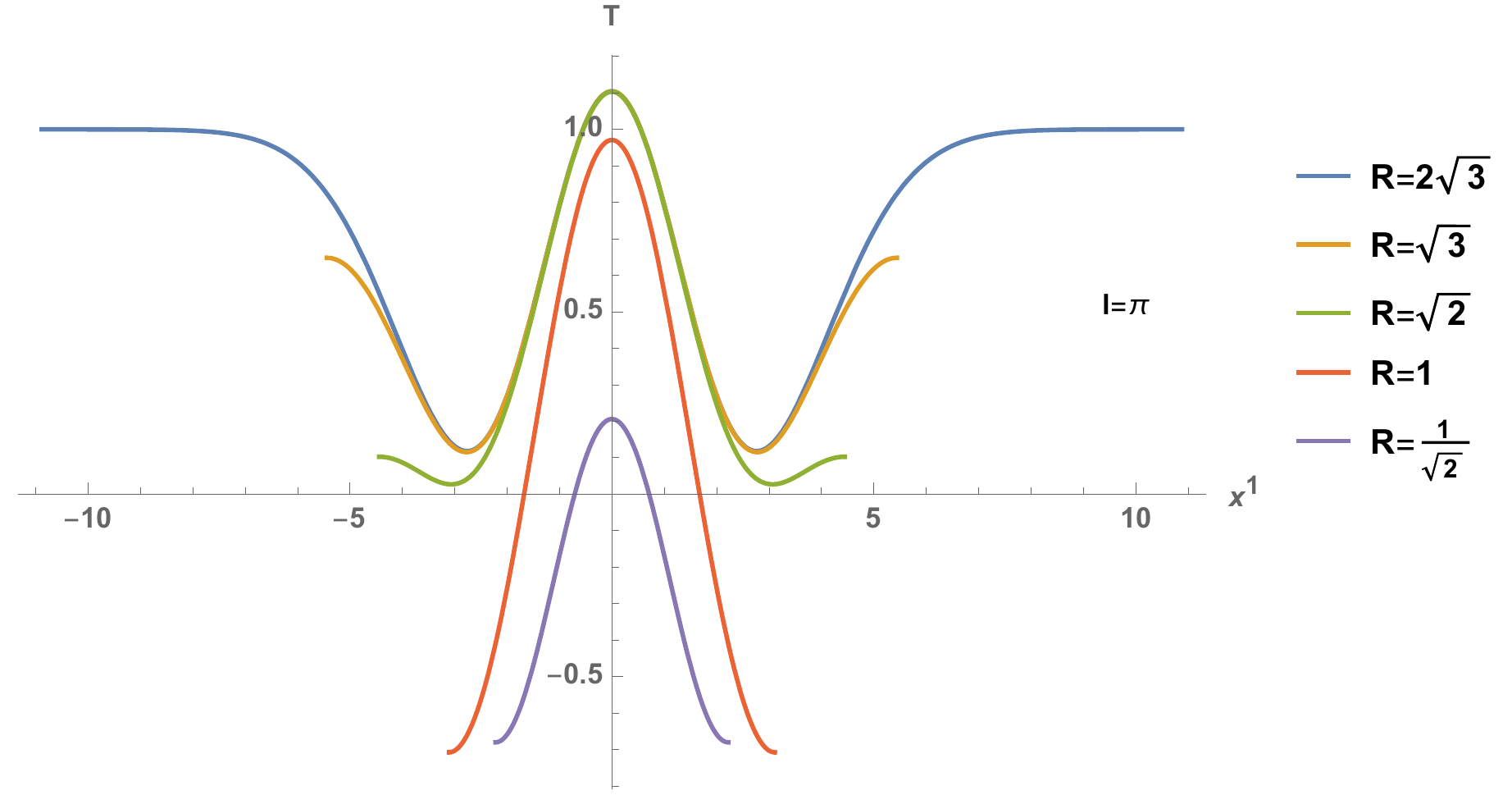}}
\end{center}
\caption{\label{fig:excited_lump} Tachyon lump profiles in the toy model derived from excited Neumann-Dirichlet twist fields of weight $9/16$, for various radii. The profile is somewhat counterintuitive as we decrease the radius, as it is closest to the tachyon vacuum at $x^1=0$, precisely where the D-brane is located. For sufficiently small radius, the lump flattens out and its expectation value turns negative, similarly to the lump of the unexcited twist fields.}
\end{figure}

\begin{figure}
\begin{center}
\resizebox{3.2in}{2in}{\includegraphics{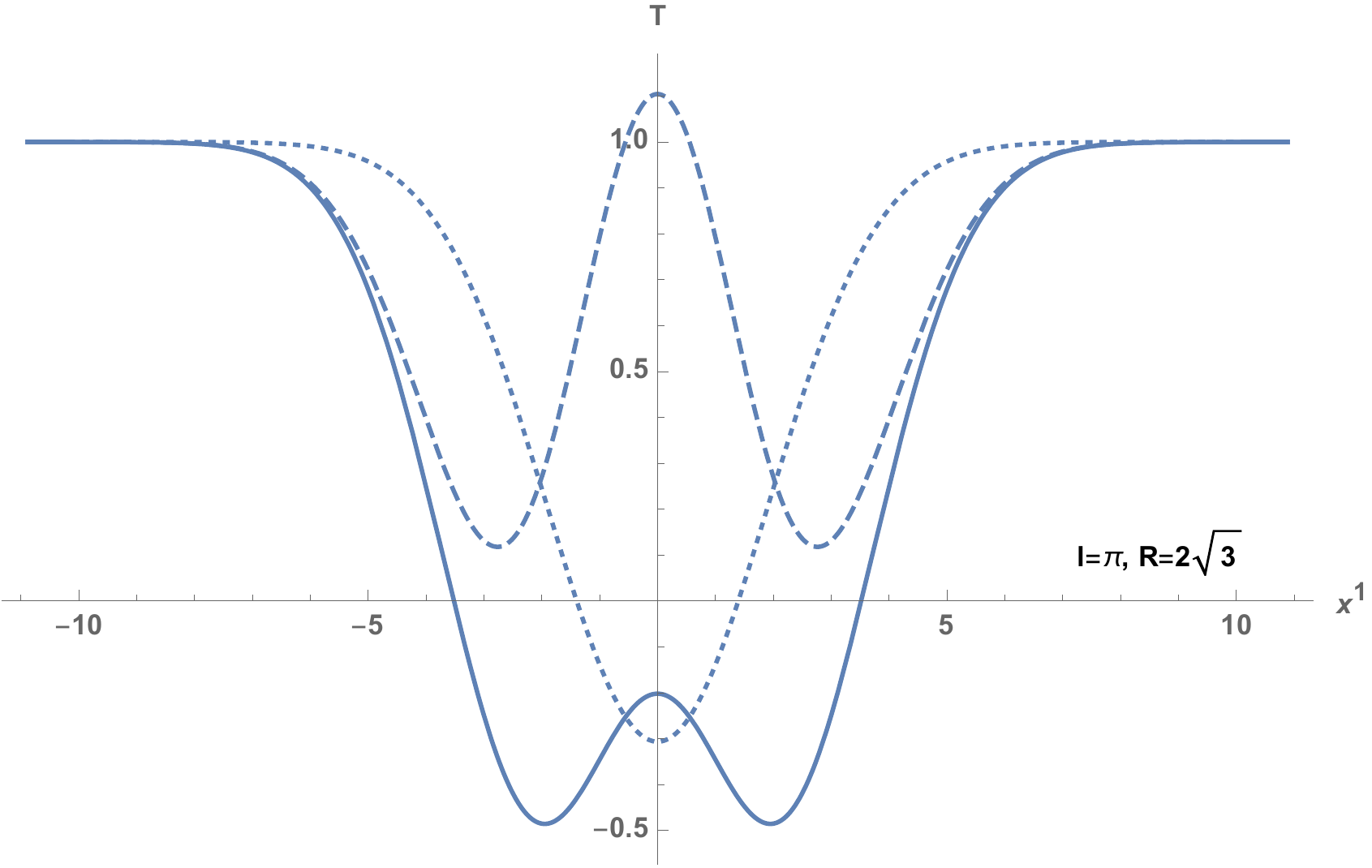}}
\end{center}
\caption{\label{fig:double_lump} Tachyon lump profile for a coincident pair of D$(p-1)$-branes in the toy model, plotted against the profiles for the single D$(p-1)$ derived from excited and unexcited twist fields at $R=2\sqrt{3}$.}
\end{figure}

It is interesting to venture into new territory and ask about lump profiles in the case that the boundary condition changing operators do not have the lowest possible conformal weight. This is a necessary step towards constructing solutions representing coincident D$(p-1)$ branes with non-abelian gauge symmetry. We consider excited Neumann-Dirichlet twist fields $\tau_\mathrm{ND},\overline{\tau}_{\mathrm{ND}}$, which are primaries of weight $\frac{9}{16}$ generated from the OPE between $\sigma_\mathrm{ND},\sigmabar_\mathrm{ND}$ and $\d X^1$. As shown in appendix \ref{app:cosine} the three point function is
\begin{equation}\Big\langle I\circ\big(\phi^{T_n^\mathrm{toy}}(0)\big)\tau_\mathrm{ND}(1)\overline{\tau}_\mathrm{ND}(0)\Big\rangle_\mathrm{UHP} = \frac{4^{-n^2/R^2}}{R}\left(1-\frac{4 n^2}{R^2}\right).
\end{equation}
The excited lump profiles are shown for various radii in figure \ref{fig:excited_lump}; the first 14 coefficients at $R=2\sqrt{3}$ can be found in in appendix \ref{app:lump}. Instead of a solitary disturbance in the tachyon condensate localized around the origin, the lump splits into two components separated by a large gap. Nothing comparable has been seen in level truncation, but the profile is fairly similar to a radial cross-section of the first excited Gopakumar-Minwalla-Strominger (GMS) soliton~\cite{GMS}. Since $\sigma_\mathrm{ND},\sigmabar_{\rm ND}$ and $\tau_\mathrm{ND},\overline{\tau}_{\rm ND}$ have different conformal weights, following the discussion of subsection \ref{subsec:comments} we can construct a solution describing a pair of coincident D$(p-1)$-brane solutions by adding the respective solutions around the tachyon vacuum. At the level of the toy model, this implies that the profile for the double lump can be obtained by superimposing the profiles generated by the excited and unexcited twist fields. This is shown in figure \ref{fig:double_lump}. The result is again a more-or-less solitary hole in the tachyon condensate but with two ``dips," perhaps suggestive of the pair of D-branes it contains.\footnote{We are not confident that the ``double dip" is stable under gauge transformation. It is much more pronounced for the solution of \cite{KOSsing} and radial cross-section of the GMS double soliton, but is absent from the profile generated from the Fock space coefficient principle (see next section) for sufficiently large $p$. What seems to be generally true, however, is that the disturbance of the tachyon vacuum is wider, but not appreciably deeper, than that of the single lump.}

Excited twist fields are not needed when the pair of D$(p-1)$-branes are separated. We can simply superimpose lump solutions based on unexcited twist fields, since in the cross terms the OPEs of the twist fields will not contain the identity operator. In Siegel gauge it is also approximately true that separated D$(p-1)$-branes can be found by superimposing solutions for a single D$(p-1)$ at different locations; this can be effectively used as an input for improving the appoximation via Newton's method in level truncation \cite{Matjej}. However, generally this approach fails to find a solution if the D-branes come closer than $2\pi$, where the ground state of the stretched string connecting the D-branes becomes massless. In particular, there is no known solution for a coincident pair of D$(p-1)$-branes in Siegel gauge. The above discussion suggests that this difficulty may be related to the absence of something analogous to excited twist field solutions in Siegel gauge. The absence of such solutions is expected, since the Siegel gauge condition should determine the solution for the D$(p-1)$-brane uniquely.\footnote{It is possible that a single background could be represented by more than one solution in Siegel gauge. In practice this has not been seen. By contrast, in Schnabl gauge many solutions exist for the same background, and this is why it is possible to find multiple D-brane solutions in the framework of \cite{KOSsing}.} This is an interesting indication that gauge fixing may be at the heart of the difficulty of constructing multiple D-bane solutions in level truncation. The flag state solution is not characterized by a gauge condition.

\subsubsection*{Cosh Deformation}

\begin{figure}
\begin{center}
\resizebox{3.2in}{2in}{\includegraphics{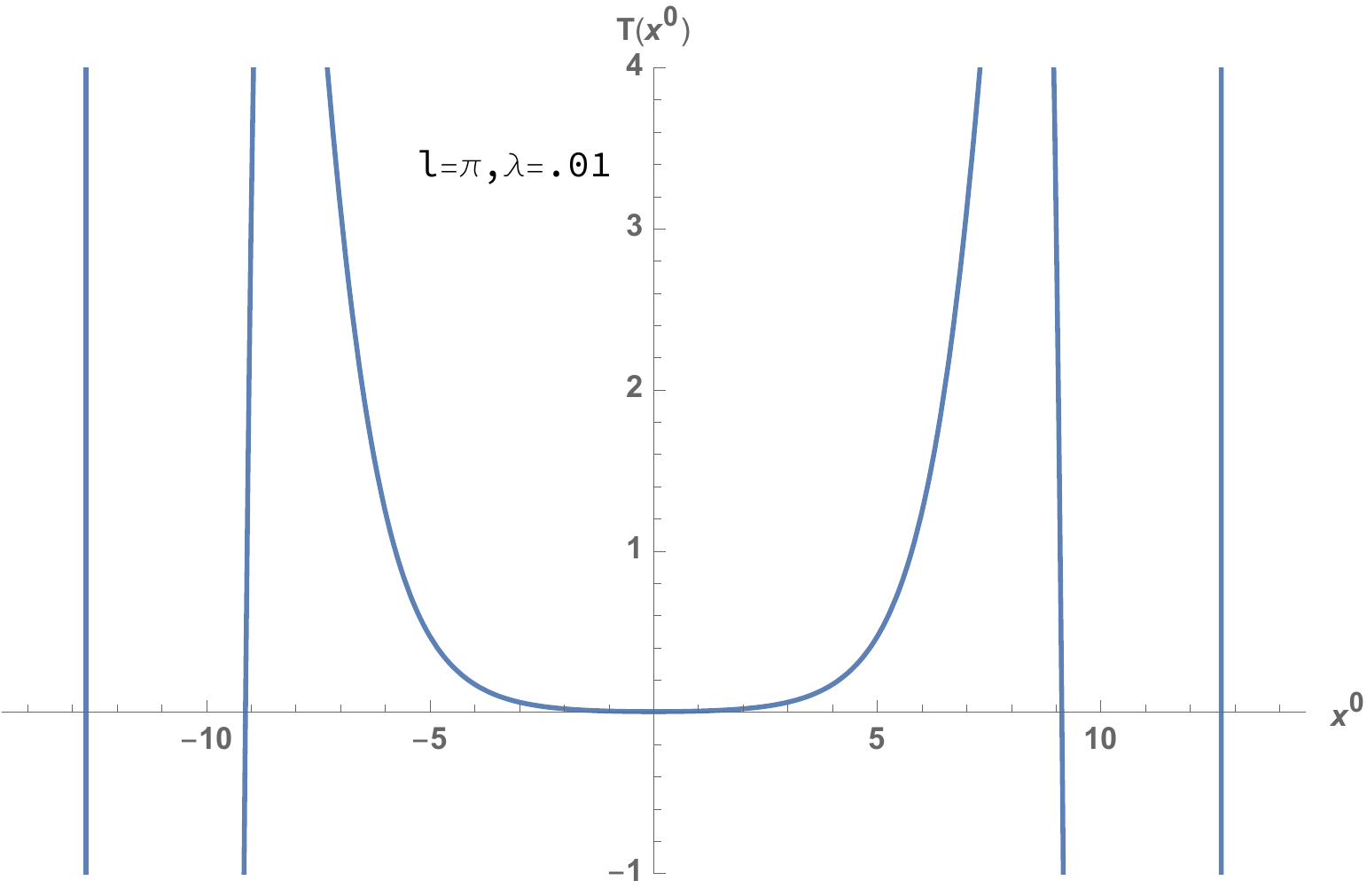}}
\end{center}
\caption{\label{fig:smalllambda} Time evolution of the tachyon in the toy model for the cosh deformation at $\lambda=.01$.}
\end{figure}

Finally, we consider the cosh rolling tachyon deformation \cite{SenRolling}, generated by a conformal boundary interaction
\begin{equation}\exp\left[\lambda\int_\mathrm{boundary}dt \cosh X^0(t)\right],\end{equation}
where $X^0(z,\overline{z})$ is the timelike free boson subject to Neumann boundary conditions. This is closely related to the cosine deformation of a D-brane with free boson $X^1(z,\overline{z})$ subject to Neumann boundary conditons on a circle at the self-dual radius; the defomations are related by Wick rotation $X^1\to i X^0$. A perturbative analysis of this background using the solution of \cite{KO} was given in~\cite{Longton},  but the flag state solution allows us to understand the background nonperturbatively in $\lambda$. Particularly interesting is the evolution of the tachyon field  for various values of the deformation parameter $\lambda$. We consider
\begin{equation}
T^\mathrm{toy}(x^0,\lambda) = \sum_{n\in \mathbb{Z}} T_n^\mathrm{toy}(\lambda) e^{n x^0},
\end{equation}
where $T_n^\mathrm{toy}(\lambda)=T_{-n}^\mathrm{toy}(\lambda)$ are coefficients of the states $e^{nX^0(0)}|0\rangle$ in the toy model. These can be extracted by contracting with the dual state
\begin{equation}|\phi^{T_n^\mathrm{toy}(\lambda)}\rangle = -\frac{1}{2\langle 0|0\rangle_\mathrm{matter}} c\d c\d^2c e^{-n X^0(0)}|0\rangle.\end{equation}
As shown in appendix \ref{app:cosine}, the relevant 3-point function is 
\begin{equation}
\Big\langle I\circ\big(\phi^{T_n^\mathrm{toy}(\lambda)}(0)\big)\sigma_{\lambda}(1)\sigmabar_{\lambda}(0)\Big\rangle_\mathrm{UHP} = (-1)^n 4^{-n^2}\frac{\mathcal{P}_n(\lambda)}{\mathcal{P}_n(\frac{1}{2})},\ \ \ \ (n\geq 0),\label{eq:cosh_def}
\end{equation}
where $\mathcal{P}_n(\lambda)$ are polynomials in $\lambda$ given by
\begin{equation}\mathcal{P}_n(\lambda) = \lambda^n \prod_{j=1}^{n-1}(j^2-\lambda^2)^{n-j}.\end{equation}
The boundary condition changing operators $\sigma_\lambda,\sigmabar_\lambda$ turn on the cosh deformation on a segment of the open string boundary with strength $\lambda$, and are primaries of weight $\frac{\lambda^2}{4}$. The cosh deformation is periodic in the coupling constant, with $\lambda$ and $\lambda+2$ representing equivalent time dependent backgrounds. However, the boundary condition changing operators and the solution are not periodic, with $\lambda$ and $\lambda+2$ representing gauge equivalent solutions constructed with boundary condition changing operators of different conformal weights. We plot the evolution of the tachyon for small positive $\lambda$ in figure \ref{fig:smalllambda}. For some time interval the tachyon lingers near the perturbative vacuum (which for the ghost number zero toy model, happens to lie at $T^\mathrm{toy}=0$) and outside this range the tachyon falls down the potential and engages in the well-known uncontrolled oscillation which has long been a source of puzzlement \cite{MoellerZwiebachRolling,HataRolling}.  

\begin{figure}
\begin{center}
\resizebox{3.2in}{2in}{\includegraphics{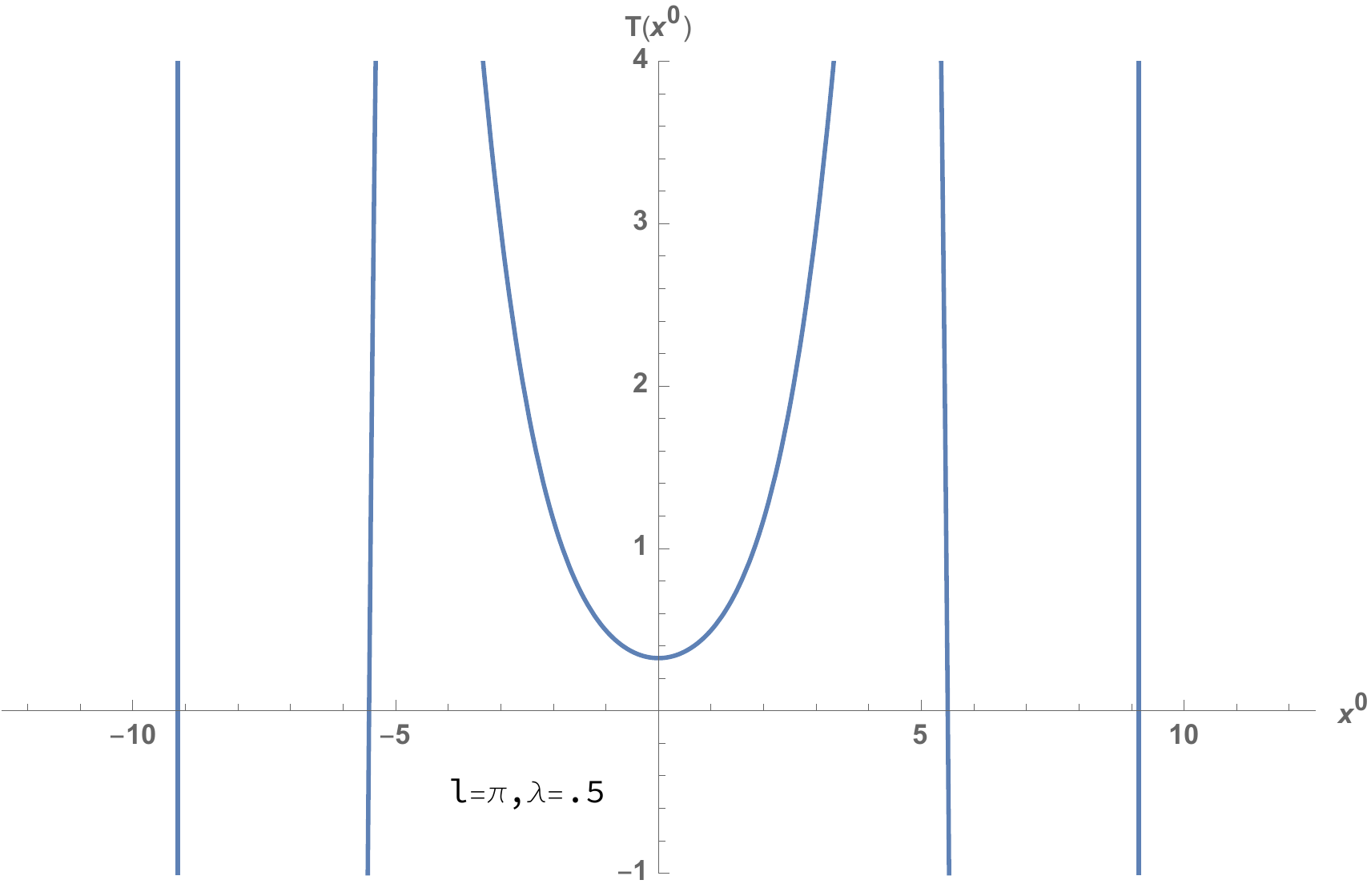}}
\end{center}
\caption{\label{fig:lambdahalf} Time evolution of the tachyon in the toy model at the critical value $\lambda=1/2$ of the cosh deformation.}
\end{figure}

What is more novel is that we can describe the evolution of the tachyon field at the critical value of the marginal parameter $\lambda=1/2$, where the boundary state vanishes and the  simplest interpretation is that the background lies at the tachyon vacuum  \cite{SenRolling}. This is shown in figure \ref{fig:lambdahalf}. Compared with small $\lambda$, the main difference is that the tachyon does not approach the perturbative vacuum quite as closely or for very long, and otherwise engages in large oscillation. This does not suggest that the background is at the tachyon vacuum. It is possible that the solution could be related to the tachyon vacuum by a time dependent gauge tansfomation. This, however, seems to contradict the fact that the solution supports cohomology, at least in the form of a marginal deformation which allows us to adjust the coupling constant of the cosh deformation. This is different from the kind of deformation of the tachyon vacuum considered in \cite{rollingvac}, which is really an effect of the nonlinear equations of motion. Some confirmation of the nontriviality of the solution comes from the characteristic projector \cite{Ellwood}, which in this case takes the form\footnote{The characteristic projector can be computed as $\lim_{N\to\infty}(-[A,\Psi_*-\Psi_\tv])^N$. To extract the limit we used $[A,\Psi_\tv] = F$ and that $F$ raised to the infinite power gives the sliver state. We also assumed that $A\Omega^\infty$ vanishes, which is true when contracted with a basis of Fock states. Under these assumptions one can readily check that the state is invariant under left and right multiplication with $-[A,\Psi_*-\Psi_\tv]$. An apparent inconsistency, however, is that the characteristic projector appears to square to zero on account of $A^2=0$. The problem is that to compute the square of the characteristic projector we need to account for contributions proportional to $A\Omega^\infty$, since while formally vanishing they can have nonzero star products with other projector-like states.}
\begin{equation}X^\infty = A\Sigma\Psi_\tv\Omega^\infty\Psi_\tv\Sigmabar A.\end{equation}
This satisfies
\begin{equation}Q_{\Psi_*}X^\infty = 0.\end{equation}
It is believed that the characteristic projector, when nonvanishing, is a representative of the BRST cohomology class of the identity operator in the target background. While we do not compute the projector in detail here, it is clearly not zero; its coefficients will oscillate in a similar way as the tachyon of the toy model. Again, this indcates that the cosh rolling tachyon at $\lambda=1/2$ supports cohomology, while the tachyon vacuum does not.  In \cite{Lambert,imaginary} it was found that closed string states produced by the cosh rolling tachyon background are present even when $\lambda=1/2$ and D-branes are absent. It is possible that this effect can be seen at the level of the purely classical open string field theory solution.  

\begin{figure}
\begin{center}
\resizebox{3.2in}{2in}{\includegraphics{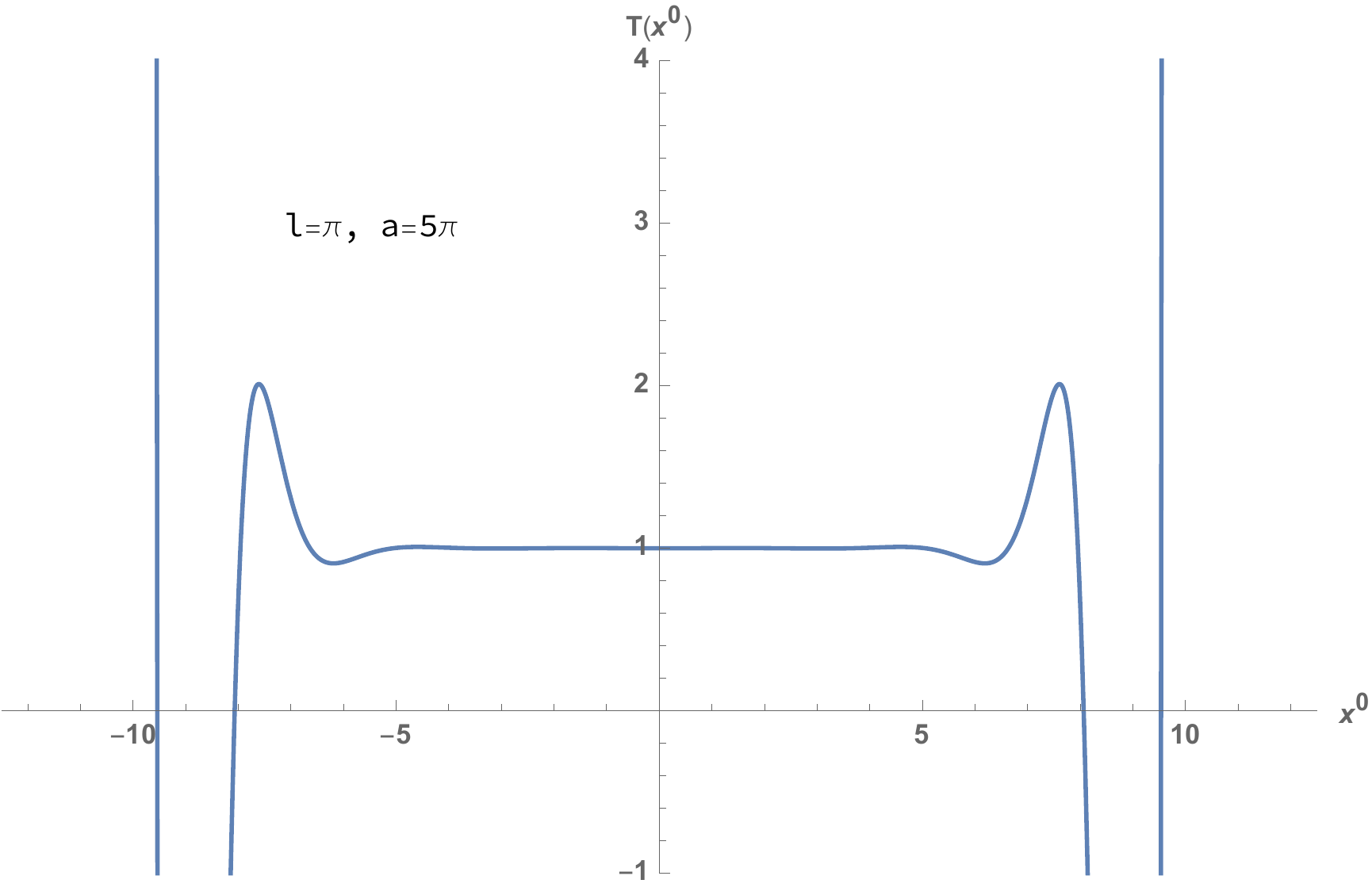}}
\end{center}
\caption{\label{fig:ainfty} Time evolution of the tachyon in the toy model for a periodic array of D-branes localized in imaginary time at $x^0 =  i a \left(n+\frac{1}{2}\right)$, with $a=5\pi$.}
\end{figure}

Through Wick rotation of the cosine deformation, the cosh deformation at $\lambda=1/2$ can be interpreted as representing a periodic array of D-branes localized in ``imaginary time" at positions 
\begin{equation}x^0 = 2\pi i\left(n+\frac{1}{2}\right),\ \ \ \ n\in\mathbb{Z}.\end{equation}
This analogy indicates that a large moduli space of backgrounds should open up at $\lambda=1/2$ corresponding to the Wick rotation of the moduli space of D-branes \cite{imaginary}. One deformation we can consider is to modify the imaginary time interval between D-branes in the periodic array, so the D-branes are located at
\begin{equation}x^0 = i a \left(n+\frac{1}{2}\right),\ \ \ \ n\in\mathbb{Z}.\end{equation}
The solution describing such a background can be obtained by Wick rotation of the corresponding periodic array of lumps. We can ask how the solution changes as we push the imaginary time interval between D-branes to infinity. For the toy model, we plot the evolution of the tachyon for large imaginary separation in figure \ref{fig:ainfty}. We see that in a large region surrounding $x^0=0$ the tachyon settles to its expectation value at the tachyon vacuum, and as the imaginary separation increases the oscillations are pushed out to $x^0=\pm\infty$. We expect that similar behavior will be seen in the characteristic projector, and the obstruction to trivializing the cohomology will gradually disappear for large imaginary separation. Thus the tachyon vacuum emerges only after increasing the coupling of the cosh deformation to $\lambda=1/2$ and sending the resulting D-branes localized in imaginary time to infinity. 

\subsection{The Fock Space Coefficient Principle}
\label{subsec:principle}

We have seen that, up to a shift from the tachyon vacuum, the coefficients of the flag state solution are given by a matter three point function times a universal factor. Since the main difficulty in extracting coefficients is in computing the universal factor, one can ask how important it is in determining the qualitative results. As it turns out, it may not be as important as one would think. This is an instance of a general observation, which we propose as a ``principle:" 
\begin{SFCP} 
Consider a solution $\Psi$ in the string field theory of $\BCFT_0$ describing some background $\BCFT_*$. The overlap of $\Psi$ with a test state $-c\d c\phi(0)|0\rangle$, where $\phi$ is a matter primary, shows a rough correspondence 
\begin{equation}\langle -c\d c \phi,\Psi\rangle \sim \delta_{h=0} - \big\langle \sigmabar(p)\phi(0)\sigma(-p)\big\rangle_\mathrm{UHP}^\mathrm{matter},\label{eq:FSCP}\end{equation}
where $\sigma$ and $\sigmabar$ are appropriately chosen boundary condition changing operators between $\BCFT_0$ and $\BCFT_*$ and the parameter $p>0$ is not too small. 
\end{SFCP}

\noindent The first term in \eq{FSCP} gives a shift for the coefficient of the zero momentum tachyon which is meant to represent the expectation value at the tachyon vacuum. The parameter $p$ can be interpreted as a gauge parameter. The sense in which the left and right hand sides of \eq{FSCP} correspond is best explained through examples:

\begin{figure}
\begin{center}
\resizebox{2in}{1.1in}{\includegraphics{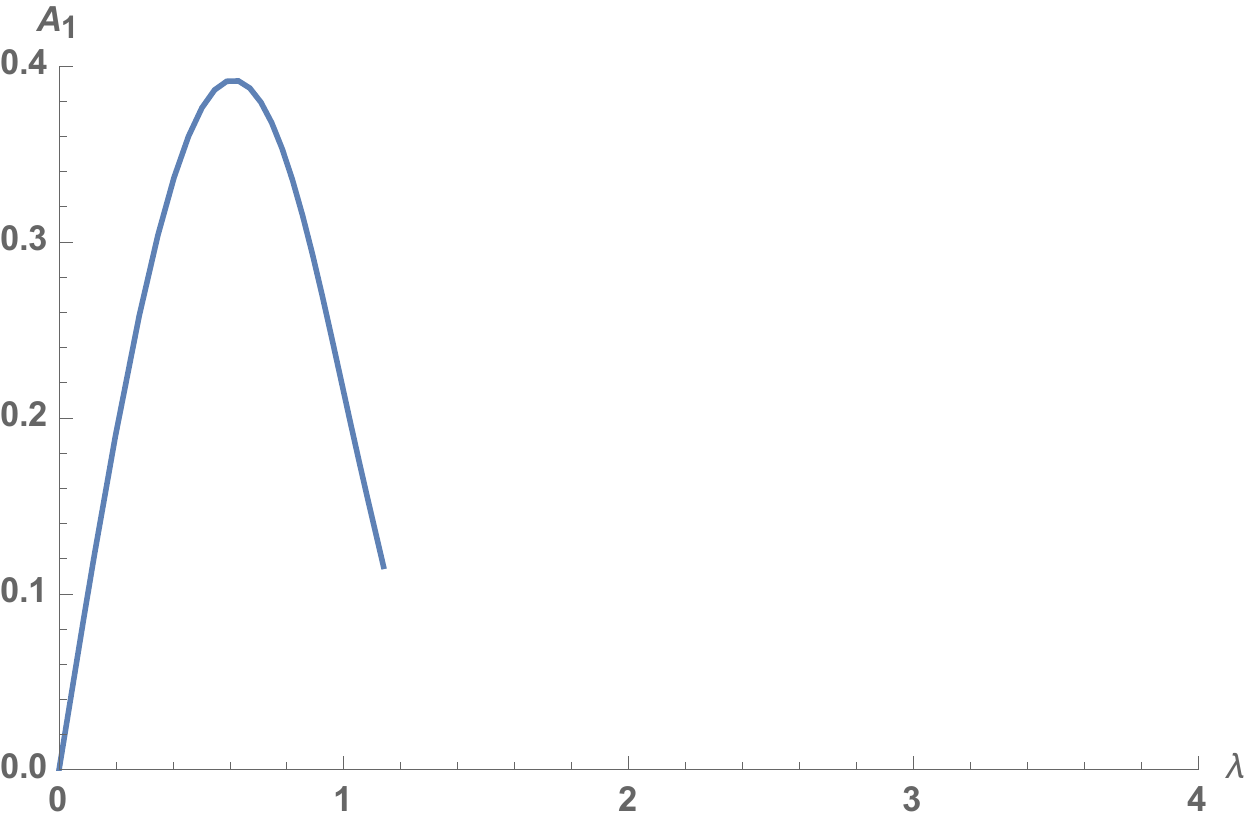}}\ \ \ \ 
\resizebox{2in}{1.1in}{\includegraphics{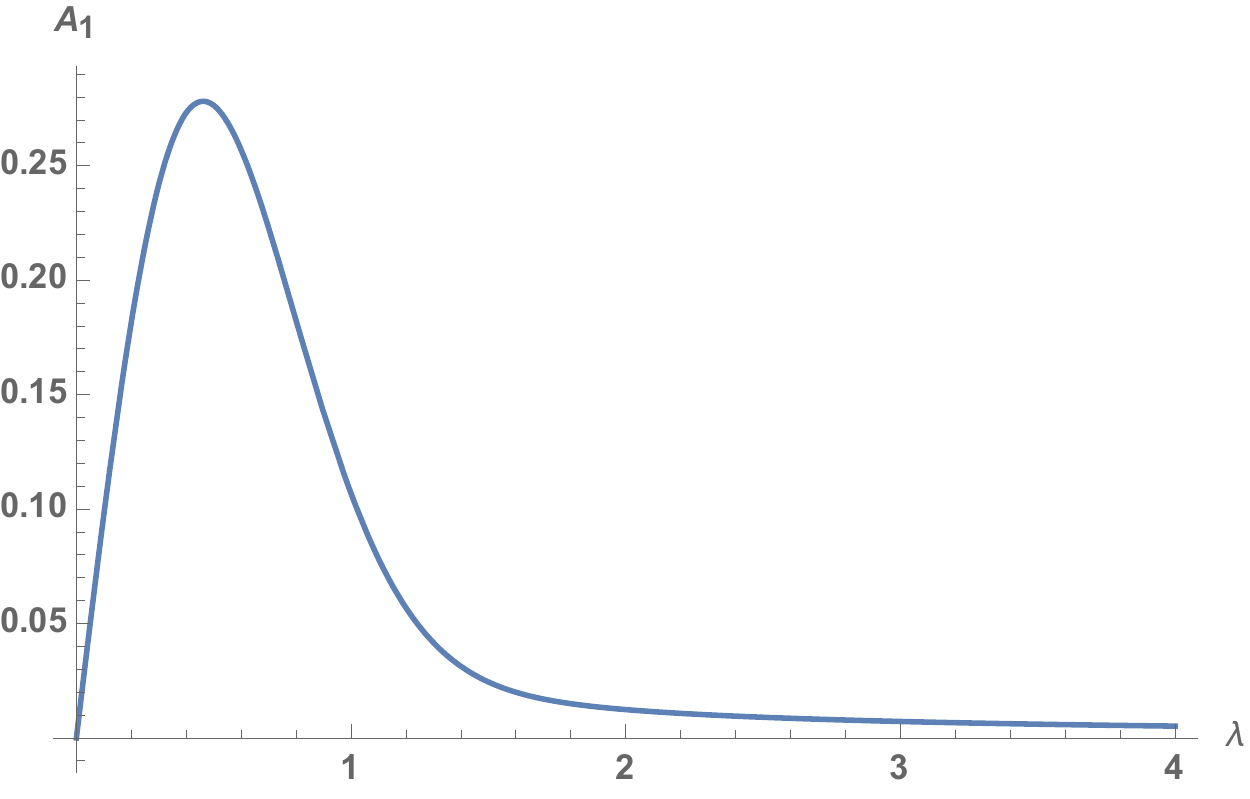}}\ \ \ \ 
\resizebox{1.7in}{1.1in}{\includegraphics{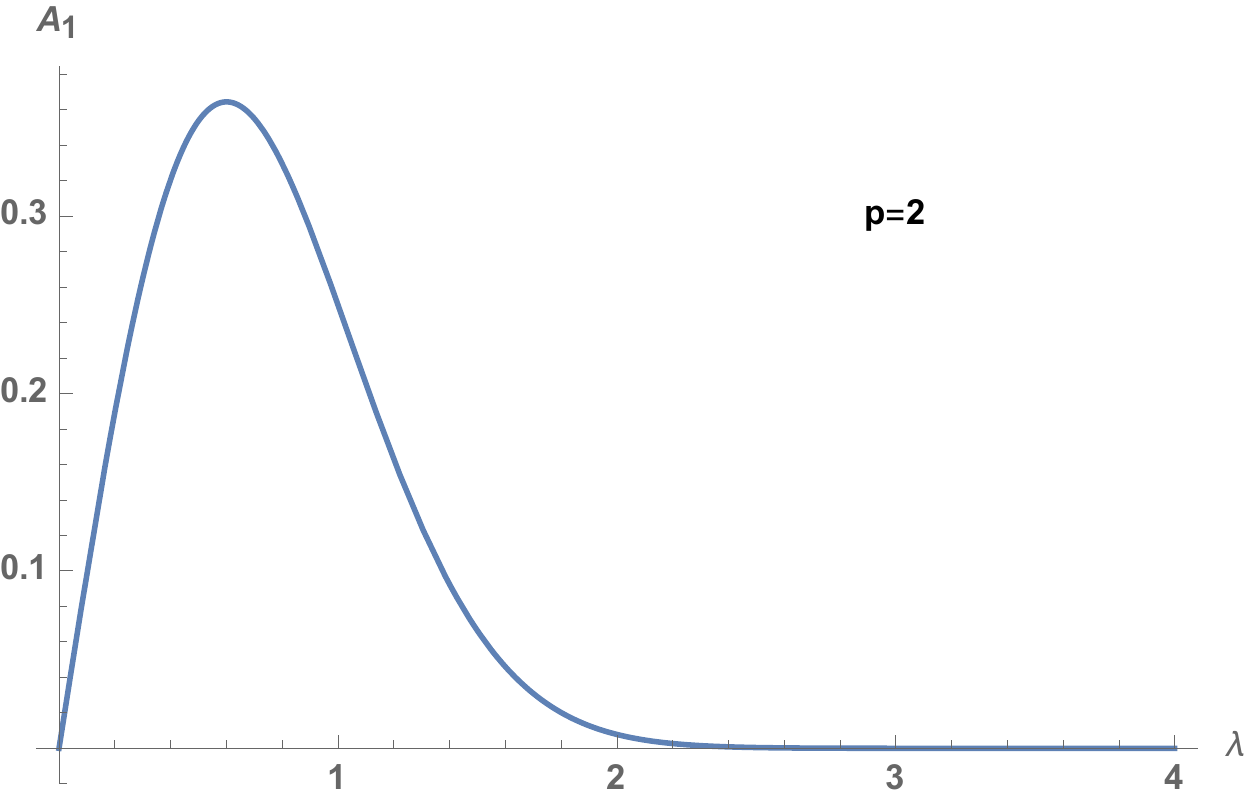}}
\end{center}
\vspace{-.7cm}
\caption{\label{fig:coefWilson} Gauge field as a function of the Wilson line coupling constant, left in Siegel gauge, center for the TT/KOS solution (with $t=5$ in \cite{Maccaferri_marg}), and right from the shifted 3-point function.}
\begin{center}
\resizebox{2in}{1.1in}{\includegraphics{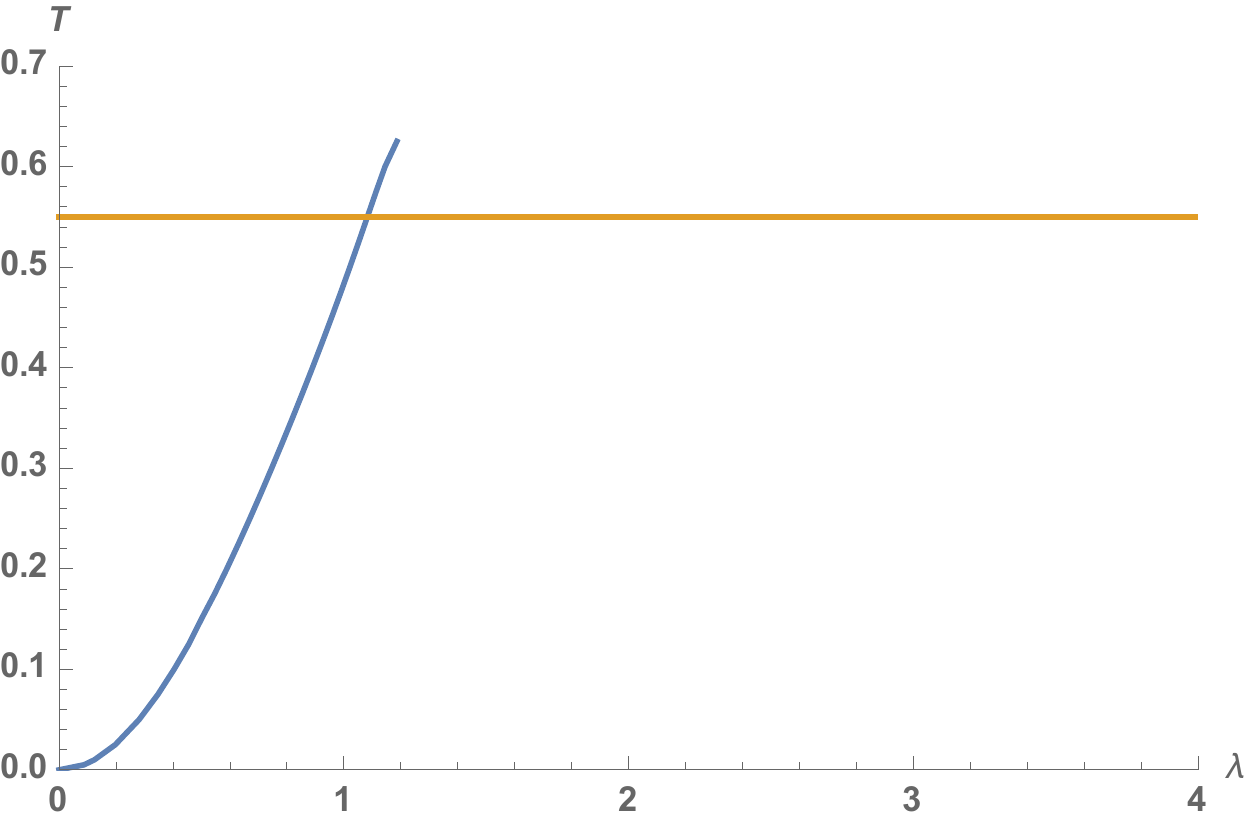}}\ \ \ \ 
\resizebox{2in}{1.1in}{\includegraphics{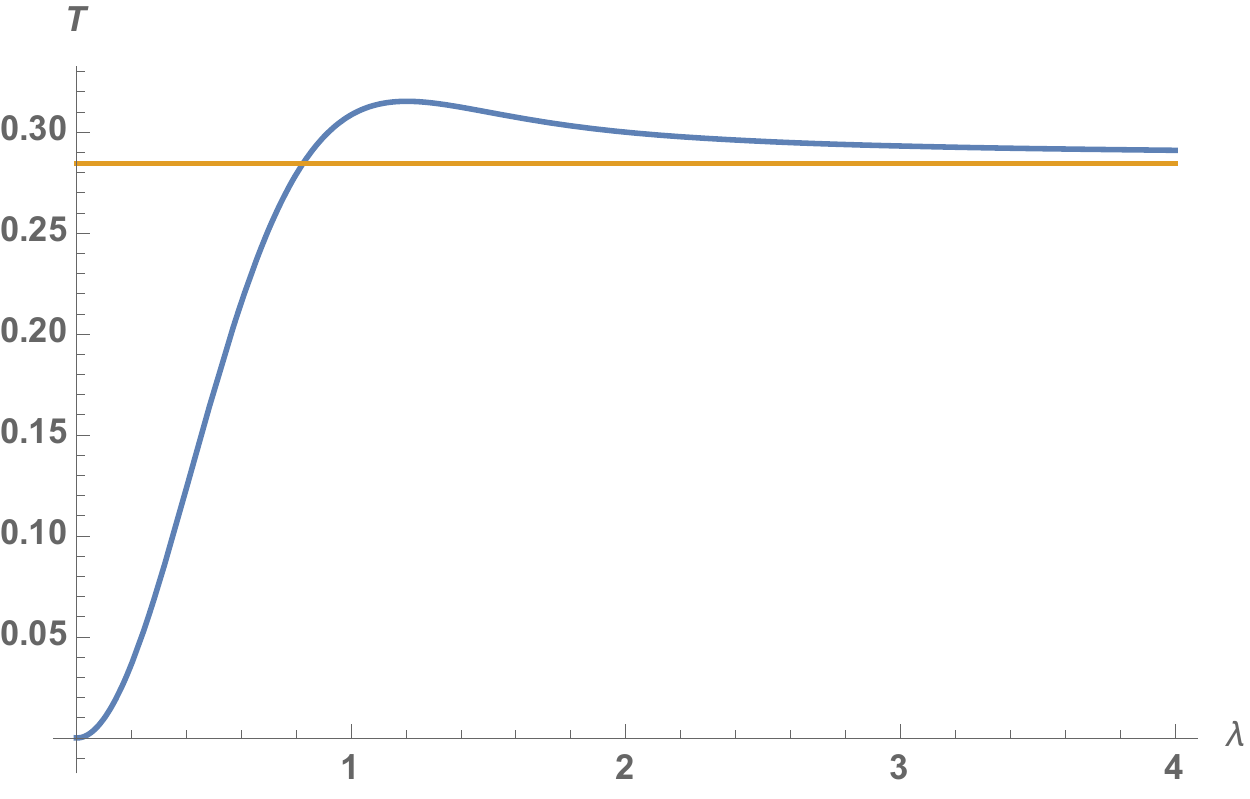}}\ \ \ \ 
\resizebox{1.7in}{1.1in}{\includegraphics{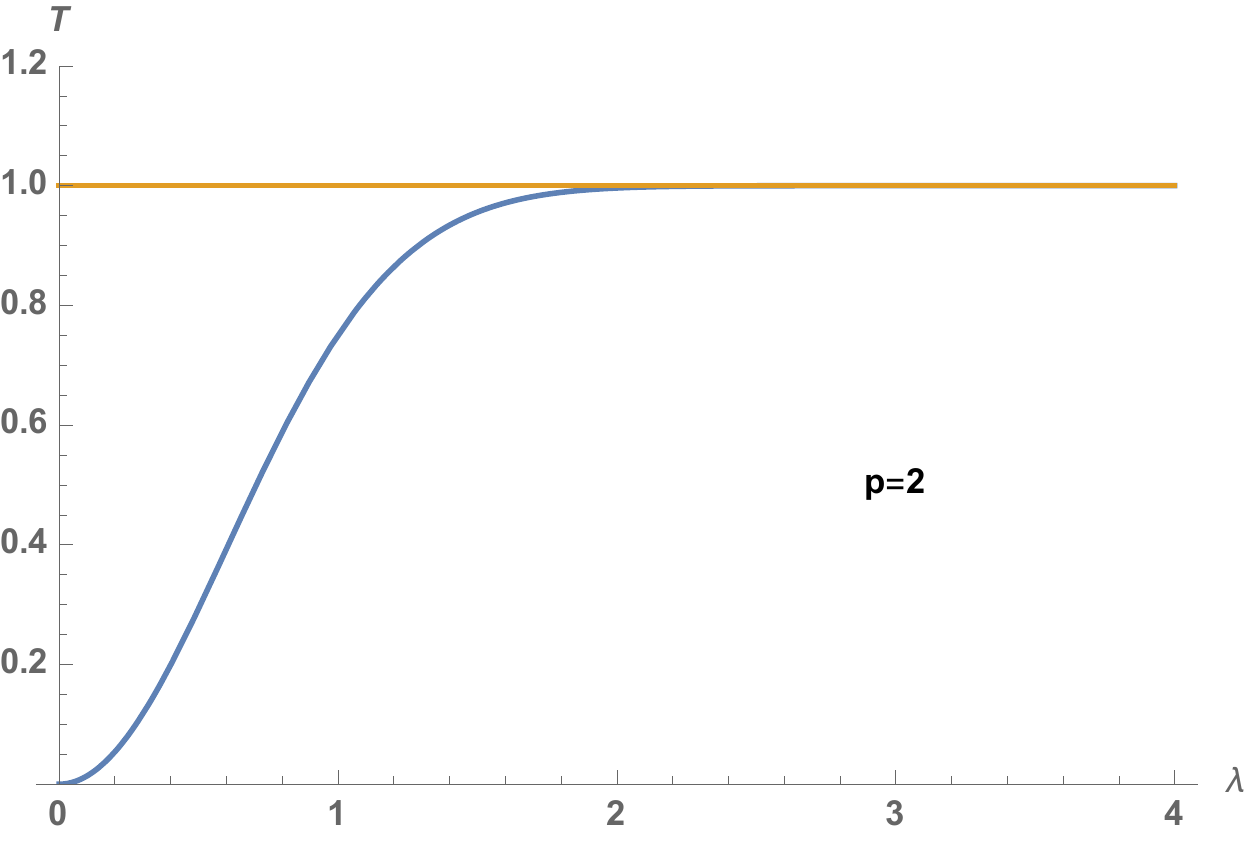}}
\end{center}
\vspace{-.7cm}
\caption{\label{fig:coefWilsontach}Tachyon as a function of the Wilson line coupling constant, left in Siegel gauge, center for the TT/KOS solution, and right from the shifted 3-point function.}
\begin{center}
\resizebox{1.7in}{1.1in}{\includegraphics{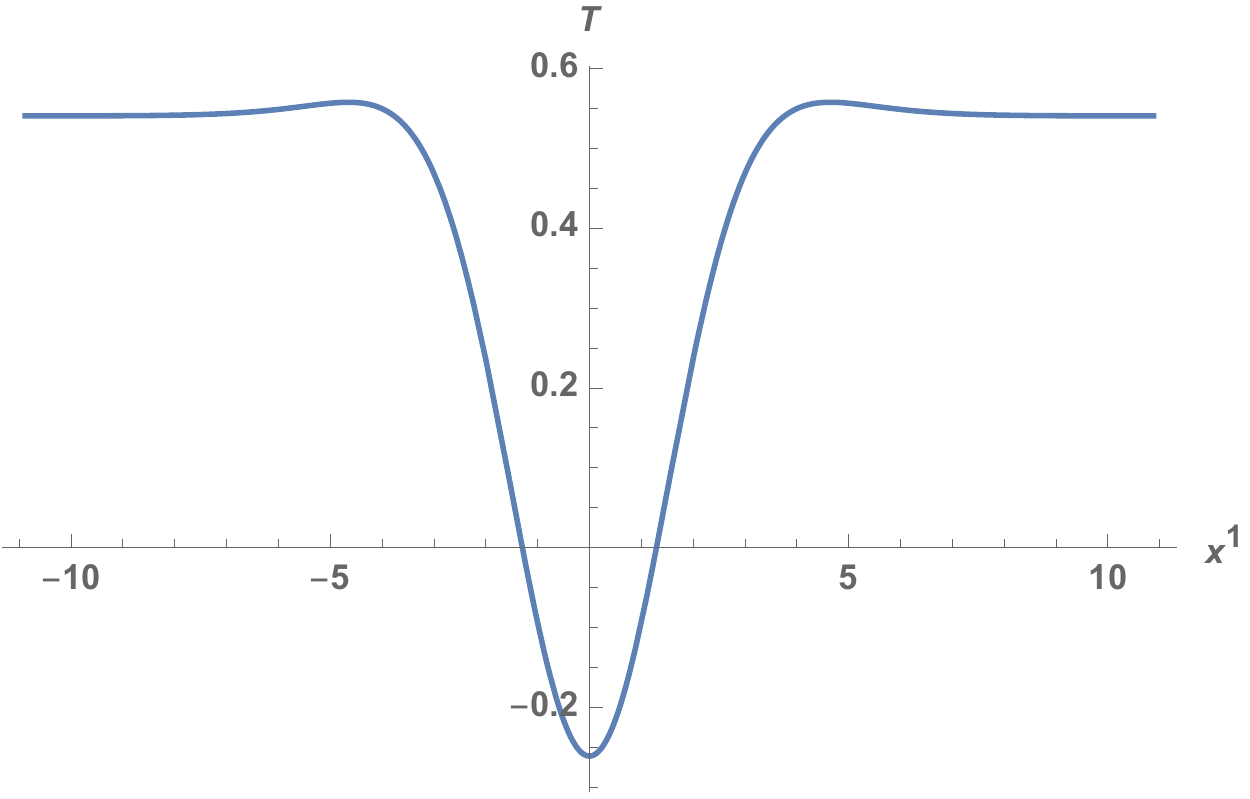}}\ \ \ \ \
\resizebox{1.7in}{1.1in}{\includegraphics{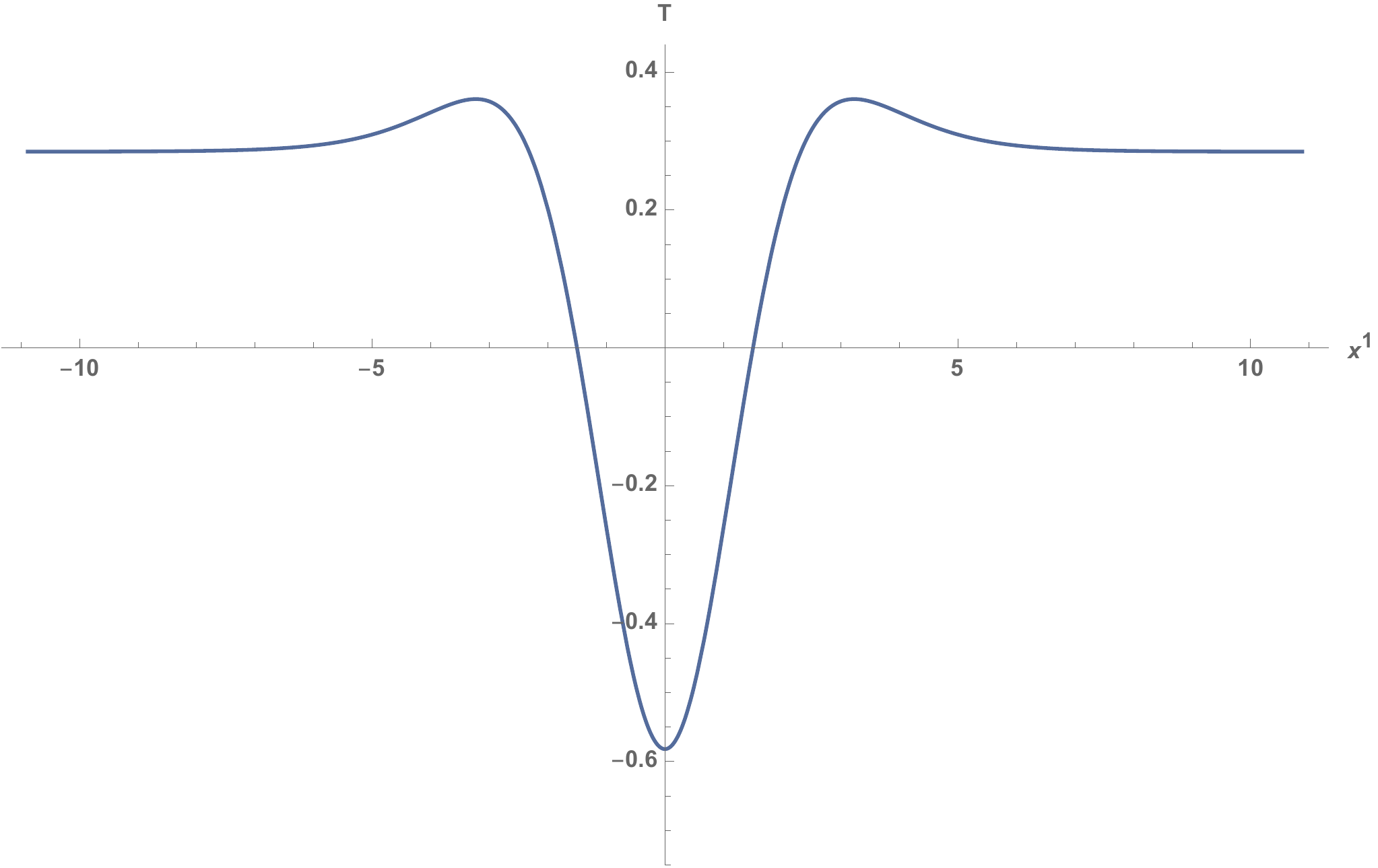}}\ \ \ \ \
\resizebox{1.7in}{1.1in}{\includegraphics{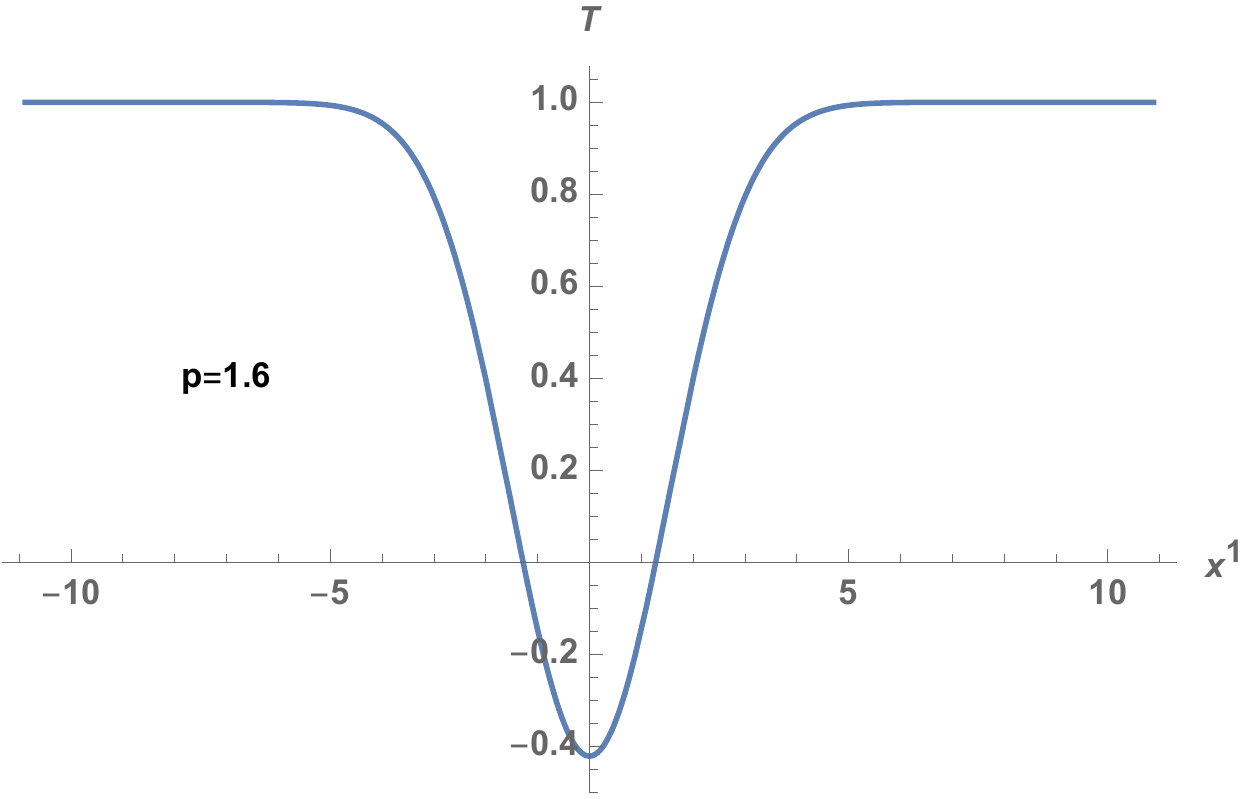}}
\end{center}
\vspace{-.7cm}
\caption{\label{fig:coeflumps}Tachyon lump, left in Siegel gauge, center from \cite{KOSsing}, and right from the shifted 3-point function.}
\begin{center}
\resizebox{2in}{1.1in}{\includegraphics{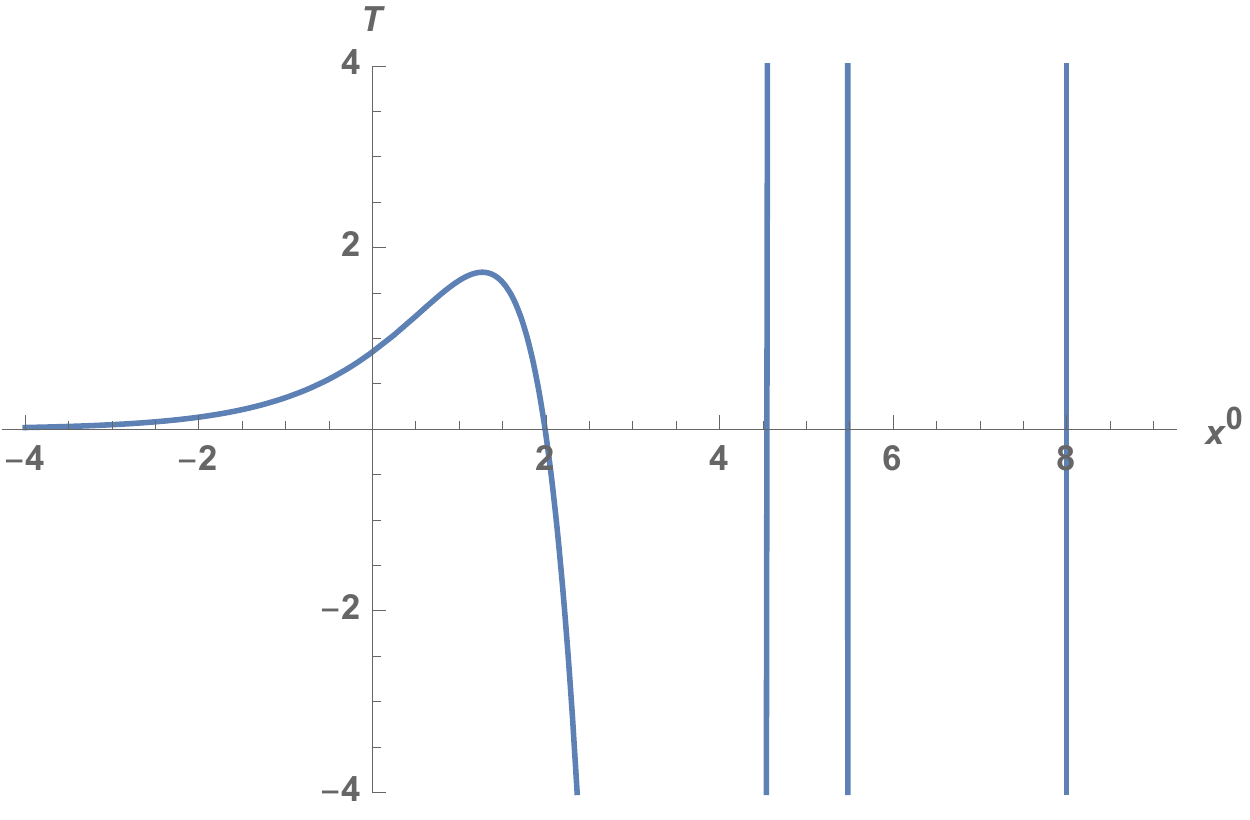}}\ \ \ \ 
\resizebox{2in}{1.1in}{\includegraphics{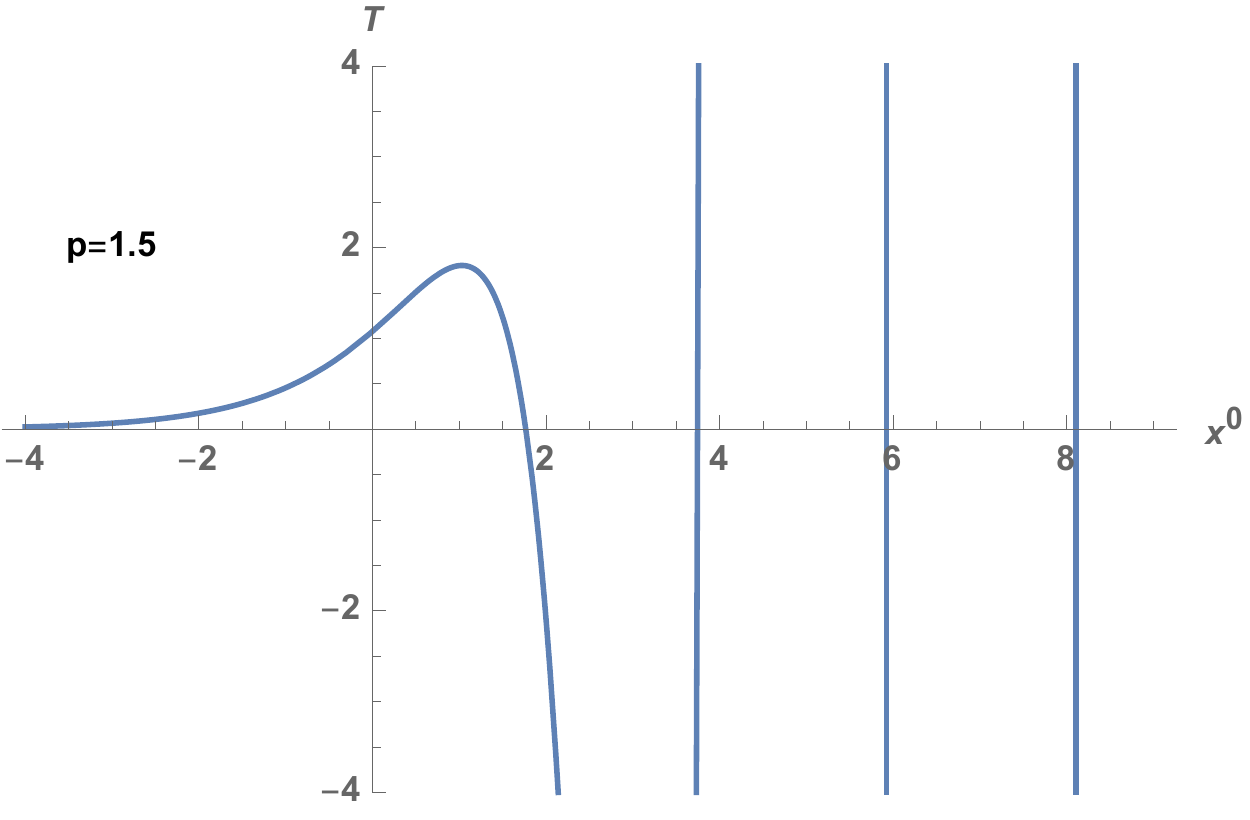}}
\end{center}
\vspace{-.7cm}
\caption{\label{fig:coefrolling}Evolution of the tachyon for the exponential rolling tachyon deformation, left as derived from the level 0 truncated action, and right from the shifted 3-point function.}
\begin{center}
\resizebox{2in}{1.1in}{\includegraphics{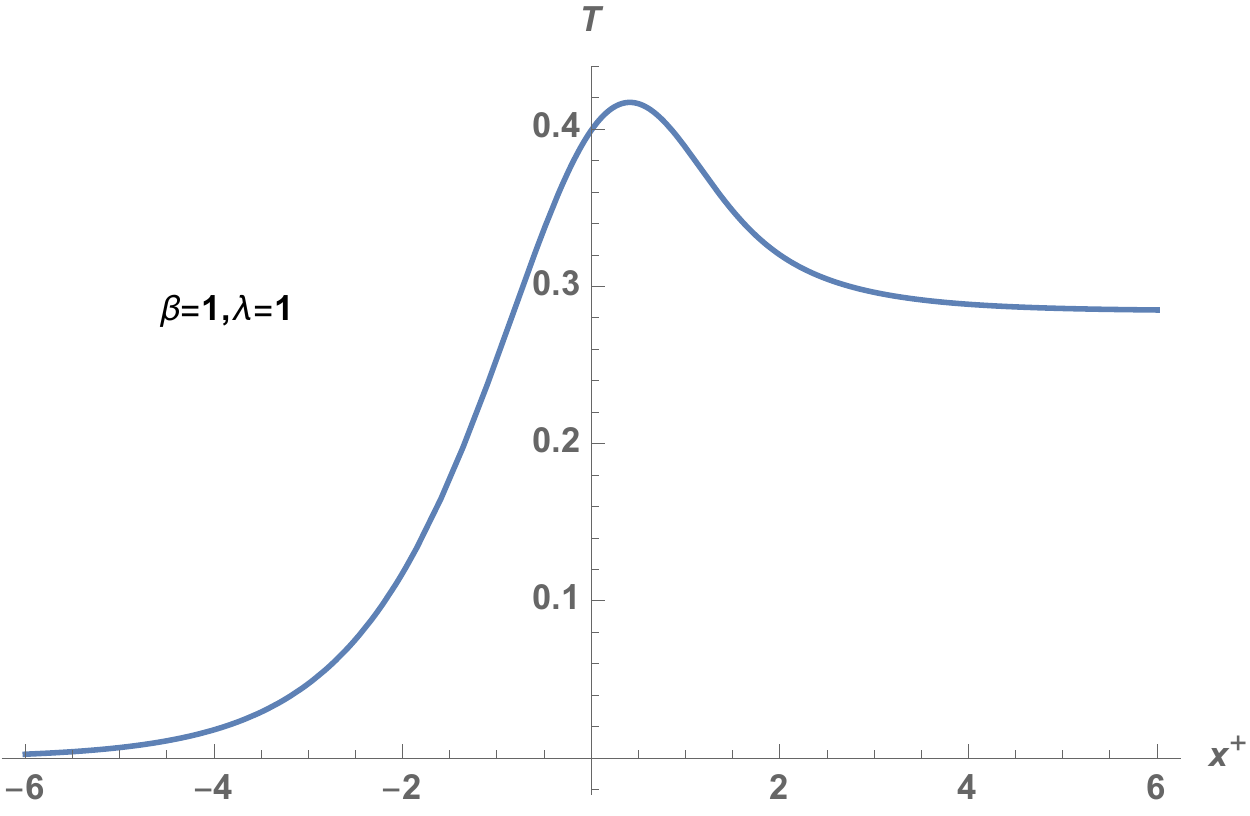}}\ \ \ \ 
\resizebox{1.7in}{1.1in}{\includegraphics{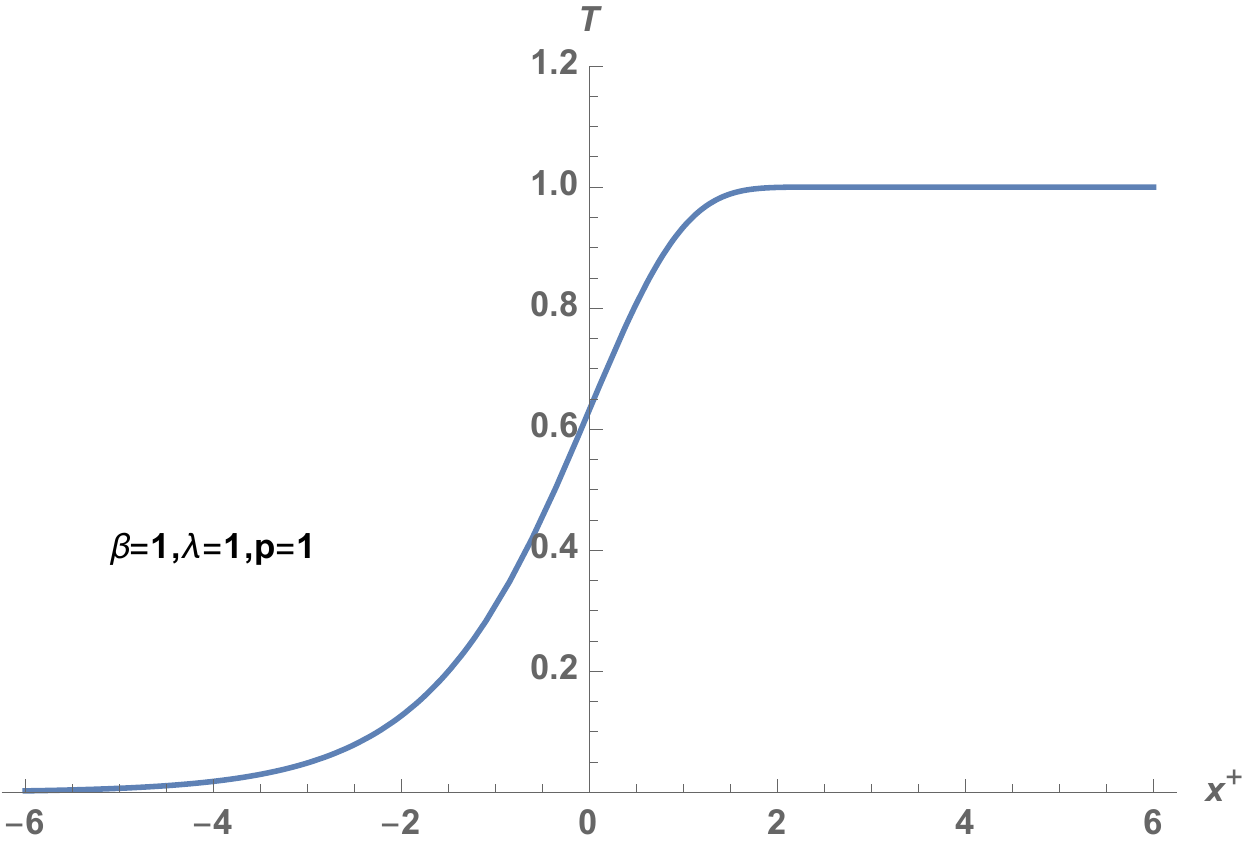}}
\end{center}
\vspace{-.7cm}
\caption{\label{fig:coeflight} Evolution of the tachyon for the lightlike rolling tachyon deformation, left as given in \cite{rollingvac}, and right from the shifted 3-point function.}
\end{figure}

\begin{description}
\item{{\it Wilson Line:}} In figure \ref{fig:coefWilson} we show the expectation value of the (ghost number 1) gauge field for the Wilson line solution in Siegel gauge \cite{MaccaferriMatjej,Matjej} and for the TT/KOS solution \cite{Maccaferri_marg,Maccaferri_large}, as a function of the coupling constant $\lambda$ which turns on the Wilson line deformation on the worldsheet. The rightmost figure shows the corresponding plot generated by the shifted 3-point function in \eq{FSCP}, with boundary condition changing operators taken from \eq{Wilsonbcc}. The Siegel gauge plot ends at $\lambda\approx.6$, since in the approach of \cite{MaccaferriMatjej} the Siegel gauge solution cannot be found past the point where the branch of the tachyon effective potential describing the Wilson line deformation terminates. Also shown in figure \ref{fig:coefWilsontach} is the tachyon expectation value as a function of $\lambda$. It is striking that the shifted three point function already captures the basic form of the dependence on $\lambda$, including the highly nontrivial predictions that $\lambda$  is not a single valued function of the gauge field expectation value, and that the solution should approach the tachyon vacuum for large $\lambda$.
\item{{\it Tachyon Lumps:}} Figure \ref{fig:coeflumps} shows the tachyon lump profile in Siegel gauge \cite{MZlump} and for the solution of \cite{KOSsing}. The rightmost figure shows the tachyon lump profile predicted by the shifted 3-point function, where the boundary condition changing operators are taken as Neumann-Dirichlet twist fields of weight $1/16$; this is given exactly by a theta function: 
\begin{equation}T^{\text{3-point}}(x^1) = 1-\frac{1}{R(2p)^{1/8}}\vartheta_{00}\left(\frac{x}{2\pi R},\frac{i}{\pi R^2}\ln(2p)\right).\end{equation}
The lump profiles are similar in all three cases. 
\item{{\it Exponential Rolling Tachyon:}} Consider the exponential rolling tachyon deformation generated by the marginal operator $e^{X^0}$. A first approximation to this background is given by truncating the open string field theory action to level 0:
\begin{equation}S_T = -\int d^{26} x\left[\frac{1}{2}\partial^\mu T\partial_\mu T-\frac{1}{2}T^2 +\frac{1}{3}\Big(K^{-\Box} T\Big)^3\right],\ \ \ \ K =\frac{4}{3\sqrt{3}},\end{equation}
and looking for a solution as a power series in $e^{x^0}$. The resulting evolution of the tachyon field is shown in figure \ref{fig:coefrolling}. The tachyon rolls off the unstable maximum and subsequently engages in uncontrolled oscillation, similar to the cosh deformation. This behavior has been confirmed convincingly by exact solutions \cite{Schnabl_marg,KORZ,KOS}. The rightmost figure in \ref{fig:coefrolling} shows the evolution of the tachyon as predicted by the shifted 3-point function. The qualitative behavior is the same. 
\item{{\it Light-like Rolling Tachyon:}} Consider the light-like rolling tachyon deformation generated by the marginal operator $e^{X^+}$ in a linear dilaton background \cite{Hellerman}. Figure \ref{fig:coeflight} shows the evolution of the tachyon field in lightcone time derived in \cite{rollingvac}, next to the evolution derived from the shifted 3-point function. The shifted 3-point function correctly predicts that the solution will approach the tachyon vacuum in the infinite lightcone future.
\end{description}
Note that many of the above solutions have no obvious relation to boundary condition changing operators. Still the shifted three point function captures the behavior fairly well. 

It is worth stepping back a moment to think about what the Fock space coefficient principle really means. It says that the coefficients of a solution reflect, in a broad sense, how a target D-brane appears through a ``window" of stretched strings extending from a reference D-brane. From the point of view of the intertwining solution this seems fairly natural, but more generally it is quite surprising. And it begins to point to an explanation of why string field theory coefficients sometimes exhibit counterintuitive behavior.

Of course there will be differences between the profiles of any specific solution and those generated by the shifted 3-point function. Some of these differences may reflect the ``approximation" implicit in the Fock space coefficient principle, and have no special meaning. Other differences can be more significant. One example concerns how solutions behave in the vicinity of the tachyon vacuum. Typically, when the tachyon field approaches the tachyon vacuum, the leading correction around the tachyon vacuum is positive. In fact, we are not aware of counterexamples. This reflects a distinctive tendency for the tachyon profiles to overshoot the tachyon vacuum and fall down from above. This implies, for example, that the tachyon coefficient will not be a global coordinatee on the moduli space of the Wilson line deformation, which may be part of the reason why the full moduli space is not seen in Siegel gauge level truncation \cite{MaccaferriMatjej}. By contrast, for tachyon profiles generated from the 3-point function, the leading correction around the tachyon vacuum tends to be {\it negative}. In fact, this behavior also appears generic in ghost number 0 toy models and field theory models. For example, the lump in $\phi^3$ theory is given by a hyperbolic secant profile which does not overshoot the nonperturbative vacuum; further, the tachyon expectation value in the field theory on the lump gives a global coordinate on the moduli space of solutions representing translations of the lump~\cite{MaccaferriMatjej}. A string field theory solution may also decay to the tachyon vacuum at a different rate than the shifted 3-point function. This is most clearly seen in the light-like rolling tachyon solution, which approaches the tachyon vacuum as $e^{-x^+}$, while the 3-point function approaches as $e^{-e^{x^+}}$. It is now understood that the exponential decay modes of the solution near the tachyon vacuum represent pure gauge fluctuations whose purpose is to ``hide" the superexponential decay $e^{-e^{x^+}}$ as a nonperturbative effect \cite{rollingvac}. All of this is to demonstrate that the coefficients of a true solution may exhibit physically interesting features which cannot be reduced to the computation of a $3$-point function. 

\begin{figure}
\begin{center}
\resizebox{3in}{2in}{\includegraphics{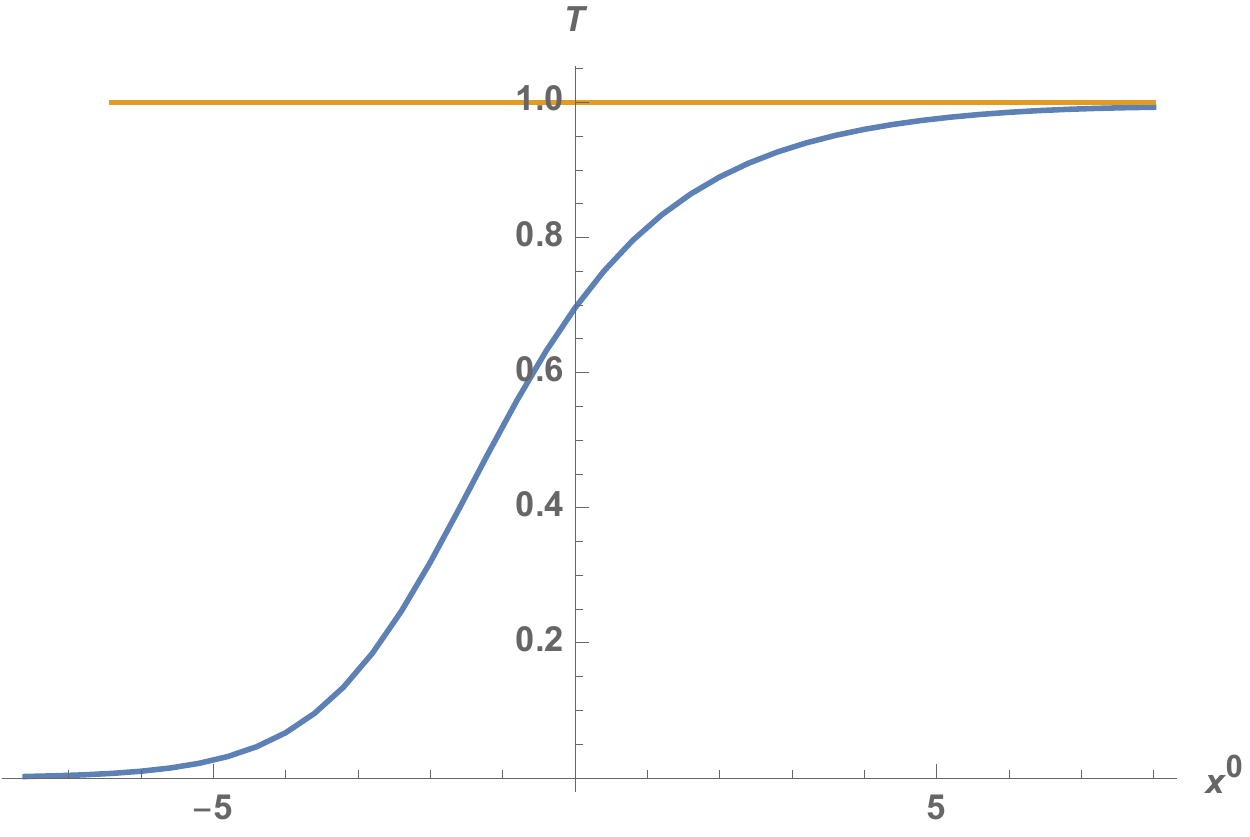} }
\end{center}
\vspace{-.7cm}
\caption{\label{fig:exp_rolling} Evolution of the tachyon for the exponential rolling tachyon deformation, as derived from the shifted 3-point function at the critical parameter $p=1/2$. In this case the sum over exponential harmonics has finite radius of convergence, and the full profile is obtained after Pad{\' e} resummation. Shown here is the Pad{\' e} approximant $P_{64}^{64}(e^{x^0})$ which reaches the tachyon vacuum in the infinite future within an accuracy of $0.5\%$.}
\end{figure}

We have mentioned that the distance between boundary condition changing operators in the 3-point function should not be too small. For small enough separation, sometimes the 3-point function will diverge with increasing conformal weights, and the behavior is markedly different from a string field theory solution. In several examples the smallest allowable separation is $p=1/2$, where interesting simplifications occur:
\begin{itemize}
\item For the Wilson line deformation, the gauge field is directly proportional to the marginal coupling constant and the tachyon field consistently vanishes, exactly as one would expect of the deformation in the low energy Yang-Mills description. 
\item For tachyon lumps, the profile is a delta function with support at the location of the Dirichlet boundary condition defining the lower dimensional D-brane. Recalling that the lump profile is described by a theta function for generic $p$, this points to an interesting relation between the parameter $p$ and time evolution in the heat equation. This gives a way to understand why lump profiles are so featureless for small radii; starting with a sufficiently dense array of delta functions, the temperature distribution reaches thermal equilibrium by the time it is ``sampled" by the string field theory solution. Related discussion of the connection between string field theory solutions and the heat equation appears in \cite{Calcagni}.
\item For the cosh rolling tachyon deformation at $\lambda =1/2$, the tachyon profile is constant in time and sits at the tachyon vacuum. This confirms that the background is at least very closely related to the tachyon vacuum.
\item For the exponential rolling tachyon deformation, the tachyon falls off the top of the potential and smoothly approaches the tachyon vacuum in the infinite future. See figure \ref{fig:exp_rolling}. This picture of the tachyon evolution was proposed in \cite{EllwoodRolling} based on formal manipulations of the solution in Schnabl gauge, but later computations confirmed the oscillatory behavior beyond doubt. The present discussion suggests that there may be more to this story. 
\end{itemize}
It is worth noting that the seemingly more reasonable behavior of the rolling tachyon profiles does not emerge in a continuous fashion as $p$ is decreased to $1/2$; the oscillations actually grow in amplitude and frequency in this limit. At $p=1/2$ the oscillations formally have infinite frequency, but can be discarded after resummation of the expansion into harmonics. It would be interesting if the behavior of tachyon and gauge field profiles at $p=1/2$ could be reproduced through an effective action.

\section{Conclusion}

In this paper we have given an analytic solution of Witten's open bosonic string field theory which, in different variations, can describe any target D-brane system from the point of view of a reference D-brane which shares the same closed string background. The solution also provides an explicit map between the field variables of different D-brane systems, demonstrating nonperturbative background independence of classical open bosonic string field theory. 

It is interesting to comment on how the flag state solution connects to old (but incomplete) ideas about how D-brane systems emerge in open string field theory. One example is open string field theory in Minkowski space with large $B$-field along a pair of coordinates. In this case, the open string star algebra factorizes into a tensor product of commuting subalgebras \cite{WittenNon}
\begin{equation}\mathcal{A}=\mathcal{A}_0\otimes \mathcal{A}_1,\end{equation}
where $\mathcal{A}_0$ is the subalgebra of states with vanishing momentum along the coordinates of the $B$-field and $\mathcal{A}_1$ consists of a Moyal plane in the directions of the $B$-field. In this setup, it was observed in~\cite{WittenNon} that codimension two D-branes in the noncommutative directions can be described with the solution
\begin{equation}\Psi_* = \Psi_\tv\otimes \mathbb{I}-\Psi_\tv\otimes\rho,\label{eq:GMS}\end{equation}
where $\rho$ is a finite rank projector in $\mathcal{A}_1$. For example $\rho$ could take the form 
\begin{equation}\rho  = |0\rangle\langle 0| + ...+ |n\rangle\langle n|,\end{equation}
which represents the formation of $n+1$ codimension 2 D-branes. We can express $\rho$ as 
\begin{equation}\rho = \Sigma\Sigmabar\end{equation}
with 
\begin{equation}\Sigma = \Big(|0\rangle\ \ ...\ \ |n\rangle\Big),\ \ \ \ \ \ \ \Sigmabar = \left(\begin{matrix}\langle 0|\\ \vdots \\ \langle n|\end{matrix}\right).\end{equation}
Therefore \eq{GMS} is an intertwining solution; apparently, the effect of the large $B$-field is to replace the intertwining fields with the projector $\rho$. It would be interesting to see how this happens more explicitly. Another connection is to the vacuum string field theory proposal \cite{VSFT}. Of course, the intertwining solution is naturally understood as a solution around the tachyon vacuum. More nontrivially, the flag state solution expanded around the tachyon vacuum is close to being a factorized functional between the left and right halves of the string, at least for the portions of the open string near the endpoints. In this sense, the flag state solution is comparable to a rank one star algebra projector, exactly the kind of state conventionally used to build classical solutions of vacuum string field theory \cite{VSFTsol}. It has been proposed that vacuum string field theory should emerge after an infinite reparameterization of Witten's theory towards the midpoint \cite{Gaiotto}. An example of such a midpoint preserving reparameterization is generated by $\mathcal{L}^-$, which adjusts the height of the flags. Indeed, as the flag height is taken to infinity, the flag state solution approaches the sliver state---the most famous rank 1 star algebra projector---with singular ghost insertions at the midpoint. From this point of view, the present work can be viewed as defining a consistent regularization of vacuum string field theory. 

The next question is how far the flag state/intertwining solution generalizes to open superstring field theory or closed string field theory. Up to a certain point, the generalization to open superstrings should be straightforward in the Wess-Zumino-Witten-like formulation \cite{Berkovits}.\footnote{The application of analytic methods to superstring field theory in the small Hilbert space  \cite{WittenSS} does not look particularly promising, though perhaps perturbative analytic solutions could be found. Using the field redefinition between the small Hilbert space and Wess-Zumino-Witten-like theories \cite{OkWB,WBsmall,WBlarge}, formally it is possible to map analytic solutions of the former into the latter; however, the field redefinition is {\it a priori} only defined perturbatively, and for nonperturbative solutions it is not clear it will produce a meaningful state. On the other hand, we expect that a tachyon vacuum should appear in level truncation of the small Hilbert space theory on a non-BPS D-brane, in a similar manner as it does in the Wess-Zumino-Witten-like theory \cite{BSZ}. So the difficulty in characterizing nonperturbative solutions {\it analytically} may not imply that such solutions do not exist.} The added complication is that tachyon vacuum solutions do not take the same form for all D-brane systems in the superstring. An analytic solution for the tachyon vacuum on wrong dimension D-branes and coincident brane-anti-brane pairs was given in \cite{supervac}. However, for BPS D-brane systems it is questionable whether tachyon vacuum solutions even exist. It is somewhat ironic that finding BPS solutions poses greater difficulty than finding solutions without any supersymmetry. The only known analytic solutions on supersymmetric D-branes represent perturbative marginal deformations. The description of BPS backgrounds is probably the most significant question in the study of analytic solutions at present. 

Regarding closed string field theory we cannot expect the same kind of fully explicit results. But it is conceivable that some kind of homotopy-intertwining solution exists that could describe nontrivial closed string backgrounds. For this to make sense we would first need some closed string version of the tachyon vacuum. A possible nonperturbative vacuum state which might have this interpretation has been identified in level truncation including up to the quartic vertex \cite{Yang}, but further calculations including higher levels \cite{MoellerYang}, and some effects of the quintic vertex~\cite{Moeller1,Moeller2}, have not done as much as might have been hoped to confirm the existence of this vacuum. The closed string analogue of the intertwining solution would also need a 1-dimensional defect operator to form an interface between reference and target bulk CFTs \cite{Bachas}. It seems clear that the defect operator cannot be purely reflective \cite{Watts}, since it is necessary to transmit physical information between the reference and target backgrounds in order to relate them by field redefinition. Such considerations might limit the range of backgrounds accessible from a given bulk CFT, but this is probably desirable. A generic matter CFT of central charge 26 can have pathological physical properties---for example spacetimes with several timelike directions---and we wouldn't necessarily want or expect all of this to be part of the same physical string theory. This line of inquiry risks going out on a limb, but in absence of other ways to gain insight into the nonperturbative vacuum structure of string theory, perhaps it would at least be interesting to formulate conjectures. 

Returning to the Witten theory, the results of this paper should not imply that the search for classical solutions is over. To obtain our results we had to make a number of compromises. The nontrivial geometry of the solution makes some calculations laborious; the open string degeneration in star products is somewhat singular; the perturbative vacuum is represented by a nonperturbative open string state. Over the years there have been a number of other ideas about how D-branes can appear as classical solutions:
\begin{itemize}
\item One can imagine constructing nontrivial solutions through a formal gauge transformation of the perturbative vacuum which, in a sense, is only defined on a ``covering space." This generalizes the observation that a Wilson line on a circle can be trivialized by a gauge transformation defined on $\mathbb{R}^1$. This idea has been used in the construction of analytic solutions for singular marginal deformations \cite{FK,KO}. 
\item A more exotic approach, which seems to be fairly specific to string field theory, is to construct nontrivial solutions using formal gauge transformations which are not invertible \cite{Ellwood,EMsingular}. This idea is especially important in understanding the structure of Schnabl's solution \cite{Schnabl,Okawa}. It is now understood that all classical solutions of open bosonic string field theory can be derived in this way \cite{phantom}, but it has not yet been successfully used to generate new solutions. 
\item Another interesting idea is to derive string field theory solutions using worldsheet renormalization group flows \cite{Ellwood,BMT}. This has not yet been consistently implemented---except in the trivial case of the tachyon vacuum---but has come tantalizingly close to working \cite{BMTanomaly}, and is  still viable.
\item Within the intertwining framework, the original solution of \cite{KOSsing} is in many ways more accessible than the flag state solution, at least in situations where it is applicable. Other realizations of the intertwining solution would be of interest.
\end{itemize}
What is perhaps the most significant compromise of the present approach is that it relies on essentially complete understanding of the BCFT of the target background. Therefore, the solution cannot be used to construct backgrounds whose worldsheet description is not already known. In principle, one can imagine a method for constructing manifestly nontrivial solutions whose physical interpretation is not known from the outset; computation of observables may then, in some cases, lead to discovery of new BCFTs. This is in fact precisely what is achieved by solving the equations of motion in  level truncation, which has proven capable of finding new BCFTs \cite{Matjej}. However, the level truncation approach is completely numerical. One might like an analytic approach which can give more substantive conceptual insight into the solution and the new BCFT. This would be great.

However, what makes a string field theory solution unique is not that it describes a BCFT; generally, solutions encode BCFT data in an obscure and inefficient way. Rather, the significance of a string field theory solution is that it represents a concrete physical relation between {\it a~pair} of BCFTs, thus establishing that they are a part of the same physical framework. This is the scope of the present contribution. The hope is that this achievement can give a new language for understanding the physics of D-branes, one which may address questions where the worldsheet point of view may not seem fully adequate. Examples include physics around the tachyon vacuum, the origin of flux quantization and topological charges, the physical significance of the massive fluctuations of the string, and the nature of locality, causality, and time evolution in string theory. We hope that the present work leads to progress in these directions. 

\subsubsection*{Acknowledgements}

We would like to thank  M. Kudrna and J. Vo{\v s}mera for helpful discussion of boundary condition changing operators for the cosine deformation, and especially M. Kudrna for providing numerical data for some of the plots given in section \ref{subsec:principle}. We also thank M. Schnabl and A. Sen for comments on an earlier draft of the paper. T.E. thanks INFN and the Physics Department of Turin University and C.M. thanks the Institute of Physics of the Czech Academy of Science for kind hospitality during part of this research.  We thank the Galileo Galilei Institute for Theoretical Physics and INFN for hospitality and partial support during the workshop "String Theory from a worldsheet perspective" where part of this work has been done. The work of C.M. is partially supported by the MIUR PRIN Contract 2015 MP2CX4 ``Non-perturbative Aspects Of Gauge Theories And Strings''.  The work of T.E. is supported by ERDF and M\v{S}MT (Project CoGraDS -CZ.02.1.01/0.0/0.0/15\_ 003/0000437) and the GA{\v C}R project 18-07776S and RVO: 67985840.

\begin{appendix}

\section{Intertwining Form for Schnabl Gauge Marginal Deformations}
\label{app:Sch}

ln this appendix we express the solution for marginal deformations in Schnabl gauge \cite{Schnabl_marg,KORZ} in intertwining form. For technical reasons it is simpler to consider the Schnabl gauge solution in a non-real form\footnote{Technically, the non-real form does not satisfy the Schnabl gauge condition $\mathcal{B}_0\Psi = \frac{1}{2}\mathcal{B}^-\Psi + \frac{1}{2}B\Psi+\frac{1}{2}\Psi B = 0$. It satisfies the condition $\frac{1}{2}\mathcal{B}^-\Psi + \Psi B = 0$. }
\begin{equation}\Psi = cB\frac{1}{1+V\frac{1-\Omega}{K}}cV\Omega.\label{eq:Schnonreal}\end{equation}
This is related to the conventional Schnabl gauge solution through similarity transformation with $\sqrt{\Omega}$. We observe that \eq{Schnonreal} can be obtained from a non-real form of the Kiermaier, Okawa, Soler solution~\cite{KOS},
\begin{equation}\Psi_\mathrm{KOS} = -cB\d \sigma \frac{1}{1+K}\sigmabar(1+K)c\frac{1}{1+K},\end{equation}
through the transformation
\begin{equation}
\Psi = \frac{1}{1+\Psi_\mathrm{KOS}B(\frac{1-\Omega}{K}-\Omega)}\Psi_\mathrm{KOS}(1+K)\Omega.
\end{equation}
This is a special instance of the Zeze map \cite{Zeze}, and can be understood as a finite gauge transformation,
\begin{equation}\Psi = U^{-1}(Q+\Psi_\mathrm{KOS})U,\end{equation}
defined by the gauge parameter
\begin{equation}
U = 1+\Psi_\mathrm{KOS}B\left(\frac{1-\Omega}{K}-\Omega\right).
\end{equation}
Since the Kiermaier, Okawa, Soler solution can be expressed in intertwining form \cite{KOSsing}, this implies that the Schnabl gauge solution can be as well: 
\begin{equation}\Psi = \Psi_\tv^{(0)} - \Sigma^{(0*)}\Psi_\tv^{(*)}\Sigmabar^{(*0)}.\end{equation}
We find explicitly
\begin{eqnarray}
\Psi_\tv^{(0)} \lineup = cB\frac{1}{1+V\frac{1-\Omega}{K}}\left[(1+K)c-(1+K+V)c\frac{1}{1+K+V}\right]\left(\frac{1-\Omega}{K}-\Omega\right)\nonumber\\ \lineup\ \ \ \ \ \ \ \ \ \ \ \ \ \ \ \ \ \ \ \ \ \ \ \ \ \ \ \ \ \ \ \ \ \ \ \ \ \ \ \ \ \ \ \ \ \ \ \ \ \ \ \ \ \ \ \ \ \ \ \ \ \ \ \ \ \ +cB\frac{1}{1+V\frac{1-\Omega}{K}}(1+K+V) c\Omega,\ \ \ \ \ \ \\
\Psi_\tv^{(*)} \lineup = cB(1+K)c\frac{1}{1+K}, \\
\Sigma^{(0*)}\lineup = \sigma +cB \d \sigma\frac{1}{1+K}-cB\frac{1}{1+V\frac{1-\Omega}{K}}V\left(\frac{1-\Omega}{K}-\Omega\right)\sigma \frac{1}{1+K},\\
\Sigmabar^{(*0)}\lineup = \sigmabar + cB\d \sigmabar \frac{1-\Omega}{K}.
\end{eqnarray}
There is significant freedom in the choice of tachyon vacuum solutions and intertwining fields, so this is not the only, or necessarily the best, way to write \eq{Schnonreal} in intertwining form. In any case, it is apparent that this is a significantly more complicated way to write the solution than \eq{Schnonreal}. Note that the tachyon vacuum solutions of the reference and target D-branes are different. For the target D-brane, it is a non-real form of the simple tachyon vacuum \cite{simple}, and for the reference D-brane it is an unfamiliar solution which does not live in the universal sector. Another thing to observe is that the rightmost matter insertion in \eq{Schnonreal} is multiplied by a wedge state $\Omega$ with strictly positive width. However, when written in intertwining form, some terms in the solution do not possess this factor. This implies that star multiplication from the right can produce OPE divergences in individual terms which are absent when multiplying the full solution. This phenomenon is ultimately a consequence of the identity $\Sigmabar\Sigma=1$. Since the identity string field has vanishing width, and the width of wedge states is additive under star multiplication, the intertwining fields cannot be made from wedge states of strictly positive width. For this reason we expect that writing a solution in intertwining form generically requires tachyon vacuum solutions and intertwining fields which are less regular than the solution itself.

\section{Repository of Formulas for Flag-anti-flag Surface}
\label{app:conformal}

In this appendix we collect several formulas related to conformal transformation between the flag-anti-flag surface and the upper half plane. The relevant conformal transformations are summarized in figure \ref{fig:conformal}.

\vspace{10pt}

\begin{itemize}

\item {\bf UHP to sliver frame:} The height of the flags is $\ell$. The target wedge state has width $\beta$, and the reference wedge state has width $\alpha$. The origin of the sliver coordinate $z$ on the flag-anti-flag surface sits on the open string boundary midway between the vertical edges of the reference wedge surface. The sliver coordinate is related to the upper half plane coordinate $u$ through
\begin{equation}
z=\mathcal{F}(u) = \frac{2\ell}{\pi}\left[\frac{p(1+s^2)}{s^2-p^2}\tan^{-1}u+\tanh^{-1}\frac{u}{p}\right],\label{eq:app:z}
\end{equation}
where $\pm s,\pm p$ are respectively the pre-images of the two slits and two punctures on the flag-anti-flag surface, and $s>p>0$.
For conformal transformation of primary operators it is useful to observe that 
\begin{equation}
\frac{dz}{du} = \mathcal{F}'(u) = \frac{2\ell}{\pi}\frac{p(1+p^2)}{s^2-p^2}\frac{s^2-u^2}{p^2-u^2}\frac{1}{1+u^2}.\label{eq:app:zt}
\end{equation}
We also introduce a dual sliver coordinate $\widetilde{z}$ on the flag-anti-flag surface, where the origin sits on the open string boundary midway between the vertical edges of the target wedge surface. This is related to a dual upper half plane coordinate $\widetilde{u}$ through
\begin{equation}
\widetilde{z}=\widetilde{\mathcal{F}}(\widetilde{u}) = \frac{2\ell}{\pi}\left(\frac{\widetilde{p}(1+\widetilde{s}^2)}{\widetilde{p}^2-\widetilde{s}^2}\tan^{-1} \widetilde{u} -\tanh^{-1}\frac{\widetilde{u}}{\widetilde{p}}\right),
\end{equation}
where $\pm \widetilde{p},\pm \widetilde{s}$ are respectively the pre-images of the two punctures and two slits on the flag-anti-flag surface, and $\widetilde{p}>\widetilde{s}>0$. We have
\begin{equation}\frac{d\widetilde{z}}{d\widetilde{u}} = \widetilde{\mathcal{F}}'(\widetilde{u})=\frac{2\ell}{\pi}\frac{\widetilde{p}(1+\widetilde{s}^2)}{\widetilde{p}^2-\widetilde{s}^2}\frac{\widetilde{s}^2-\widetilde{u}^2}{\widetilde{p}^2-\widetilde{u}^2}\frac{1}{1+\widetilde{u}^2}.\end{equation}
The coordinates and dual coordinates are related by
\begin{equation}\widetilde{u} = -\frac{1}{u},\ \ \ \ \ \widetilde{z} = z - \frac{\alpha+\beta}{2}\mathrm{sign}(\mathrm{Re}(z))-i\ell.\end{equation}

\item {\bf UHP moduli to wedge parameters:} 
\begin{eqnarray}
\alpha \lineup = \frac{4\ell}{\pi}\left[\frac{p(1+s^2)}{s^2-p^2}\tan^{-1}s+\tanh^{-1}\frac{p}{s}\right],\label{eq:app:alpha}\\
\beta \lineup = \frac{4\ell}{\pi}\left[\frac{p(1+s^2)}{s^2-p^2}\tan^{-1}\frac{1}{s}-\tanh^{-1}\frac{p}{s}\right],\label{eq:app:beta}\\
\alpha+\beta\lineup = \frac{2\ell  p(1+s^2)}{s^2-p^2}.
\end{eqnarray}

\item {\bf UHP to puncture coordinates on the flags:} The local coordinate around the puncture of the flag state is $\rho$; the local coordinate around the puncture of the anti-flag state is $\widetilde{\rho}$.
\begin{eqnarray}
\rho = \mathcal{R}(u)\lineup = -\frac{(s+p)(u+p)}{(s-p)(u-p)}\exp\left[\frac{2p(1+s^2)}{s^2-p^2}\Big(\tan^{-1}s+\tan^{-1}u\Big)\right],\label{eq:R}\\
\widetilde{\rho}=\widetilde{\mathcal{R}}(u)\lineup = \frac{(s+p)(u-p)}{(s-p)(u+p)}\exp\left[\frac{2p(1+s^2)}{s^2-p^2}\Big(\tan^{-1}s-\tan^{-1}u\Big)\right].
\end{eqnarray}
Note that $\mathcal{R}(-p)=\widetilde{\mathcal{R}}(p)=0$ since $u=\pm p$ maps to the origin of the respective local coordinates. For conformal transformation of primary operators at the origin of the respective local coordinates it is useful to define 
\begin{equation}
\rho'\equiv \widetilde{\mathcal{R}}'(p)=\mathcal{R}'(-p)=\frac{s+p}{2p(s-p)}\exp\left[\frac{2p(1+s^2)}{s^2-p^2}\Big(\tan^{-1}s-\tan^{-1}p\Big)\right].\label{eq:rhop}
\end{equation}

\item{\bf $b$-ghost residue:} Commuting $B$ through flag states is described with the formulas:
\begin{eqnarray}
\Big[\flag{\mathcal{O}},B\Big] \lineup =  \flag{\mathcal{O}}b- (-1)^\mathcal{O}\frac{\pi}{\ell}\flag{b_0\mathcal{O}},  \\
\Big[B,\Aflag{\widetilde{\mathcal{O}}}\Big] \lineup =  b\Aflag{\widetilde{\mathcal{O}}} +\frac{\pi}{\ell}\Aflag{b_0\widetilde{\mathcal{O}}} .
\end{eqnarray}
The states $b\Aflag{\widetilde{\mathcal{O}}}$ and $\flag{\mathcal{O}}b$ may be characterized through correlators on the upper half plane~as
\begin{eqnarray}
\Tr\Big[\big(...\big)\flag{1}b\Big]\lineup = \mathcal{F}_b\Big\langle\big(...\big)b(-s)\Big\rangle_\mathrm{UHP},\\
\Tr\Big[\big(...\big)b\Aflag{1}\Big]\lineup = \mathcal{F}_b\Big\langle\big(...\big)b(s)\Big\rangle_\mathrm{UHP},
\end{eqnarray}
where $\mathcal{F}_b$ is the $b$-ghost residue:
\begin{equation}\mathcal{F}_b = \frac{1}{\mathcal{F}''(s)}=\frac{\pi}{4\ell ps}\frac{1+s^2}{1+p^2}( s^2-p^2)^2. \label{eq:app:Fb}\end{equation}
In terms of the dual upper half plane coordinate $\widetilde{u}$ we have
\begin{eqnarray}
\Tr\Big[\big(...\big)\flag{1}b\Big]\lineup = \widetilde{\mathcal{F}}_b\Big\langle\big(...\big)b(\widetilde{s})\Big\rangle_{\widetilde{\mathrm{UHP}}},\\
\Tr\Big[\big(...\big)b\Aflag{1}\Big]\lineup = \widetilde{\mathcal{F}}_b\Big\langle\big(...\big)b(-\widetilde{s})\Big\rangle_{\widetilde{\mathrm{UHP}}}
,\end{eqnarray}
where $\widetilde{\mathcal{F}}_b$ takes the same form as $\mathcal{F}_b$ after replacing $p,s$ with $\widetilde{p},\widetilde{s}$. 

\begin{figure}
\begin{center}
\resizebox{5in}{2.4in}{\includegraphics{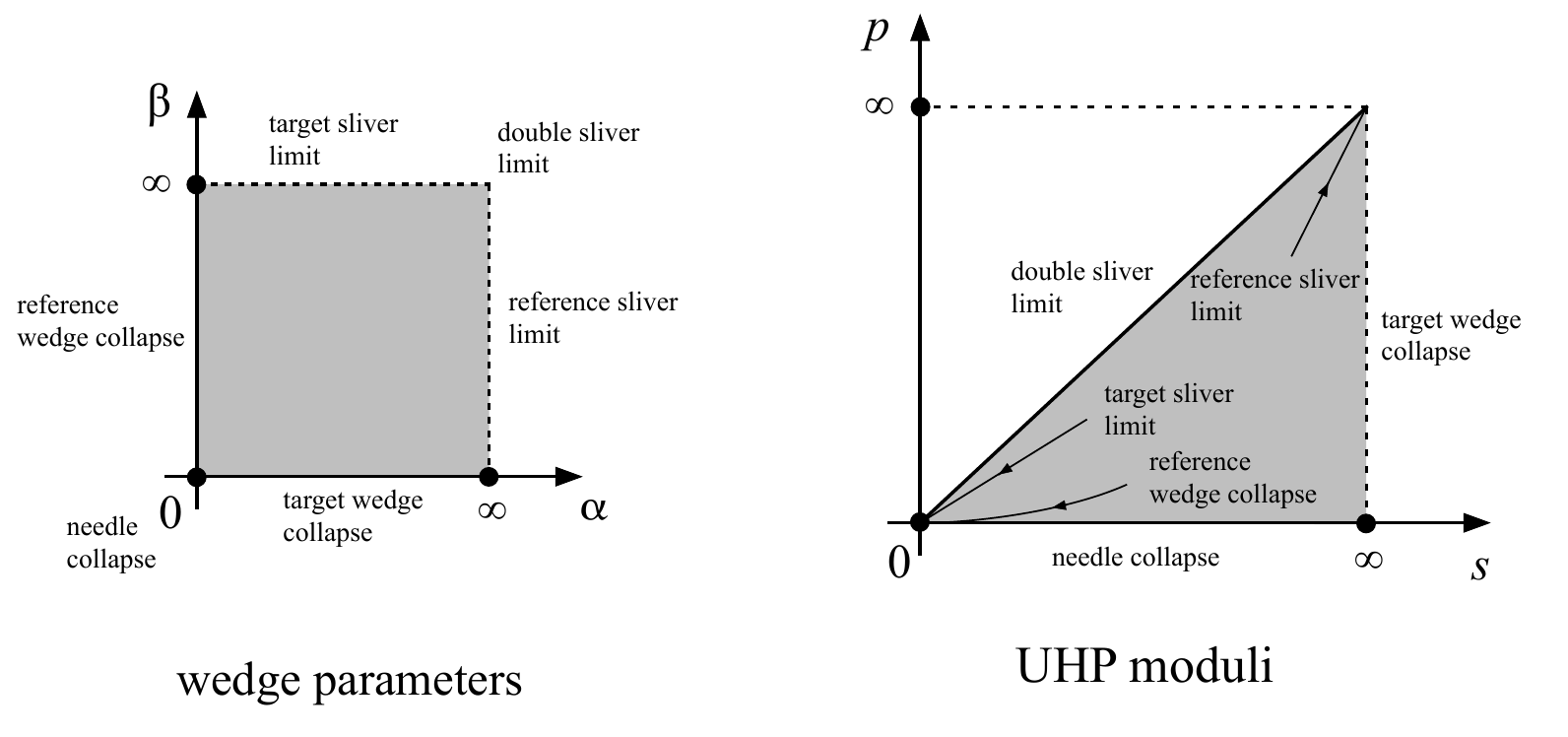}}
\end{center}
\caption{\label{fig:degeneration} Limits of flag-anti-flag surface.}
\end{figure}

\item{\bf Limits:} Various limits of the flag-anti-flag surface, and their realization in terms of upper half plane moduli, are summarized in figure \ref{fig:degeneration}
\begin{enumerate}
\item Reference wedge collapse: 
\begin{equation}\alpha\to 0, \beta = \mathrm{fixed};\ \ \ \ p,s\to 0,\ \frac{p}{s^2} = \chi = \mathrm{fixed}.\end{equation}
\item Target wedge collapse: 
\begin{equation}\alpha=\mathrm{fixed},\ \beta\to 0;\ \ \ \  p=\mathrm{fixed}, \ s\to\infty; \ \ \ \  \widetilde{p} =\mathrm{fixed},\ \widetilde{s}\to 0.\end{equation}
\item Needle collapse: 
\begin{equation}\alpha,\beta\to 0,\  \frac{\alpha}{\alpha+\beta}=\mathrm{fixed};\ \ \ \  p\to 0,\ s=\mathrm{fixed}.\end{equation}
\item Flag collapse: 
\begin{equation}\ell\to 0,\ \alpha,\beta = \mathrm{fixed};\ \ \ \  \ell\to 0,\ p\to s, \ s=\mathrm{fixed},\ \frac{s-p}{\ell}=\mathrm{fixed},.\end{equation}
\item Reference sliver limit:
\begin{equation}\alpha\to \infty,\ \beta= \mathrm{fixed};\ \ \ \ p,s\to\infty,\ \frac{p}{s}  = \mathrm{fixed}.\end{equation}
\item Target sliver limit
\begin{equation}\alpha = \mathrm{fixed},\ \beta\to\infty;\ \ \ \ p,s\to 0,\ \frac{p}{s}= \mathrm{fixed};\ \ \ \ \widetilde{p},\widetilde{s}\to \infty,\ \frac{\widetilde{p}}{\widetilde{s}}=\mathrm{fixed}.\end{equation}
\item Double sliver limit
\begin{equation}\alpha,\beta\to\infty,\ \frac{\alpha}{\alpha+\beta}=\mathrm{fixed};\ \ \ \ p\to s,\ s=\mathrm{fixed}.\end{equation}
\item Flag sliver limit
\begin{equation}\ell\to\infty,\ \alpha,\beta=\mathrm{fixed};\ \ \ \ \ell\to \infty,\ p\to 0,\ s=\mathrm{fixed},\ \ell p = \mathrm{fixed}.\end{equation}
\end{enumerate}

\item{\bf Flag-Anti-flag Jacobian:} The flag-anti-flag Jacobian is defined 
\begin{equation}d\alpha \, d\beta = dp\,ds\, \Delta(p,s)\end{equation}
where
\begin{equation}
\Delta(p,s) = \frac{16\ell^2}{\pi}\frac{s^2 p(1+p^2)}{(s^2-p^2)^3}.
\end{equation}

\end{itemize}

\section{Ghost Correlators}
\label{app:ghost}

In this appendix we give formulas for correlators of the $c$ ghost and vertical line integral insertions of the $b$ ghost on the flag-anti-flag surface. These generalize the perennial formulas for ghost correlators on the cylinder derived in \cite{Schnabl,Okawa}. We use the doubling trick, and normalize ghost correlators in the upper half plane according to
\begin{equation}\langle c(u_1)c(u_2)c(u_3)\rangle_\mathrm{UHP}^{gh} = u_{12}u_{23}u_{13},\end{equation}
where $u_{ij}\equiv u_i-u_j$. The three-point function of $c$-ghosts on the flag-anti-flag surface is easily seen to be
\begin{equation}
\langle c(z_1)c(z_2)c(z_3)\rangle_{\text{flag-anti-flag}}^{gh}=\mathcal{F}'(u_1)\mathcal{F}'(u_2)\mathcal{F}'(u_3)u_{12}u_{23}u_{13},\label{eq:c_cor}
\end{equation}
where the upper half plane coordinates $u_1,u_2,u_3$ and moduli $p,s$ on the right hand side are related to the sliver coordinates $z_1,z_2,z_3$ and wedge angles $\alpha,\beta$ on the left hand side through \eq{app:z}-\eq{app:beta}. The correlator cannot be written as an explicit function of the sliver coordinates and wedge angles due to the usual difficulty of inverting the Schwarz-Christoffel map. 

There are two kinds of vertical line integral insertion of the $b$ ghost which can appear in correlators on the flag-anti-flag surface, which we write
\begin{eqnarray}
B_\inn\lineup = \int_{\frac{\alpha+\beta}{2}+i\ell}^{\frac{\alpha+\beta}{2}+i\infty}\frac{dz}{2\pi i}b(z) + \int_{\frac{\alpha+\beta}{2}-i\infty}^{\frac{\alpha+\beta}{2}-i\ell}\frac{dz}{2\pi i}b(z),\label{eq:Bin}\\
B_\out\lineup =\int_{-i\infty}^{i\infty}\frac{dz}{2\pi i} b(z).
\end{eqnarray}
The operator $B_\inn$ corresponds to the string field $B$ when it is embedded in the target wedge state, and the operator $B_\out$ corresponds to the string field $B$ embedded in the reference wedge state. The two terms of $B_\inn$ define an uninterrupted contour on the holomorphically doubled flag-anti-flag surface, as can be seen from figure \ref{fig:double}. Note that we specifically define $B_\out$ so that the real part of the integration contour is constant and equal to $0$ in the sliver coordinate $z$ on the flag-anti-flag surface, and we define $B_\inn$ so that the real part of the integration contour is constant and equal to $\frac{\alpha+\beta}{2}$. The contours can be deformed without changing the correlation function as long as we do not cross a $c$-ghost or a slit. By explicitly carrying out the integration, we can find the correlator with an insertion of $B_\out$:
\begin{eqnarray}
\lineup  \Big\langle c(z_1)c(z_2)c(z_3)c(z_4)B_\out\Big\rangle_\text{flag-anti-flag}^{gh}\nonumber\\
\lineup\ \ \ \ \ \ \ \ \ \  =\ \mathcal{F}'(u_1)\mathcal{F}'(u_2)\mathcal{F}'(u_3)u_{12}u_{23}u_{13}C_\out(u_4)-\mathcal{F}'(u_1)\mathcal{F}'(u_2)\mathcal{F}'(u_4)u_{12}u_{24}u_{14}C_\out(u_3)\nonumber\\
\lineup\ \ \ \ \ \ \ \ \ \  \ \ \ \ +\mathcal{F}'(u_1)\mathcal{F}'(u_3)\mathcal{F}'(u_4)u_{13}u_{34}u_{14}C_\out(u_1)-\mathcal{F}'(u_2)\mathcal{F}'(u_3)\mathcal{F}'(u_4)u_{23}u_{34}u_{24}C_\out(u_1),\ \ \ \ \ \ \label{eq:Bout_cor}
\end{eqnarray}
where
\begin{equation}
C_\out(u)\equiv -\frac{1}{\pi}\frac{1}{u^2-p^2}\frac{u}{1+u^2}\left(\frac{(s^2-p^2)(1+s^2)}{s}\tan^{-1}\frac{1}{s}-\frac{(u^2-p^2)(1+u^2)}{u}\tan^{-1}\frac{1}{u}\right).
\end{equation}
The inverse tangent implies that $C_\out(u)$ has a branch cut extending through the origin on the imaginary axis from $-i$ to $i$. The branch cut coincides with the integration contour of $B_\out$ in the upper half plane, and represents a discontinuity of the correlation function as a $c$ ghost jumps to the other side of the $b$ ghost contour. The correlator with an insertion of $B_\inn$ is given by
\begin{eqnarray}
\lineup  \Big\langle c(z_1)c(z_2)c(z_3)c(z_4)B_\inn\Big\rangle_\text{flag-anti-flag}^{gh}\nonumber\\
\lineup\ \ \ \ \ \ \ \ \ \  =\ \mathcal{F}'(u_1)\mathcal{F}'(u_2)\mathcal{F}'(u_3)u_{12}u_{23}u_{13}C_\inn(u_4)-\mathcal{F}'(u_1)\mathcal{F}'(u_2)\mathcal{F}'(u_4)u_{12}u_{24}u_{14}C_\inn(u_3)\nonumber\\
\lineup\ \ \ \ \ \ \ \ \ \  \ \ \ \ +\mathcal{F}'(u_1)\mathcal{F}'(u_3)\mathcal{F}'(u_4)u_{13}u_{34}u_{14}C_\inn(u_1)-\mathcal{F}'(u_2)\mathcal{F}'(u_3)\mathcal{F}'(u_4)u_{23}u_{34}u_{24}C_\inn(u_1),\ \ \ \ \ \ 
\end{eqnarray}
where
\begin{equation}
C_\inn(u) \equiv \frac{1}{\pi}\frac{1}{u^2-p^2}\frac{u}{1+u^2}\left(\frac{(s^2-p^2)(1+s^2)}{s}\tan^{-1}s-\frac{(u^2-p^2)(1+u^2)}{u}\tan^{-1}u\right).
\end{equation}
The branch cut extends on the imaginary axis from $i$ out to $+i\infty$, and then from $-i\infty$ to $-i$. This coincides with the integration contour of $B_\inn$ in the upper half plane. Note that when integrating over various wedge parameters of the solution, it may be convenient to deform the $b$ ghost contours in such a way that they cannot be returned to their standard position without crossing a $c$ ghost. This can be dealt with by correspondingly adjusting the branch cuts.  Incidentally, one may confirm that 
\begin{eqnarray}
\lineup \Big\langle c(z_1)c(z_2)c(z_3)c(z_4)B_\out\Big\rangle_\text{flag-anti-flag}-\Big\langle c(z_1)c(z_2)c(z_3)c(z_4)B_\inn\Big\rangle_\text{flag-anti-flag}\nonumber\\
\lineup\ \ \ \ \ \ \ \ \ \ \ =\mathcal{F}'(u_1)\mathcal{F}'(u_2)\mathcal{F}'(u_3)\mathcal{F}'(u_4)\mathcal{F}_b\Big\langle c(u_1)c(u_2)c(u_3)c(u_4)b(-s)\Big\rangle_\text{UHP},\label{eq:Bin_cor}
\end{eqnarray}
provided that the real parts of $u_1,...,u_4$ are positive, so that only the slit contributes to the contour deformation from $B_\out$ to $B_\inn$. 

We may also compute the correlator containing both insertions $B_\out$ and $B_\inn$:
\begin{eqnarray}
\lineup \Big\langle c(z_1)c(z_2)c(z_3)c(z_4)c(z_5)B_\out B_\inn\Big\rangle_\text{flag-anti-flag}^{gh} \nonumber\\
\lineup \ \ =\!\! \sum_{{i_1<i_2<i_3\atop i_4<i_5}}\!\!(-1)^{i_4+i_5}\mathcal{F}'(u_{i_1})\mathcal{F}'(u_{i_2})\mathcal{F}'(u_{i_3})u_{i_1i_2}u_{i_2i_3}u_{i_1i_3}\Big(C_\out(u_{i_4})C_\inn(u_{i_5})-C_\out(u_{i_5})C_\inn(u_{i_4})\Big),\ \ \ \ \ \ \ \ \ \ \label{eq:cccccBinBout}
\end{eqnarray}
where the sum is over all partitions of $1,...,5$ into sets $\{i_1,i_2,i_3\}$ and $\{i_4,i_5\}$ with  $i_1<i_2<i_3$ and $i_4<i_5$. There are 10 terms.
This is the most general correlator involving insertions of the $c$-ghost and vertical line integrals of the $b$-ghost on the flag-anti-flag surface. The other correlators \eq{c_cor}, \eq{Bout_cor} and \eq{Bin_cor} are special cases when two $c$  ghosts pinch together on a $b$-ghost contour.

A number of other special cases occur with correlators involving $\d c$, correlators with $c$ at the midpoint (appearing in computations of the Ellwood invariant), and correlators with $c$ at the puncture (arising from BRST variations of boundary condition changing operators). In these cases it is useful to make additional definitions. For correlators with $\d c$ we introduce
\begin{eqnarray}
D_\out(u) \equiv \frac{\d}{\d u}\frac{C_\out(u)}{\mathcal{F}'(u)}\lineup =\frac{s^2-p^2}{2\ell p(1+p^2)}\left[-\frac{u^2-p^2}{u^2-s^2}+\frac{(s^2-p^2)(1+s^2)}{s}\frac{u^2+s^2}{(u^2-s^2)^2}\tan^{-1}\frac{1}{s}\right.\phantom{\Bigg(}\nonumber\\
\lineup\ \ \ \ \ \ \ \ \ \ \ \ \ \ \ \ \ \ \ \ \ \ \left.+2 \left(1-\frac{(1+s^2)(s^2-p^2)}{(u^2-s^2)^2}\right)u\tan^{-1}\frac{1}{u}\right],\phantom{\Bigg(}\\
D_\inn(u) \equiv \frac{\d}{\d u}\frac{C_\inn(u)}{\mathcal{F}'(u)}\lineup = -\frac{s^2-p^2}{2\ell p(1+p^2)}\left[\frac{u^2-p^2}{u^2-s^2}+\frac{(s^2-p^2)(1+s^2)}{s}\frac{u^2+s^2}{(u^2-s^2)^2}\tan^{-1}s\right.\phantom{\Bigg(}\nonumber\\
\lineup\ \ \ \ \ \ \ \ \ \ \ \ \ \ \ \ \ \ \ \ \ \ \left.+2 \left(1-\frac{(1+s^2)(s^2-p^2)}{(u^2-s^2)^2}\right)u\tan^{-1}u\right].\phantom{\Bigg(}\ \ \ \ \ \ \ \ \ \ 
\end{eqnarray}
For correlators with $c$ at the midpoint we introduce
\begin{eqnarray}
I_\out \lineup \equiv\frac{C_\out(i)}{\mathcal{F}'(i)} = \frac{i}{2\ell}\frac{(s^2-p^2)^2}{sp(1+p^2)}\tan^{-1} \frac{1}{s},\\
I_\inn \lineup \equiv\frac{C_\inn(i)}{\mathcal{F}'(i)} =-\frac{i}{2\ell}\frac{(s^2-p^2)^2}{sp(1+p^2)}\tan^{-1} s,
\end{eqnarray}
and for correlators with $c$ at the puncture we introduce
\begin{eqnarray}
P_\out\lineup \equiv \frac{C_\out(p)}{\mathcal{F}'(p)}=\frac{1}{2\ell}\frac{(s^2-p^2)(1+s^2)}{s(1+p^2)}\tan^{-1}\frac{1}{s},\\
P_\inn\lineup \equiv \frac{C_\inn(p)}{\mathcal{F}'(p)}=-\frac{1}{2\ell}\frac{(s^2-p^2)(1+s^2)}{s(1+p^2)}\tan^{-1}s.
\end{eqnarray}
A correlator which appears in the analysis of the Ellwood invariant in subsection \ref{subsec:Ellwood} is
\begin{eqnarray}
\lineup  \Big\langle c\d c (z_1) \mathcal{F}\circ\big(c(i)c(-i)\big) B_\out\Big\rangle_\text{flag-anti-flag}^{gh}\nonumber\\
\lineup\ \ \ \ \ \ \ \ \ \ \ \ \ \ \  =-2\mathcal{F}'(u_1) (1-u_1^2)I_\out +2 i\mathcal{F}'(u_1)(1+u_1^2)D_\out(u_1)-4i u_1 C_\out(u_1).
\end{eqnarray}

\section{Lump Coefficients}
\label{app:lump}

\renewcommand{\arraystretch}{2}

\begin{table}
\begin{center}
\begin{tabular}{|c|c|c|c|c|c|}
\hline
\raisebox{3pt}{harmonic} & level & \raisebox{3pt}{${\text{Flag state lump,}\atop\text{gh\# 0 toy model }(\ell=\pi)}$} & \raisebox{3pt}{${\text{Lump of \cite{KOSsing},}\atop\text{gh\# 0 toy model}}$} &  \raisebox{3pt}{Lump of \cite{KOSsing}} &  \raisebox{3pt}{Siegel gauge lump} \\
\hline
\raisebox{3pt}{0} & \raisebox{3pt}{$0$} & \raisebox{3pt}{$0.71%(2581)
$} & \raisebox{3pt}{$0.711324$} &  \raisebox{3pt}{$0.202297$} & \raisebox{3pt}{$0.414218$} \\
\hline
\raisebox{3pt}{1} & \raisebox{3pt}{$\frac{1}{12} $} & \raisebox{3pt}{$-0.25%(6881)
$} & \raisebox{3pt}{$-0.280999$} & \raisebox{3pt}{$-0.0878951$} & \raisebox{3pt}{$-0.120192$} \\
\hline
\raisebox{3pt}{2} & \raisebox{3pt}{$\frac{1}{3}$} & \raisebox{3pt}{$-0.16%(6458)
$} & \raisebox{3pt}{$-0.261895$} & \raisebox{3pt}{$-0.0926562$} & \raisebox{3pt}{$-0.0987233$} \\
\hline
\raisebox{3pt}{3} & \raisebox{3pt}{$\frac{3}{4}$} & \raisebox{3pt}{$-0.073%(1512)
$} & \raisebox{3pt}{$-0.234590$} & \raisebox{3pt}{$-0.0818861$} & \raisebox{3pt}{$-0.0650643$} \\
\hline
\raisebox{3pt}{4} & \raisebox{3pt}{$1\frac{1}{3}  $} & \raisebox{3pt}{$-0.025%(2259)
$} & \raisebox{3pt}{$-0.198024$} & \raisebox{3pt}{$-0.0592841$} & \raisebox{3pt}{$-0.0336245$} \\
\hline
\raisebox{3pt}{5} & \raisebox{3pt}{$2\frac{1}{12} $} & \raisebox{3pt}{$-0.0064%(2073)
$} & \raisebox{3pt}{$-0.154011$} & \raisebox{3pt}{$-0.0360931$} & \raisebox{3pt}{$-0.0137873$} \\
\hline
\raisebox{3pt}{6} & \raisebox{3pt}{$3 $} & \raisebox{3pt}{$-0.0012%(0371)
$} & \raisebox{3pt}{$-0.109000$} & \raisebox{3pt}{$-0.0190752$} & \raisebox{3pt}{$-0.00456357$} \\
\hline
\raisebox{3pt}{7} & \raisebox{3pt}{$4\frac{1}{12}$} & \raisebox{3pt}{$-0.00017%(6373)
$} & \raisebox{3pt}{$-0.0699578$} & \raisebox{3pt}{$-0.00900014$} & \raisebox{3pt}{$-0.00123586$} \\
\hline
\raisebox{3pt}{8} &\raisebox{3pt}{$5\frac{1}{3} $} & \raisebox{3pt}{$-0.000017%(9255)
$} & \raisebox{3pt}{$-0.0407627$} & \raisebox{3pt}{$-0.00387693$} & \raisebox{3pt}{$-0.000276564$} \\
\hline
\raisebox{3pt}{9} & \raisebox{3pt}{$6\frac{3}{4} $} & \raisebox{3pt}{$-1.3%(7258)
\times 10^{-6}$} & \raisebox{3pt}{$-0.0216228$} & \raisebox{3pt}{$-0.00154912$} & \raisebox{3pt}{$-0.0000515764$} \\
\hline
\raisebox{3pt}{10} & \raisebox{3pt}{$8\frac{1}{3} $} & \raisebox{3pt}{$-7.0%(2859)
\times 10^{-8}$} & \raisebox{3pt}{$-0.0104735$} & \raisebox{3pt}{$-0.000579571$} & \raisebox{3pt}{$-8.06609\times 10^{-6}$} \\
\hline
\raisebox{3pt}{11} & \raisebox{3pt}{$10\frac{1}{12}$} & \raisebox{3pt}{$-2.9%(5874)
\times 10^{-9}$} & \raisebox{3pt}{$-0.00464500$} & \raisebox{3pt}{$-0.00020383$} & \raisebox{3pt}{$-1.06546\times 10^{-6}$} \\
\hline
\raisebox{3pt}{12} & \raisebox{3pt}{$12 $} & \raisebox{3pt}{$-8.5%(4786)
\times 10^{-11}$} & \raisebox{3pt}{$-0.00189051$} & \raisebox{3pt}{$-0.0000674125$} & \raisebox{3pt}{$-1.19126\times 10^{-7}$} \\
\hline
\raisebox{3pt}{13} & \raisebox{3pt}{$14\frac{1}{12} $} & \raisebox{3pt}{$-1.9%(8212)
\times 10^{-12}$} & \raisebox{3pt}{$-0.000707411$} & \raisebox{3pt}{$-0.0000209381$} & \raisebox{3pt}{$-1.14000\times 10^{-8}$} \\
\hline
\raisebox{3pt}{14} & \raisebox{3pt}{$16\frac{1}{3} $} & \raisebox{3pt}{$-3.0%(396)
\times 10^{-14}$} & \raisebox{3pt}{$-0.000243728$} & \raisebox{3pt}{$-6.09597\times 10^{-6}$} & \raisebox{3pt}{$-7.36039\times 10^{-10}$} \\
\hline
\end{tabular}
\end{center}
\caption{\label{tab:lump1} Coefficients $T_n$ of the $n$th harmonic $e^{in x/R}$ of the tachyon lump when $R=2\sqrt{3}$. We assume $T_n=T_{-n}$ so the modes of the cosine appear with an additional factor of $2$ when $n\neq 0$. %Parentheses surround digits which cannot be trusted based on the estimated accuracy of numerical integration.
}
\end{table}

\begin{table}
\begin{center}
\begin{tabular}{|c|c|c|c|}
\hline
\raisebox{3pt}{${\text{Flag state lump,}\atop\text{gh\# 0 toy model}}$} & \raisebox{3pt}{${\text{Lump of \cite{KOSsing},}\atop\text{gh\# 0 toy model}}$} &  \raisebox{3pt}{Lump of \cite{KOSsing}} &  \raisebox{3pt}{Siegel gauge lump} \\
\hline
\raisebox{3pt}{$e^{-1.8 L}$} & \raisebox{3pt}{$e^{-0.47 L}$} & \raisebox{3pt}{$e^{-0.55 L}$} & \raisebox{3pt}{$e^{-1.2 L}$} \\
\hline
\end{tabular}
\end{center}
\caption{\label{tab:lump2} Rate of decay of lump coefficients at $R=2\sqrt{3}$ as a function of level $L$.}
\end{table}

\begin{table}
\begin{center}
\begin{tabular}{|c|c|c|c|c|}
\hline
\raisebox{3pt}{harmonic} & level & \raisebox{3pt}{${\text{Flag state excited lump,}\atop\text{gh\# 0 toy model }(\ell=\pi)}$} & \raisebox{3pt}{${\text{Excited lump of \cite{KOSsing},}\atop\text{gh\# 0 toy model}}$} & \raisebox{3pt}{Excited Lump of \cite{KOSsing}} \\
\hline
\raisebox{3pt}{0} & \raisebox{3pt}{$0 $} & \raisebox{3pt}{$0.75%(7403)
$} & \raisebox{3pt}{$0.711325$} & \raisebox{3pt}{$0.202297$}\\
\hline
\raisebox{3pt}{1} & \raisebox{3pt}{$\frac{1}{12}  $} & \raisebox{3pt}{$-0.14%(4548)
$} & \raisebox{3pt}{$-0.187332$}  & \raisebox{3pt}{$-0.0585967$}\\
\hline
\raisebox{3pt}{2} & \raisebox{3pt}{$\frac{1}{3} $} & \raisebox{3pt}{$0.046%(7088)
$} & \raisebox{3pt}{$0.0872984$}  & \raisebox{3pt}{$0.0308854$}\\
\hline
\raisebox{3pt}{3} & \raisebox{3pt}{$\frac{3}{4} $} & \raisebox{3pt}{$0.13%(7780)
$} & \raisebox{3pt}{$0.469181$}  & \raisebox{3pt}{$0.163772$}\\
\hline
\raisebox{3pt}{4} & \raisebox{3pt}{$1\frac{1}{3}$} & \raisebox{3pt}{$0.095%(0361)
$} & \raisebox{3pt}{$0.858103$}  & \raisebox{3pt}{$0.25690$}\\
\hline
\raisebox{3pt}{5} & \raisebox{3pt}{$2\frac{1}{12}  $} & \raisebox{3pt}{$0.041%(6662)
$} & \raisebox{3pt}{$1.12942$}  & \raisebox{3pt}{$0.264683$}\\
\hline
\raisebox{3pt}{6} & \raisebox{3pt}{$3 $} & \raisebox{3pt}{$0.011%(3592)
$} & \raisebox{3pt}{$ 1.19900$}  & \raisebox{3pt}{$0.209827$}\\
\hline
\raisebox{3pt}{7} & \raisebox{3pt}{$4\frac{1}{12}$} & \raisebox{3pt}{$0.0022%(7132)
$} & \raisebox{3pt}{$1.07269$}  & \raisebox{3pt}{$0.138002$}\\
\hline
\raisebox{3pt}{8} & \raisebox{3pt}{$5\frac{1}{3} $} & \raisebox{3pt}{$0.00029%(2670)
$} & \raisebox{3pt}{$0.828841$}  & \raisebox{3pt}{$0.0788309$}\\
\hline
\raisebox{3pt}{9} & \raisebox{3pt}{$6\frac{3}{4} $} & \raisebox{3pt}{$0.000028%(7921)
$} & \raisebox{3pt}{$0.562192$}  & \raisebox{3pt}{$0.0402771$}\\
\hline
\raisebox{3pt}{10} & \raisebox{3pt}{$8\frac{1}{3} $} & \raisebox{3pt}{$1.9%(8916)
\times 10^{-6}$} & \raisebox{3pt}{$0.338643$}  & \raisebox{3pt}{$0.0187395$}\\
\hline
\raisebox{3pt}{11} & \raisebox{3pt}{$10\frac{1}{12} $} & \raisebox{3pt}{$9.2%(4059)
\times 10^{-8}$} & \raisebox{3pt}{$0.182703$}  & \raisebox{3pt}{$0.00801731$}\\
\hline
\raisebox{3pt}{12} & \raisebox{3pt}{$12 $} & \raisebox{3pt}{$3.3%(6578)
\times 10^{-9} $} & \raisebox{3pt}{$ 0.0888540$}  & \raisebox{3pt}{$0.00316839$}\\
\hline
\raisebox{3pt}{13} & \raisebox{3pt}{$ 14\frac{1}{12} $} & \raisebox{3pt}{$8.4%(6261)
\times 10^{-11} $} & \raisebox{3pt}{$0.0391434$}  & \raisebox{3pt}{$0.00115857$}\\
\hline
\raisebox{3pt}{14} & \raisebox{3pt}{$16\frac{1}{3}  $} & \raisebox{3pt}{$1.5%(5512)
\times 10^{-12}$} & \raisebox{3pt}{$0.0156798$}  & \raisebox{3pt}{$0.000392174$}\\
\hline
\end{tabular}
\end{center}
\caption{\label{tab:lump3} Coefficients $T_n$ of the $n$th exponential harmonic of the lump constructed from excited twist fields at $R=2\sqrt{3}$.}
\end{table}

To get a first impression of how the flag state solution behaves in the level expansion, here we give the coefficients of the harmonics of the toy model lump, constructed from Neumann-Dirichlet twist fields of weight $\frac{1}{16}$. We set $\ell=\pi$ and the compatification radius $R=2\sqrt{3}$. Since the numbers do not mean much without a baseline for comparison, we also give coefficients for the harmonics of several other lumps. This is shown in Table \ref{tab:lump1}. First we give the coefficients for the ghost number 0 toy model of \cite{KOSsing}:
\begin{equation}\Gamma_{\cite{KOSsing}} = 1-\sqrt{1+K}\sigma\frac{1}{1+K}\sigmabar\sqrt{1+K}\end{equation}
which is most closely analogous to what we presently have for the flag state solution. The coefficients of this state are given by
\begin{equation}
\langle \phi,\Gamma_{\cite{KOSsing}}\rangle = \delta_{h_\phi=0}-\big\langle \big(I\circ\phi(0)\big)\sigma(1)\sigmabar(0)\big\rangle_\mathrm{UHP}\int_0^\infty dt_1 dt_2 dt_3\, f_{\Gamma_*}(t_1,t_2,t_3)\left(\frac{2}{L}\frac{\sin\theta_{t_2}}{\sin\theta_{\frac{1}{2}+t_1}\sin\theta_{1/2+t_3}}\right)^{h_\phi},\label{eq:13toy}
\end{equation}
where 
\begin{equation}L=t_1+t_2+t_3,\ \ \ \ \theta_t = \frac{\pi t}{t_1+t_2+t+3}.\end{equation}
For the lump, we must choose $\phi = \phi^{T_n^\mathrm{toy}}$ from \eq{phiTntoy} and $\sigma,\sigmabar = \sigma_{\mathrm{ND}},\sigmabar_{\mathrm{ND}}$. In principle the boundary condition changing operators should contain a Wilson line factor to cancel their conformal weight, but in \eq{13toy} this only contributes a factor given by the two-point function of plane wave vertex operators, which evaluates to unity. Second, we give the coefficients of the true lump solution of \cite{KOSsing} at ghost number 1, which were calculated in \cite{KOSsing} for the non-real form of the solution. Third, we give the coefficients of the lump solution in Siegel gauge, calculated up to level 18 and extrapolated to infinite level.\footnote{We thank M. Kudrna for providing this data.}.

The first thing to note is that the results for the flag state toy model are clearly less accurate than in other cases. This is due to the estimated 1\% errors in numerical integration. The difficulty does not seem to originate in the integrand, but in determining the precise shape of the ``arrowhead" integration region. In practice, we computed the coefficients by integrating over all upper half plane moduli $p<s$ and $|x|<p$, relying on Heaviside step functions to project the integrand onto the arrowhead region. Computing even the volume of a portion of the arrowhead in this way produced 1\% errors.

\begin{figure}
\begin{center}
\resizebox{3in}{1.8in}{\includegraphics{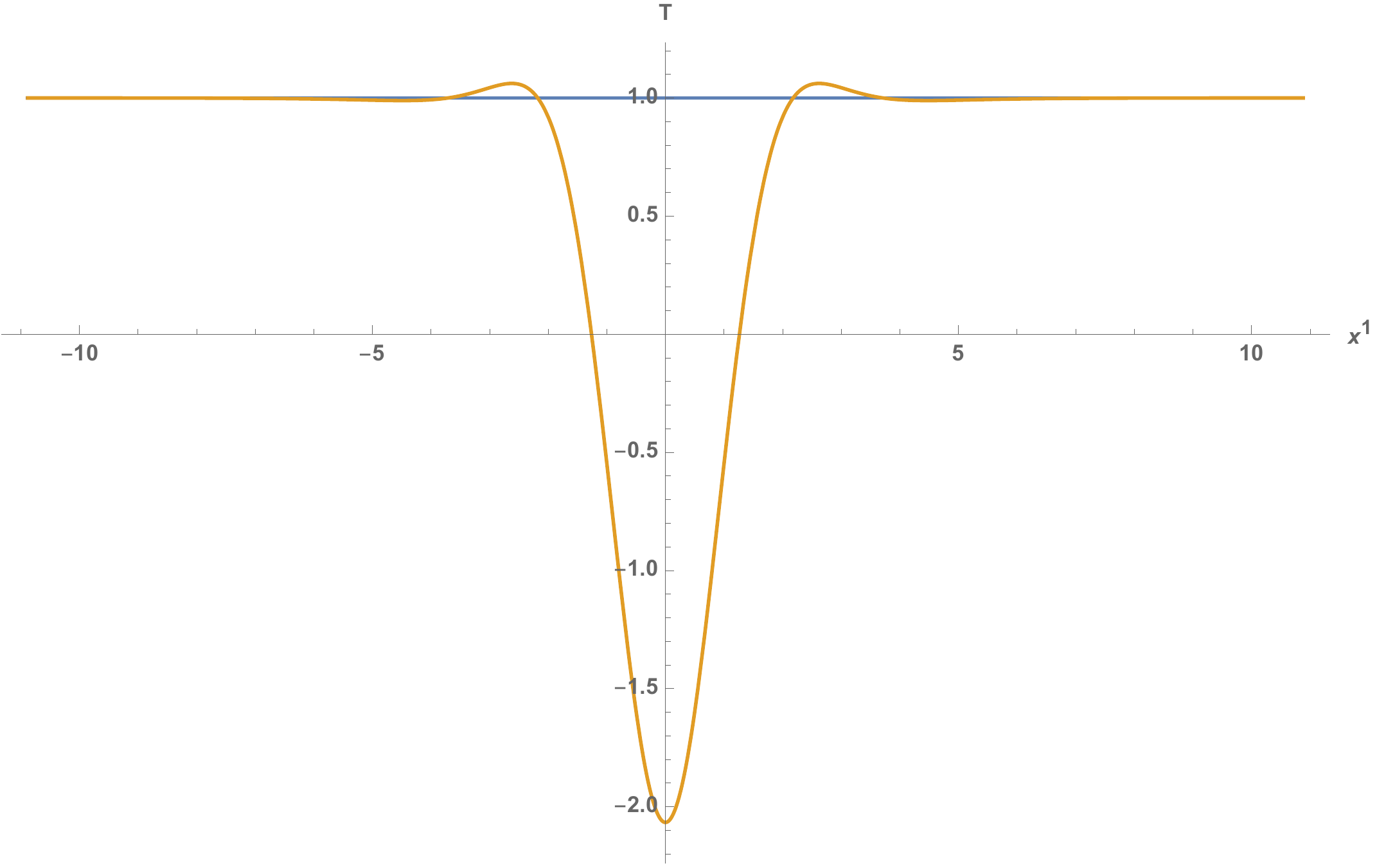}}\ 
\resizebox{3in}{1.8in}{\includegraphics{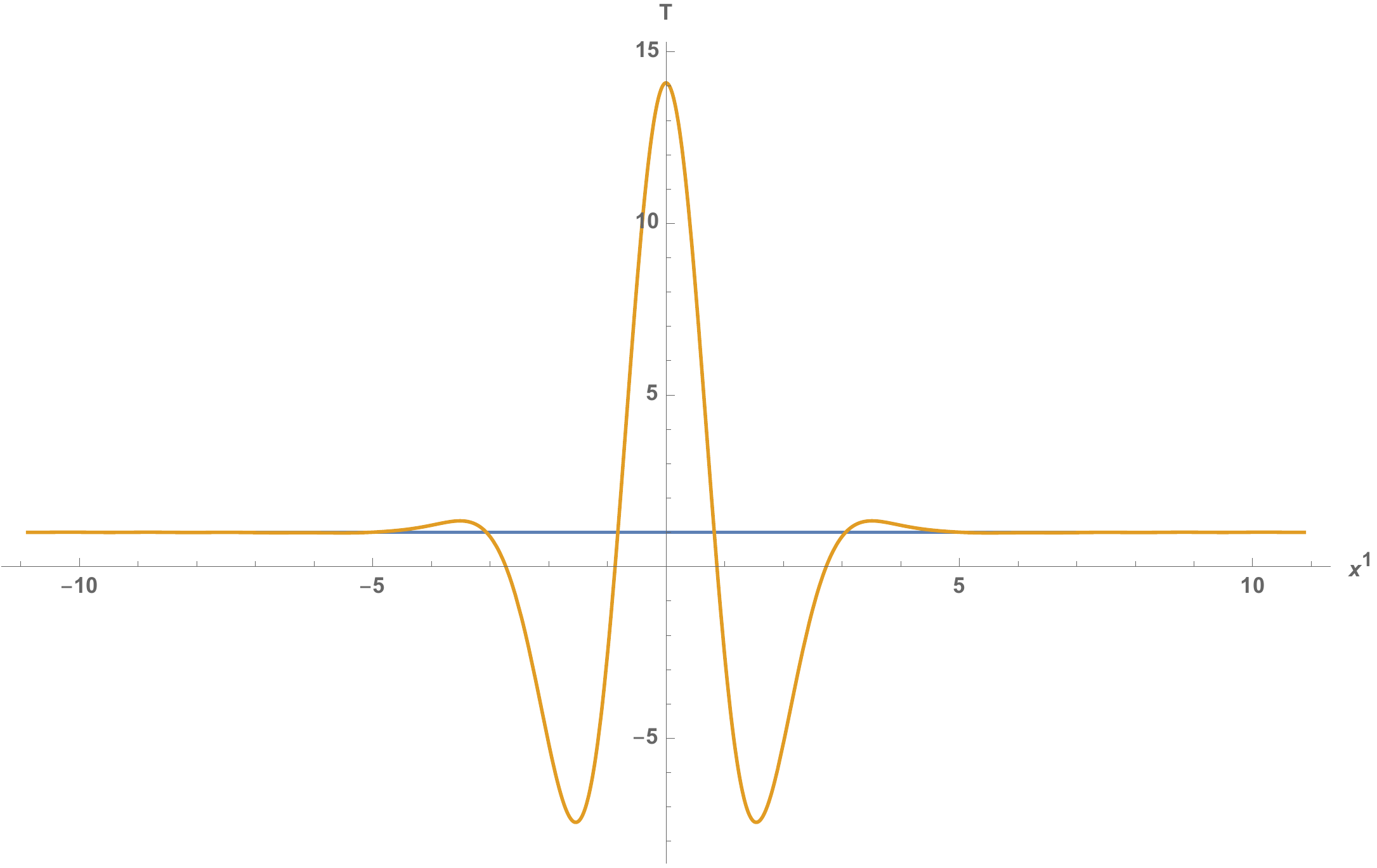}}
\end{center}
\caption{\label{fig:13lumps} Tachyon profiles for the lump (left) and excited lump (right) for the ghost number 0 toy model of the solution of \cite{KOSsing}.}
\end{figure}

Still, the data is good enough to make useful comparisons. Of particular interest is the rate of decay of the coefficients at high levels, shown in table \ref{tab:lump2}. In all cases the lump coefficients decay roughly exponentially with the level. The coefficients for the flag state toy model decay especially rapidly, even more so than for the Siegel gauge lump. The decay rate varies a lot with the height of the flag; we have chosen $\ell=\pi$, and for smaller $\ell$ the coefficients decay more slowly. But the comparison to Siegel gauge is encouraging. By contrast, the decay of lump coefficients for the solution and toy model of \cite{KOSsing} is slow. This is likely due to the identity-like nature of the solution. While the flag state toy model is also in a sense identity-like, towards the endpoints of the string it contains a fair amount of surface from  the flags. Another thing to observe is that the coefficients of the true solution of \cite{KOSsing} decay somewhat faster than those of the toy model of \cite{KOSsing}. This gives hope that coefficients of the flag state solution at ghost number 1 will not show different or worse behavior than the flag state toy model.

Out of interest, we also give coefficients for lumps constructed from excited Neumann-Dirichlet twist fields of weight $\frac{9}{16}$. This is shown in table \ref{tab:lump3}. Again we assume $\ell=\pi$ and $R=2\sqrt{3}$. We list coefficients of the flag state toy model, the toy model of \cite{KOSsing}, and the full solution of \cite{KOSsing}. The coefficients are rather different from the ordinary lump. Beyond the first harmonic, they are all positive, and the first handful of harmonics are quite large, especially for the toy model of \cite{KOSsing}. At sufficiently high level the coefficients decay at a similar (or perhaps identical) rate to the ordinary lump. Perhaps this data can give a hint as to how an excited lump would appear in Siegel gauge level truncation, if such a solution exists.

\begin{figure}
\begin{center}
\resizebox{3in}{1.8in}{\includegraphics{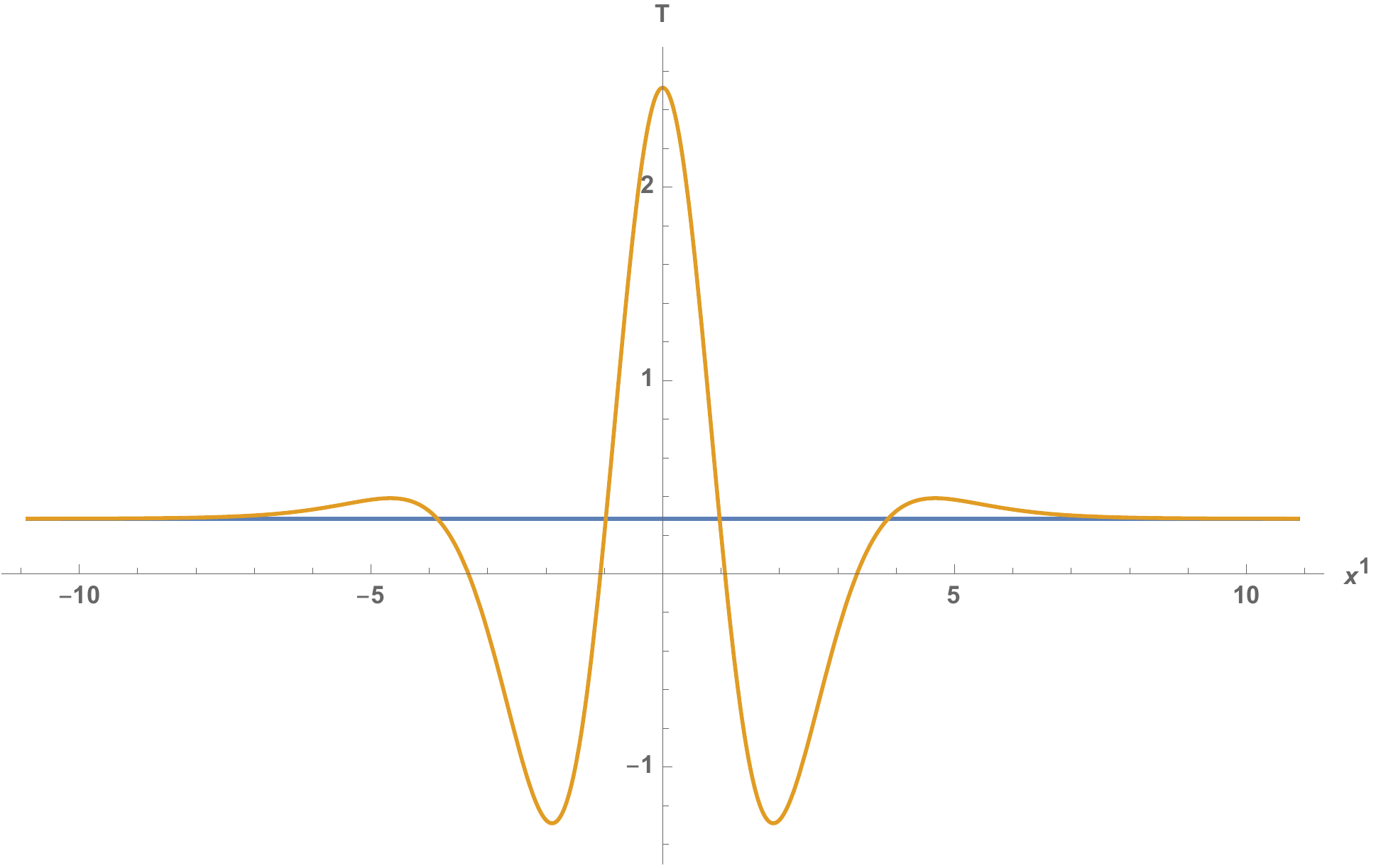}}
\end{center}
\caption{\label{fig:13lumps2} Tachyon profile for the excited lump solution of \cite{KOSsing}.}
\end{figure}

Finally, in figure \ref{fig:13lumps} we give the profiles for the toy model lump and toy model excited lump of~\cite{KOSsing}. The amplitude of the profiles is quite large, especially for the excited lump, as a consequence of the fact that the coefficients are comparatively large. One surprising feature is that the tachyon appears to overshoot the tachyon vacuum, in a similar way as it does for solutions at ghost number~1. However, a careful look reveals that the profiles swing back underneath. This is consistent with the general pattern observed in subsection \ref{subsec:principle} that toy model tachyon profiles tend to asymptotically approach the tachyon vacuum from below. In figure \ref{fig:13lumps2} we give the profile for the true excited lump solution of \cite{KOSsing}. Again, the tachyon overshoots the tachyon vacuum, and it can be checked that it does not swing back underneath. Note that the ``gap" between the two peaks of the excited lump is much more pronounced than for the flag state toy model, shown in figure \ref{fig:excited_lump}.

\section{Three Point Function for the Cosine Deformation}
\label{app:cosine}

In this appendix we compute the boundary three point function
\begin{eqnarray}
\Big\langle\bs_\lambda(u_1)\,e^{i nX^1(u)} \s_\lambda(u_2)\Big\rangle_{\rm UHP}\label{three-cos}
\end{eqnarray}
in the BCFT of a free boson $X^1$ compactified on a circle of unit radius with initially Neumann boundary conditions which are then modified by boundary condition changing operators $\s_\lambda$ and $\bs_\lambda$ which switch on the cosine deformation on a segment of the open string boundary with strength~$\lambda$. The corresponding formulas for the cosh rolling tachyon deformation are obtained by replacing $X^1$ with $i X^0$, with $X^0$ a timelike free boson. Our definition of the marginal coupling constant $\lambda$ agrees with $\widetilde{\lambda}$ of \cite{SenRolling}, and in particular the moduli space of the cosine deformation is a circle with $\lambda\sim\lambda+2$ representing equivalent boundary conditions. The change of boundary condition, however, does not completely specify the boundary condition changing operators. In a fundamental domain $-1< \lambda\leq 1$, we will assume that $\s_\lambda$ and $\bs_\lambda$ are the primaries of the minimal conformal weight implementing the change of boundary condition given by $\lambda$. This implies that the conformal weights are given by $h_\lambda=\left(\frac{\lambda}{2}\right)^2$, but note that this is not invariant under the identification $\lambda\sim\lambda+2$. Therefore, outside the fundamental domain $\s_\lambda$ and $\bs_\lambda$ do not have the lowest conformal weight consistent with the prescribed change of boundary condition.

One can try to compute the three point function by replacing the boundary condition changing operators with the appropriately renormalized exponential boundary interaction,
\begin{eqnarray}
\s_\lambda(a)\bs_\lambda(b)=\exp\left(\lambda\int_a^b dt\, \cos X^1(t)\right),
\end{eqnarray}
and using Wick's theorem. However, a more efficient approach is to take advantage of the symmetry implied by the $SU(2)$ current algebra \cite{Ludwig,Schomerus}
\begin{eqnarray}
j_a(z)j_b(w)=\frac{\frac12 \delta_{ab}}{(z-w)^2}+i \epsilon_{abc}\frac{j_c(z)}{z-w}+(\rm regular).
\end{eqnarray}
The relevant currents are given by 
\begin{eqnarray}
j_1(z)&=&\cos X^1(z),\\
j_2(z)&=&\sin X^1(z),\\
j_3(z)&=&\frac i2 \partial X^1(z).
\end{eqnarray}
The corresponding conformally invariant charges,
\begin{eqnarray}
J_a=\oint \frac{dz}{2\pi i} j_a(z),
\end{eqnarray}
realize the Lie algebra of $SU(2)$:
\begin{eqnarray}
[J_a,J_b]=i \epsilon_{abc}J_c.
\end{eqnarray}
To compute \eqref{three-cos} we use the $SU(2)$ generators to rotate $j_1=\cos X^1$ into the simpler current $j_3=\frac i2 \partial X^1$.
Geometrically this is a clockwise rotation of $\pi/2$ around the $y$-axis (corresponding to $j_2=\sin X^1$) which is realized by the operator
 \begin{eqnarray}
 {\mathbb D}=\exp\left[-i\frac\pi2 J_2\right].\label{D-rot}
 \end{eqnarray}
At the self dual radius $R=1$, the states and operators of the theory decompose into a direct sum of $SU(2)$ irreducible representations of spin $j$. Taking the usual orthonormal basis $|j,m\rangle$ for this representation, where $m$ satisfying $-j\leq m\leq j$ is the component of the ``angular momentum" along the $z$ axis, we have 
 \begin{eqnarray}
 L_0|j,m\rangle&=&j^2|j,m\rangle,\\
 J_3|j,m\rangle&=&m |j,m\rangle.
 \end{eqnarray}
Note that in the present context the eigenvalue $m$ represents the momentum of the state. The vertex operators which create the states $|j,m\rangle$ will be denoted $\phi_{j,m}(z)$. We have the identifications
\begin{equation}e^{\pm i n X^1(z)} = \phi_{n,\pm n}(z),\ \ \ \ j_3(z) = \frac{1}{\sqrt{2}}\phi_{1,0}(z),\ \ \ \ j_1(z) = \frac{1}{2}\big(\phi_{1,1}(z)+\phi_{1,-1}(z)\big).
\end{equation}
The operator ${\mathbb D}$ acts on the state $|j,m\rangle$ through the Wigner $D$-matrix\footnote{In the present case (clockwise rotation by $\pi/2$ around the $y$ axis) the $D$-matrix  elements in the given representation 
 are given in Mathematica as $D^{(j)}_{m,m'}=e^{i\pi/2 j}{\rm WignerD}[\{j,m,m'\},0,\pi/2,\pi/2]$, where the overall phase in front (which is independent of the rotated state in the given representation) is needed to preserve the reality condition  after the rotation.}
\begin{eqnarray}
{\mathbb D}|j,m\rangle=\sum_{m'=-j}^{j} D^{(j)}_{m,m'}|j,m'\rangle.
\end{eqnarray}
In particular this implies 
\begin{eqnarray}
{\mathbb D}\,\cos X^1(z)=\frac i2 \partial X^1(z),
\end{eqnarray}
and therefore 
\begin{eqnarray}
{\mathbb D}(\s_\lambda(b)\,\bs_\lambda(a))&=&{\mathbb D}\exp\left(\lambda\int_a^b ds\, \cos X^1(s)\right)\nonumber\\
&=&\exp\left(\lambda\int_a^b ds\, \frac i2 \partial  X^1(s)\right)=e^{i\frac\lambda 2 X^1}(b)\, e^{-i\frac\lambda 2 X^1}(a).\label{rot-twist}
\end{eqnarray}
This manipulation is somewhat formal since we have not described how to implement the appropriate renormalization of the exponential boundary interactions. But in fact the result can be taken as a definition of $\s_\lambda$ and $\bs_\lambda$, and it follows that they are primaries of weight $h_\lambda=\left(\frac{\lambda}{2}\right)^2$, just like the plane wave vertex operators. Similarly rotating the vertex operator $e^{i n X^1}$ gives 
\begin{eqnarray}
{\mathbb D}\,e^{i n X^1(z)}=\sum_{m'=-n}^{n} D^{(n)}_{n,m'}\phi_{n,m'}(z).\label{rot-test}
\end{eqnarray}
$SU(2)$ invariance of correlators at the self dual radius then implies
\begin{eqnarray}
\Big\langle e^{i n X^1(u)}  \s_\lambda(u_2) \,\bs_\lambda(u_1)\Big\rangle_\mathrm{UHP} &=& \Big\langle{\mathbb{D}}\big( e^{i n X^1(u)}  \s_\lambda(u_2) \,\bs_\lambda(u_1)\big)\Big\rangle_\mathrm{UHP}\nonumber\\
&=&\sum_{m'=-n}^{n} D^{(n)}_{n,m'}\Big\langle \phi_{n,m'}(u)e^{i\frac\lambda 2 X^1(u_2)}\, e^{-i\frac\lambda 2 X^1(u_1)}\Big\rangle_\mathrm{UHP}\nonumber\\
&=&D^{(n)}_{n,0}\,\Big\langle \phi_{n,0}(u)e^{i\frac\lambda 2 X^1(u_2)}\, e^{-i\frac\lambda 2 X^1(u_1)}\Big\rangle_\mathrm{UHP}.
\label{eq:preladder}\end{eqnarray}
In the last line we have used the fact that the rotated boundary condition changing operators together carry zero momentum, and therefore only the $m=0$ component of the sum can contribute.  The operators $\phi_{n,0}(u)$ are the zero-momentum primaries of the free boson CFT. For present purposes the most useful representation of these primaries is given by acting $SU(2)$ ladder operators on the highest weight state\footnote{We thank  Jakub Vo{\v s}mera for this suggestion.}
\begin{eqnarray}
\phi_{n,0}(u)=\frac1{\sqrt{(2n)!}}(J_-)^n e^{i nX^1}(u),
\end{eqnarray}
where 
\begin{eqnarray}
J_\pm=J_1\pm i J_2=\oint \frac{dz}{2\pi i} e^{\pm i X^1}(z).
\end{eqnarray}
Plugging into \eq{preladder} gives 
\begin{eqnarray}\lineup \left\langle\bs_\lambda(u_1)\,e^{i nX^1(u)} \s_\lambda(u_2)\right\rangle_{\rm UHP}\nonumber\\
\lineup\ \ \ \ \ \ =\frac{D^{(n)}_{n,0}}{\sqrt{(2n)!}}\oint_w \frac{dz_1}{2\pi i}\cdots\oint_w \frac{dz_n}{2\pi i}\left\langle e^{-i\frac\lambda 2 X^1}(u_1)\,\prod_{j=1}^n e^{-i X^1}(z_j)e^{i nX^1}(u)\ e^{i\frac\lambda 2 X^1}(u_2)\right\rangle_{\rm UHP}.\ \ \ \ \ \ 
\end{eqnarray}
Evaluating the correlator with Wick's theorem and extracting the residues leads to the formula 
\begin{eqnarray}
\left\langle\bs_\lambda(u_1)\,e^{i nX^1}(u) \s_\lambda(u_2)\right\rangle_{\rm UHP}=\frac{(-1)^n 2\pi\frac{D^{(n)}_{n,0}}{\sqrt{2n!}}\, \left(\prod_{j=0}^{n-1}\frac{(j+1)!}{(n+j)!}\right)
{\cal P}_n(\lambda)}{(u_1-u_2)^{\frac{\lambda^2}{2}-n^2}(u_1-u)^{n^2}(u-u_2)^{n^2}},
\end{eqnarray}
where
\begin{eqnarray}
 {\cal P}_n(\lambda)=\lambda^n\prod_{j=1}^{n-1}(j^2-\lambda^2)^{n-j}
\end{eqnarray}
are polynomials in $\lambda$. Interestingly the roots of this polynomial are at integer $\lambda$ and correspond to the points in the moduli space where the deformed theory has Neumann boundary conditions. A useful consistency check of the above result is that, for $\lambda=1/2$,  it should reduce to the three point function with Neumann-Dirichlet twist operators of weight $1/16$ with the Dirichlet boundary condition at $x^1=\pi$ and at the self dual radius $R=1$:
\begin{eqnarray}
\left\langle\bs_{\lambda=1/2}(u_1)\,e^{i nX^1}(u) \s_{\lambda=1/2}(u_2)\right\rangle_{\rm UHP}=\frac{2\pi e^{in \pi}\,4^{-n^2} }{(u_1-u_2)^{\frac1{8}-n^2}(u_1-u)^{n^2}(u-u_2)^{n^2}}.
\end{eqnarray}
This indeed happens as a consequence of the identity
\begin{eqnarray}
4^{n^2}\,\frac{D^{(n)}_{n,0}}{\sqrt{2n!}}\,{\cal P}_n(1/2)=\prod_{j=0}^{n-1}\frac{(n+j)!}{(j+1)!}.
\end{eqnarray}
This allows us to write the three point function in a slightly more compact form: 
\begin{eqnarray}
\left\langle\bs_\lambda(u_1)\,e^{i nX^1}(u) \s_\lambda(u_2)\right\rangle_{\rm UHP}=\frac{(-1)^n 2\pi 4^{-n^2}\frac
{{\cal P}_n(\lambda)}{{\cal P}_n(1/2)}}{(u_1-u_2)^{\frac{\lambda^2}{2}-n^2}(u_1-u)^{n^2}(u-u_2)^{n^2}}.
\end{eqnarray}
Note that the three-point function is clearly not periodic under $\lambda\sim\lambda+2$. Therefore winding around the moduli space must produce ``excited" boundary condition changing operators which implement the same change of boundary condition but possess higher conformal weight. For example, at $\lambda=-1/2$ we find Neumann-Dirichlet twist operators  $(\s_{\rm ND},\bs_{\rm ND})$ of weight $1/16$ which impose Dirichlet boundary conditions at $x^1=0$. At the equivalent point $\lambda=3/2$ in the moduli space we instead find the excited twist fields $(\tau_{\rm ND},\overline\tau_{\rm ND})$ of weight $9/16$. This observation allows us to extract the three-point function with excited Neumann-Dirichlet twist fields at the self dual radius
\begin{eqnarray}
\left\langle\overline\tau_{\rm ND}(u_1)\,e^{i nX^1}(u)\tau_{\rm ND}(u_2)\right\rangle_{\rm UHP}=\frac{2\pi 4^{-n^2}(1-4n^2) }{(u_1-u_2)^{\frac9{8}-n^2}(u_1-u)^{n^2}(u-u_2)^{n^2}}.
\end{eqnarray}
This readily generalizes to a generic compactification radius $R$ 
\begin{eqnarray}
\left\langle\overline\tau_{\text{ND}}(u_1)\,e^{i \frac nRX^1}(u)\tau_{\text{ND}}(u_2)\right\rangle_{\rm UHP}=\frac{2\pi 4^{-\left(\frac nR\right)^2}(1-4\left(\frac nR\right)^2) }{(u_1-u_2)^{\frac9{8}-(n/R)^2}(u_1-u)^{\left(\frac nR\right)^2}(u-u_2)^{\left(\frac nR\right)^2}}.
\end{eqnarray}
This three point function is used in subsection \ref{subsec:toy} to construct lump solutions based on excited Neumann-Dirichlet twist fields.

\end{appendix}

\end{document}